\documentclass[sn-mathphys,Numbered]{sn-jnl}

\usepackage[utf8]{inputenc}
\usepackage{graphicx}
\usepackage{amsmath,amssymb,amsfonts,bm}
\usepackage{mathtools}
\usepackage{multirow}
\usepackage{booktabs}
\usepackage{array}
\usepackage{tabularx}
\usepackage[title]{appendix}
\usepackage{xcolor}
\usepackage{textcomp}
\usepackage{manyfoot}
\usepackage{siunitx}
\usepackage{float}
\usepackage{subcaption}
\usepackage{geometry}
\usepackage[section]{placeins}
\usepackage{enumitem}
\usepackage{color,soul}
\usepackage{algorithm}
\usepackage{algorithmic}
\usepackage{tikz}
\usetikzlibrary{arrows.meta,calc,positioning}

\usepackage{setspace}
\graphicspath{{figures/}}

\DeclareUnicodeCharacter{00A0}{~}
\DeclareUnicodeCharacter{202F}{\,}
\DeclareUnicodeCharacter{2011}{-}
\DeclareUnicodeCharacter{2013}{--}
\DeclareUnicodeCharacter{2014}{---}
\DeclareUnicodeCharacter{2019}{'}
\DeclareUnicodeCharacter{201C}{``}
\DeclareUnicodeCharacter{201D}{''}
\DeclareUnicodeCharacter{2212}{-}
\DeclareUnicodeCharacter{FB01}{fi}
\DeclareUnicodeCharacter{FB02}{fl}

\sethlcolor{yellow}

\begin{document}

\title[Two-way actuator-line coupling of a finite-volume beam]{Two-way actuator-line coupling of a geometrically exact finite-volume beam with incompressible and free-surface flow}

\author[1]{\fnm{Amirhossein} \sur{Taran}}
\author[1]{\fnm{Seevani} \sur{Bali}}
\author[2]{\fnm{\v{Z}eljko} \sur{Tukovi\'{c}}}
\author[1]{\fnm{Vikram} \sur{Pakrashi}}
\author*[1]{\fnm{Philip} \sur{Cardiff}}\email{philip.cardiff@ucd.ie}

\affil[1]{\orgdiv{School of Mechanical and Materials Engineering}, \orgname{University College Dublin}, \orgaddress{\country{Ireland}}}
\affil[2]{\orgdiv{Faculty of Mechanical Engineering and Naval Architecture}, \orgname{University of Zagreb}, \orgaddress{\country{Croatia}}}

\abstract{\unboldmath
Slender flexible structures such as mooring lines, aquaculture netting, and aquatic vegetation are too thin to be resolved economically by a body-fitted mesh, yet they exchange momentum with the flow in both directions. This paper presents a finite volume actuator-line formulation for the fluid--structure interaction of such structures, in which a geometrically exact Simo--Reissner beam, discretised by a cell-centred finite volume method, is coupled two-way to an incompressible finite volume flow solver through conservative and parallel-consistent transfer operators. The fluid velocity is sampled upstream of each beam control volume, a quasi-steady Morison-type drag closure evaluates the line force, and the equal and opposite reaction is projected back into the fluid momentum equation through a normalised Gaussian kernel that conserves the transferred force. The two subsystems share one discretisation, one time loop, and one domain decomposition, so no external structural code is required, and the OpenFOAM implementation is released publicly. The accuracy of the drag-only closure, and the range of conditions over which it holds, are then quantified against two channel benchmarks and a wave-flume experiment, with all drag coefficients fixed in advance. In a steady confined channel flow the predicted tip deflection of a wall-mounted flexible cantilever falls within the range of the published continuum solutions. For a flexible leaflet in a sinusoidal channel flow the peak tip excursion departs from the reference at the longer forcing period. The leaflet then moves further relative to the surrounding fluid, so the added-mass and history forces, which the closure omits, become important. In the wave flume the simulated stem reproduces the measured mode shape, the intra-wave asymmetry, and the base force, and recovers $71$--$83\%$ of the oscillation range. A quasi-static reconstruction test places the remaining shortfall in the omitted fluid inertia rather than in the structural model.
}


\maketitle

\section{Introduction}
\label{sec:intro}

\noindent
Mooring lines, risers, aquaculture netting, textile yarns, and aquatic vegetation all have a cross-section orders of magnitude smaller than the flow domain they occupy, and they all deform under the flow that loads them. Resolving such a body with a boundary-fitted mesh imposes severe local refinement, and the mesh must then be deformed or regenerated at every time step as the structure moves. The computational cost becomes prohibitive for a canopy of vegetation stems or a net of many strands.

\noindent
Distributed forcing methods avoid that cost by representing the structure through a lower-dimensional description and introducing its action on the fluid as regularised momentum sources on a fixed Eulerian mesh. The immersed boundary method \citep{peskin_flow_1972,peskin_immersed_2002,mittal_immersed_2005,uhlmann_immersed_2005} established the idea for surfaces, and the actuator line method of \citet{sorensen_numerical_2002} reduced it further for slender members by concentrating the hydrodynamic interaction at actuator points along a curve. Developed for wind-turbine wakes, where it captures the momentum exchange on meshes far coarser than blade-resolved calculations \citep{troldborg_numerical_2010}, it has since been refined through improved sampling, smoothing, and generalised actuator-curve representations \citep{churchfield_advanced_2017,jha_actuator_2018,navarro_diaz_actuator_2023}. Nothing in the concept restricts it to rotating blades. \citet{bral_development_2025} adapted it to flexible yarns in air-jet weaving and then extended it into a two-way coupled framework driving a beam model \citep{bral_modeling_2025}, while \citet{schmitt_coupled_2022} coupled an actuator line to a finite element solver in OpenFOAM for flexible turbine blades and fish-farm nets.

\noindent
Two operators govern the accuracy of such a coupling, namely velocity sampling and force regularisation. Sampling may be done at a point, along a line, or as a volume average \citep{sorensen_numerical_2002,merabet_parametric_2019,zormpa_effect_2025}. The choice matters most for drag-dominated members, because the regularised source acts as a local sink that slows the fluid in the same region where the velocity would be sampled, and the drag scales with the square of that velocity, so an underestimate feeds back on itself. \citet{bral_development_2025} broke this feedback by sampling at a fixed distance upstream of the actuator point, outside the momentum-deficit region, and the same concern underlies the sampling and smearing corrections of \citet{churchfield_modeling_2015}, \citet{martinez_tossas_optimal_2017} and \citet{muscari_effective_2024}. Regularisation is needed because a singular point force cannot be applied to any discretisation of finite resolution, whether a finite volume or finite element mesh, a finite difference grid, or a set of particles, without a severe dependence on that resolution. The Gaussian kernel of width $\varepsilon$ that replaces it leaves the source grid-localised when too narrow and weakens the local effect of the line when too wide \citep{martinez_tossas_optimal_2017,navarro_diaz_actuator_2023}. The two operators also fix which discrete properties the coupling preserves, namely the conservation of exchanged momentum, the resolution of the kernel relative to the local grid, and invariance under parallel domain decomposition \citep{bral_development_2025}.

\noindent
On the structural side, the geometrically exact beam theory of \citet{reissner_finite_1981}, extended to three dimensions by Simo, treats finite rotations without linearisation and retains shear and torsion alongside bending and axial stretch. Most implementations of it are finite element, and in coupled simulations the structure is usually solved by a different numerical method from the flow. Numerical studies of flexible vegetation illustrate this, with the stems represented by finite element elastic rods \citep{chen_eulerianlagrangian_2019,elrahi_numerical_2023}, immersed-boundary soft bodies \citep{kim_steady_2024}, direct-forcing immersed-boundary solvers \citep{pruter_comprehensive_2025}, or articulated-body methods \citep{dickinson_modelling_2025}, each coupled to a separate flow solver. The two sides then hold their data differently, are decomposed differently across processors, and must each be verified in their own setting first. One way to remove that separation is to discretise the flow and the structure within the same numerical framework, whether finite element or finite volume. The present work takes the finite volume route, using the cell-centred finite volume discretisation of the same beam theory developed by \citet{bali_cell-centered_2022,bali_finite_2024,bali_beamfoam_2025}, so that the two subsystems share one set of conservation statements, one mesh and field infrastructure, one domain decomposition, and one time loop. Two-way actuator-line coupling has been demonstrated for flexible marine structures, but on finite element or external structural solvers \citep{schmitt_coupled_2022,bral_modeling_2025}, so the sampling, regularisation, and parallel exchange have not been established for a geometrically exact finite volume beam.

\noindent
The hydrodynamic closure adopted here is a quasi-steady Morison-type drag law with no added-mass or history term, and a closure of this kind has an applicability boundary that can be stated in advance. \citet{leclercq_reconfiguration_2018} showed that the wave-induced response of a flexible blade is governed by the combination $CaL/KC$, and that below a threshold of approximately $0.2$ the blade responds quasi-statically in a single spatial mode, while above it the added mass of the surrounding water amplifies the motion and excites higher modes. \citet{jacobsen_waveinduced_2019} confirmed this experimentally for single vegetation mimics and derived drag coefficients by fitting a relative-velocity Morison formulation to the measured base force. Those fitted coefficients carry a documented spread between repeat tests, so a validation that fixes the coefficient in advance should propagate that band through the simulation rather than tune a single value against the response being validated.

\noindent
This paper presents a finite volume actuator-line formulation for the fluid--structure interaction of flexible slender structures, in which a cell-centred finite volume Simo--Reissner beam is coupled two-way to an incompressible finite volume flow solver through conservative and parallel-consistent transfer operators, and in which the accuracy limits of a drag-only Morison closure are quantified against benchmark and wave-flume data. No external structural code is required, because the beam shares the mesh and field infrastructure, the time loop, and the domain decomposition of the flow solver. The sampling and projection operators are formulated to be parallel consistent on a domain-decomposed fluid mesh, and the projection is normalised so that it conserves the transferred resultant force independently of the mesh, the kernel width, and the kernel truncation. The formulation does not depend on the number of fluid phases, and is used unchanged in single-phase channel flow and in a two-phase wave flume. It is verified and validated against three reference problems of increasing difficulty, with every drag coefficient fixed in advance, the two operator settings with a physical rather than a purely numerical influence on the response identified, and the accuracy of the closure together with the range of conditions over which it holds quantified rather than assumed. Verification and validation are used throughout in the standard sense \citep{roache_quantification_1997,oberkampf_roy_2010}, the first established against independent numerical solutions and the second against experiment.

\noindent
Section~\ref{sec:mathmodel} presents the governing equations and the hydrodynamic closure, and Section~\ref{sec:numericalmodel} the finite volume beam discretisation and the coupling operators. Sections~\ref{sec:cantilever} to~\ref{sec:leaflet} verify the structural solver and then the coupling, against a flexible beam in a steady confined channel and a leaflet in a sinusoidal channel flow. Section~\ref{sec:flexveg} validates the coupling against the wave-flume experiment of \citet{jacobsen_waveinduced_2019} on a flexible vegetation stem, and Section~\ref{sec:conclusion} draws the conclusions.


\section{Mathematical model}
\label{sec:mathmodel}

\noindent
The coupled problem consists of a slender flexible structure, represented by a geometrically exact beam, embedded inside an incompressible flow domain. There is no body-fitted mesh around the structure. The beam exchanges momentum with the flow through a distributed volumetric source, and the surrounding Eulerian fluid mesh remains stationary throughout. Boldface symbols denote vector or, where stated, second-order tensor quantities.

\subsection{Fluid subsystem}
\label{sec:fluidSubsystem}

\noindent
The fluid is modelled as an incompressible Newtonian flow governed by
\begin{equation}
\nabla \cdot \mathbf{U} = 0,
\label{eq:continuity}
\end{equation}
\begin{equation}
\frac{\partial (\rho \mathbf{U})}{\partial t}
+ \nabla \cdot (\rho \mathbf{U}\mathbf{U})
= -\nabla p
+ \nabla \cdot \left[ \mu \left(\nabla \mathbf{U} + \nabla \mathbf{U}^{T}\right)\right]
+ \rho \mathbf{g}
+ \mathbf{S}_{m},
\label{eq:momentum}
\end{equation}
where $\mathbf{U}$ is the velocity, $\rho$ the density, $\mu$ the dynamic viscosity, $p$ the pressure, and $\mathbf{g}$ the gravitational acceleration. The term $\mathbf{S}_{m}$ is the volumetric momentum source through which the beam acts on the flow, derived in Section~\ref{sec:almProjection}. The pressure--velocity coupling is handled by the segregated PIMPLE algorithm, which combines the pressure-correction steps of the PISO algorithm \citep{issa_solution_1986} with the outer iterations of the SIMPLE algorithm \citep{patankar_calculation_1972} within each time step. The source $\mathbf{S}_{m}$ is updated once per time step rather than at every outer iteration, as set out in Section~\ref{sec:almParallelSearch}.

\noindent
The coupling operator is formulated independently of the number of fluid phases. The channel-flow benchmarks of Sections~\ref{sec:beamtunnel} and~\ref{sec:leaflet} are single-phase and the wave-flume validation of Section~\ref{sec:flexveg} is two-phase, with the same implementation used unchanged in both. For the two-phase case a volume-of-fluid formulation transports a phase fraction $\alpha$ to distinguish the two immiscible fluids, the system is treated as a single mixture with phase-averaged properties $\rho = \alpha \rho_1 + (1-\alpha)\rho_2$ and $\mu = \alpha \mu_1 + (1-\alpha)\mu_2$, and the phase fraction is advanced by
\begin{equation}
\frac{\partial \alpha}{\partial t}
+ \nabla \cdot (\mathbf{U}\alpha)
+ \nabla \cdot \left(\mathbf{U}_r\,\alpha(1-\alpha)\right) = 0,
\label{eq:alphaTransportCompressed}
\end{equation}
in which the third term is the interface-compression flux that counteracts numerical smearing, with $\mathbf{U}_r$ a compression velocity active only near the interface. Boundedness of $\alpha$ is enforced by the MULES limiter \citep{damian_extended_2014}, and the momentum equation is solved for the modified pressure $p_{\mathrm{rgh}} = p - \rho\,\mathbf{g}\cdot\mathbf{x}$, from which the hydrostatic contribution has been removed.

\subsubsection{Regular wave theories, generation, and absorption}
\label{sec:theory_waves}
The free-surface case of Section~\ref{sec:flexveg} is driven by regular waves prescribed at an inlet boundary. Progressive waves of permanent form on a finite depth $d$ are described by the Stokes expansion, a perturbation series in the steepness $ka$ with $k = 2\pi/\lambda$ the wavenumber and $a = H/2$ the amplitude of a wave of height $H$, whose frequency $\omega = 2\pi/T$ follows from $\omega^2 = g k \tanh kd$ \citep{dean_water_1991}. The adequacy of a given order is judged by the steepness together with the Ursell number $\mathrm{Ur} = H\lambda^2/d^3$, second-order theory being adequate for $\mathrm{Ur} \lesssim 10$ \citep{dean_water_1991}, so where the simulated conditions approach $\mathrm{Ur} \approx 20$ the fifth-order theory of \citet{fenton_fifth_1985} is used instead. The second harmonic of the expansion sharpens the crests, flattens the troughs, and makes the crest-phase velocity larger in magnitude but shorter in duration than the trough-phase velocity. This asymmetry matters for drag-driven structures, because a quadratic drag force rectifies it into a net forcing asymmetry over the cycle.

\noindent
The chosen theory is imposed at the inlet as time-dependent boundary values of velocity and phase fraction, ramped smoothly from rest, combined with active absorption that cancels waves reflected back toward the inlet \citep{higuera_realistic_2013}. At the outlet, a shallow-water absorption condition removes the outgoing wave using the linear long-wave celerity $\sqrt{gd}$. Because that celerity is exact only in the long-wave limit, the outlet absorption degrades as the relative depth $kd$ approaches unity, which is relevant to one of the validation conditions of Section~\ref{sec:flexveg}.

\subsection{Beam subsystem}
\label{sec:beamSection}

\noindent
The structure is modelled as a one-dimensional \emph{Simo--Reissner} beam whose centreline can undergo arbitrarily large three-dimensional translations and rotations without any small-rotation linearisation, and which retains shear deformation and torsion in addition to bending and axial stretch. A \emph{total Lagrangian} description is adopted, in which every kinematic quantity is referred back to the fixed reference configuration through the arc-length coordinate $s \in [0,L]$. The initial stress-free centreline is $\mathbf{r}_0(s)$ and the current deformed centreline is $\mathbf{r}(s,t) = \mathbf{r}_0(s) + \mathbf{w}(s,t)$, and the orientation of the cross-section at each $s$ is represented by a rotation tensor $\boldsymbol{\Lambda}(s,t) \in SO(3)$. Finite rotations are parameterised through a rotation vector and the exponential map $\boldsymbol{\Lambda}(\boldsymbol{\psi}) = \exp(\widehat{\boldsymbol{\psi}})$, with $\widehat{(\cdot)}$ the skew-symmetric matrix representing the cross product, and the Newton iteration updates them multiplicatively so that every iterate stays exactly on $SO(3)$ \citep{dai_eulerrodrigues_2015,spurrier_comment_1978}.

\noindent
The strain measures are the co-rotated translational strain $\boldsymbol{\Gamma} = \boldsymbol{\Lambda}^T \mathbf{r}' - \boldsymbol{\Lambda}_0^T \mathbf{r}_0'$, whose components are axial extension and shear, and the curvature $\mathbf{K}$, whose components are torsion and bending, with $(\cdot)' = \partial(\cdot)/\partial s$ and $\boldsymbol{\Lambda}_0(s)$ the rotation tensor of the initial configuration. Both are objective, since rotating each quantity back into the local body-attached basis removes any superposed rigid-body motion. A hyperelastic model quadratic in the strain measures gives the material resultants $\mathbf{N} = \mathbf{C}_N \boldsymbol{\Gamma}$ and $\mathbf{M} = \mathbf{C}_M \mathbf{K}$, with the diagonal stiffness matrices $\mathbf{C}_N = \mathrm{diag}(EA,\,GA_2,\,GA_3)$ and $\mathbf{C}_M = \mathrm{diag}(GJ,\,EI_2,\,EI_3)$ appropriate for the bi-symmetric cross-sections used in every case reported here, and the spatial resultants follow by push-forward, $\mathbf{n} = \boldsymbol{\Lambda}\mathbf{N}$ and $\mathbf{m} = \boldsymbol{\Lambda}\mathbf{M}$. The kinematics, the strain measures, and the linearisation of the resultants are set out in full by \citet{bali_cell-centered_2022,bali_finite_2024}, and only what the coupling needs is repeated here.

\noindent
The balance of linear and angular momentum along the centreline reads
\begin{equation}
\mathbf{n}' + \mathbf{f} \;=\; \rho_{b}A\,\ddot{\mathbf{w}},
\label{eq:ForceBalanceDynamic}
\end{equation}
\begin{equation}
\mathbf{m}' + \mathbf{r}'\times\mathbf{n} + \mathbf{t} \;=\;
\boldsymbol{\Lambda}\,\bigl(\mathbf{I}_{\rho}\,\dot{\boldsymbol{\Omega}}
\;+\; \boldsymbol{\Omega}\times\bigl(\mathbf{I}_{\rho}\,\boldsymbol{\Omega}\bigr)\bigr),
\label{eq:MomentBalanceDynamic}
\end{equation}
where $\mathbf{f}$ and $\mathbf{t}$ are the external distributed force and torque per unit length, $\rho_{b}A$ is the mass per unit reference length, $\ddot{\mathbf{w}}$ the translational acceleration of the centreline, $\boldsymbol{\Omega}$ the material angular velocity defined through $\widehat{\boldsymbol{\Omega}} = \boldsymbol{\Lambda}^{T}\dot{\boldsymbol{\Lambda}}$, and $\mathbf{I}_{\rho} = \rho_{b}\,\mathrm{diag}\left(J,\;I_{2},\;I_{3}\right)$ the material rotary-inertia tensor. The right-hand side of Eq.~\eqref{eq:MomentBalanceDynamic} is the push-forward of the material angular-momentum rate and contains both the rotary-inertia and the gyroscopic term. Setting the right-hand sides to zero recovers the quasi-static form used for the static benchmark of Section~\ref{sec:cantilever}. No physical structural damping is added in any simulation reported here, and the hydrodynamic drag loading provides the physical damping of the coupled response.

\subsection{Hydrodynamic closure on the beam}
\label{sec:beamExternalLoading}

\noindent
The distributed external force per unit length in Eq.~\eqref{eq:ForceBalanceDynamic} comprises a hydrodynamic drag force and a net buoyancy force, $\mathbf{f} = \mathbf{f}_{d} + \mathbf{f}_{b}$. The drag follows the quasi-steady Morison formulation \citep{morison_force_1950}, split into tangential and normal components relative to the local beam tangent,
\begin{equation}
\mathbf{f}_{d} \;=\; \tfrac{1}{2}\,\rho_{f}\,d\,
\bigl(C_{dt}\,\lVert\mathbf{V}_{r,t}\rVert\,\mathbf{V}_{r,t}
\;+\; C_{dn}\,\lVert\mathbf{V}_{r,n}\rVert\,\mathbf{V}_{r,n}\bigr),
\label{eq:fext_drag}
\end{equation}
where $\rho_{f}$ is the fluid density, $d$ the cross-section drag reference length, $C_{dt}$ and $C_{dn}$ the tangential and normal drag coefficients, and $\mathbf{V}_{r,t}$ and $\mathbf{V}_{r,n}$ the tangential and normal projections of the relative velocity between the beam and the surrounding fluid. That surrounding velocity is the upstream-sampled velocity provided by the operator of Section~\ref{sec:almCouplingTheory}, and the discrete form of the line force is given in Section~\ref{sec:almLineForce}. The net effect of gravity and buoyancy on a submerged segment is
\begin{equation}
\mathbf{f}_{b} \;=\; \rho_{b}\,A\,\!\left(\frac{\rho_{b}-\rho_{f}}{\rho_{b}}\right)\mathbf{g}.
\label{eq:fext_buoyancy}
\end{equation}

\noindent
The closure carries no added-mass term. When a beam accelerates through the surrounding fluid, the displaced fluid exerts an additional inertial reaction, which the classical Morison formulation represents as
\begin{equation}
\mathbf{f}_{a} \;=\; \rho_{f}\,A\,\bigl(
C_{Mt}\,\mathbf{a}_{r,t}
\;+\; C_{Mn}\,\mathbf{a}_{r,n}
\;+\; \mathbf{a}_{f}\bigr),
\label{eq:fext_addedmass}
\end{equation}
with $\mathbf{a}_{r} = \mathbf{a}_{b} - \mathbf{a}_{f}$ the relative acceleration, $C_{Mt}$ and $C_{Mn}$ the added-mass coefficients, and the third term the Froude--Krylov contribution carried by the undisturbed fluid acceleration. Equation~\eqref{eq:fext_addedmass} is not evaluated in the present model, so $C_{Mt} = C_{Mn} = 0$ throughout and the closure is quasi-steady and drag-only.

\noindent
Equations~\eqref{eq:fext_drag} and~\eqref{eq:fext_buoyancy} are therefore a modelling approximation of the hydrodynamic load rather than an exact expression for it, since a body-fitted calculation obtains the force by integrating the resolved pressure and viscous traction over the wetted surface and so carries the added-mass and history contributions that this closure omits. That omission is a deliberate scope decision, and its size, together with the range of conditions in which the approximation remains acceptable, is quantified against benchmark and experimental data in Sections~\ref{sec:leaflet} and~\ref{sec:flexveg} rather than assumed.


\section{Numerical model}
\label{sec:numericalmodel}

\subsection{Finite volume discretisation of the beam}
\label{sec:beamFVDiscretisation}

\noindent
The beam domain $s\in[0,L]$ is discretised into non-overlapping one-dimensional control volumes (CVs). The centreline position and the rotation parameters are stored at the CV centres, the spatial resultants $(\mathbf{n},\mathbf{m})$ are evaluated at the faces by linear interpolation of the cell-centred kinematic quantities, and the external loading is collocated at the cell centre \citep{bali_cell-centered_2022}. Integrating Eqs.~\eqref{eq:ForceBalanceDynamic}--\eqref{eq:MomentBalanceDynamic} over a cell $P$ of length $L_{P}$, bounded by the faces $w$ and $e$ and neighboured by the cells $W$ and $E$, gives the locally conservative discrete form
\begin{align}
\mathbf{n}_{e} - \mathbf{n}_{w} + \mathbf{f}_{P}\,L_{P} &= \rho_{b}A\,L_{P}\,\ddot{\mathbf{w}}_{P}, \label{eq:fvForceBalance} \\
\mathbf{m}_{e} - \mathbf{m}_{w} + \bigl(\mathbf{r}'\times\mathbf{n}\bigr)_{P}\,L_{P} + \mathbf{t}_{P}\,L_{P} &= L_{P}\,\boldsymbol{\Lambda}_{P}\bigl(\mathbf{I}_{\rho}\dot{\boldsymbol{\Omega}}_{P} + \boldsymbol{\Omega}_{P}\times(\mathbf{I}_{\rho}\boldsymbol{\Omega}_{P})\bigr), \label{eq:fvMomentBalance}
\end{align}
with the cell-centre source terms already using the midpoint-rule approximation of the volume integral. The face resultants are nonlinear functions of the kinematic unknowns at cells $P$, $W$, and $E$.

\noindent
The unknowns at each CV centre are organised as the six-component increment $\boldsymbol{\phi}_P = (\Delta\mathbf{w}_P,\,\Delta\boldsymbol{\psi}_P)^{T}$ between successive Newton iterates, so the translational and rotational rows are solved simultaneously. Linearising about the current iterate gives the three-point finite volume form
\begin{equation}
\mathbf{A}_{W}\,\boldsymbol{\phi}_{W} \;+\; \mathbf{A}_{P}\,\boldsymbol{\phi}_{P} \;+\; \mathbf{A}_{E}\,\boldsymbol{\phi}_{E} \;=\; \mathbf{b}_{P},
\label{eq:fvDiscreteSystem}
\end{equation}
in which the coefficients are dense $6\times 6$ blocks mixing the translation and rotation rows, which the moment-balance cross-product $\mathbf{r}'\times\mathbf{n}$ couples tightly, and $\mathbf{b}_{P}$ is the residual of cell $P$. Assembling over all CVs yields a sparse block-tridiagonal system of size $6N_b\times 6N_b$, where $N_b$ is the number of beam cells. It is solved by a Newton--Raphson iteration in which the kinematic fields are updated cell by cell,
\begin{equation}
\mathbf{w}^{(k+1)} \;=\; \mathbf{w}^{(k)} + \Delta\mathbf{w}^{(k)},
\qquad
\boldsymbol{\Lambda}^{(k+1)} \;=\; \exp\!\left(\widehat{\Delta\boldsymbol{\psi}^{(k)}}\right)\,\boldsymbol{\Lambda}^{(k)},
\label{eq:beamNewtonUpdate}
\end{equation}
until both the increment norm and the residual norm fall below prescribed tolerances. The tangent coefficients and the verification of the discretisation are given by \citet{bali_cell-centered_2022,bali_finite_2024}, and its OpenFOAM implementation by \citet{bali_beamfoam_2025}. All transient results reported here integrate the acceleration terms with a first-order backward scheme, in which the current-level unknown appears linearly, so the inertia term contributes a diagonal-block coefficient of $\rho_{b}A\,L_{P}/\Delta t^{2}$ to $\mathbf{A}_{P}$, the history terms enter the residual, and the block-tridiagonal structure is unchanged. Time steps between $10^{2}$ and $4\times10^{4}$ steps per forcing period are used, and each transient case is started from rest.

\subsection{Actuator-line beam--fluid coupling}
\label{sec:almCouplingTheory}

\noindent
The beam centreline carries a set of \emph{actuator points} $\mathbf{P}_{a,i}$, one per beam control volume, at which the kinematic state of the beam is sampled and the hydrodynamic interaction is concentrated. Each actuator point follows the beam-cell centre as the structure deforms, the surrounding Eulerian mesh remains stationary, and the hydrodynamic exchange is realised through the volumetric source $\mathbf{S}_{m}$ of Eq.~\eqref{eq:momentum}. The strategy follows the centreline-forcing approach of \citet{bral_development_2025,bral_modeling_2025}, and extends it in two ways. The structural side is the cell-centred finite volume beam of Section~\ref{sec:beamFVDiscretisation} rather than a finite element discretisation, so the beam unknowns reside on the same data structures as the fluid unknowns and are updated within the same outer time loop, and the sampling and projection operators are implemented in a parallel-consistent form that scales to domain-decomposed fluid meshes (Section~\ref{sec:almParallelSearch}).

\noindent
Two transfer operations are needed at each time step. The first samples the local fluid velocity near each actuator point and uses it to evaluate the hydrodynamic line force on the beam. The second projects the equal and opposite reaction back onto the fluid mesh as a regularised volumetric source. They are treated differently by design. Velocity sampling is confined to a single fluid cell per beam cell to avoid contamination from the modified flow induced by the source itself, whereas force projection is smoothed over neighbouring cells through a Gaussian kernel, because a singular line force cannot be imposed on a discretisation of finite resolution without severe grid dependence.

\subsubsection{Centreline coupling conditions}
In body-fitted FSI the kinematic and dynamic coupling conditions are imposed at a resolved fluid--solid surface, where velocities match and tractions are equal and opposite. The centreline formulation replaces this surface exchange with two transfer operations on the actuator points. Kinematically, each actuator point tracks the position and velocity of its beam cell,
\begin{equation}
\mathbf{P}_{a,i}(t) \;=\; \mathbf{P}_{a,i}^{\,0} + \mathbf{W}_{i}(t),
\qquad
\dot{\mathbf{P}}_{a,i}(t) \;=\; \mathbf{U}_{b,i}(t),
\label{eq:almKinematicCoupling}
\end{equation}
where $\mathbf{P}_{a,i}^{\,0}$ is the initial position, $\mathbf{W}_{i}$ the translational displacement of beam cell $i$, and $\mathbf{U}_{b,i}$ the beam-cell velocity. No no-slip condition is imposed, because the beam surface is not represented in the fluid mesh. The rotation field enters only implicitly, through the local tangent used to decompose the relative velocity. Dynamically, the sampled relative velocity gives a line force $\mathbf{f}_{i}$ per unit beam length acting on the beam, and the fluid receives the equal and opposite force distributed as a regularised source, so the discrete coupling satisfies the action--reaction condition in integral form,
\begin{equation}
\sum_{i} \mathbf{f}_{i}\,L_{i} \;+\; \sum_{j} \mathbf{S}_{m,j}\,V_{j} \;\approx\; \mathbf{0},
\label{eq:almActionReaction}
\end{equation}
with $L_{i}$ the beam-cell length and $V_{j}$ the volume of fluid cell $j$.

\noindent
Equation~\eqref{eq:almActionReaction} is a statement about the transferred load alone, and not a momentum balance over a region of material. Each subsystem carries its own inertia in its own momentum equation, and those terms are untouched by the exchange, so the relation plays the same role as the equality of tractions across a resolved interface in body-fitted FSI, where the interface itself carries no mass. For the \emph{resultant} force the exchange is conservative, because the normalisation of Section~\ref{sec:almProjection} rescales the projected source so that its discrete volume integral matches the force carried by the actuator segments. What remains approximate is the spatial distribution. The kernel spreads the reaction over a finite region rather than along the centreline, so the moment of the applied source about a given point differs from that of the line load by an amount that scales with the kernel width, and the angular-momentum transfer differs with it.

\begin{figure}[!ht]
\centering
\resizebox{0.80\textwidth}{!}{%
\begin{tikzpicture}[x=1cm,y=1cm,>=Latex,font=\normalsize]
  \tikzstyle{block}=[draw,rounded corners=3pt,minimum width=7.5cm,minimum height=1.55cm,align=center,very thick]
  \tikzstyle{steplabel}=[draw,circle,thick,fill=white,inner sep=0.5pt,minimum size=0.62cm,font=\small\bfseries]

  \node[block,fill=blue!8]    (fluid) at (0, 4.0) {\textbf{Fluid finite volume solver}\\Eulerian mesh, Eqs.~(\ref{eq:continuity})--(\ref{eq:momentum})};
  \node[block,fill=gray!10]   (alm)   at (0, 0.0) {\textbf{Actuator-line coupling operator}\\velocity sampling and force projection};
  \node[block,fill=orange!10] (beam)  at (0,-4.0) {\textbf{Finite volume Simo--Reissner beam solver}\\centreline mesh, Eqs.~(\ref{eq:fvForceBalance})--(\ref{eq:fvMomentBalance})};

  \draw[->,very thick,blue!70!black]
       ([xshift=-1.3cm]fluid.south) -- ([xshift=-1.3cm]alm.north)
       node[midway,left,font=\small,align=right]
       {sampled fluid velocity\\$\mathbf{U}_{c(\mathbf{P}_{s,i})}$};
  \node[steplabel] at (-1.3, 3.0) {3};
  \draw[->,very thick,red!75!black]
       ([xshift=1.3cm]alm.north) -- ([xshift=1.3cm]fluid.south)
       node[midway,right,font=\small,align=left]
       {reaction source\\$\mathbf{S}_{m,j}$};
  \node[steplabel] at (1.3, 1.0) {4};

  \draw[->,very thick,orange!80!black]
       ([xshift=-1.3cm]beam.north) -- ([xshift=-1.3cm]alm.south)
       node[midway,left,font=\small,align=right]
       {actuator-point kinematics\\$\mathbf{P}_{a,i},\;\dot{\mathbf{P}}_{a,i}$};
  \node[steplabel] at (-1.3,-3.0) {2};
  \draw[->,very thick,purple!80!black]
       ([xshift=1.3cm]alm.south) -- ([xshift=1.3cm]beam.north)
       node[midway,right,font=\small,align=left]
       {line load\\$\mathbf{f}_{i}\,L_{i}$};
  \node[steplabel] at (1.3,-1.0) {5};

  \node[draw,thick,fill=yellow!25,rounded corners=2pt,
        minimum width=2.8cm,minimum height=0.75cm,font=\small,align=center]
       (start) at (-7.0,-4.0) {\textbf{start of}\\\textbf{time step}};
  \draw[->,very thick,black] (start.east) -- (beam.west);
  \node[steplabel] at (-5.0,-4.0) {1};

  \node[draw,thick,fill=yellow!25,rounded corners=2pt,
        minimum width=2.8cm,minimum height=0.75cm,font=\small,align=center]
       (next) at ( 7.0, 4.0) {\textbf{next}\\\textbf{time step}};
  \draw[->,very thick,black] (fluid.east) -- (next.west);
  \node[steplabel] at ( 5.0, 4.0) {6};
\end{tikzpicture}
}
\caption{Centreline coupling drawn as a per-time-step data-flow diagram between the three solver components. \protect\textcircled{1} The beam solver advances the structural state and produces the current actuator-point positions and velocities; \protect\textcircled{2} these pass to the coupling operator; \protect\textcircled{3} the fluid solver supplies the upstream-sampled velocity at every actuator point; the operator evaluates the hydrodynamic line force and returns it as \protect\textcircled{4} a regularised volumetric reaction source to the fluid solver and \protect\textcircled{5} an equal and opposite line load to the beam solver; \protect\textcircled{6} the fluid pressure--velocity system is advanced before the next time step. The sequence runs once per time step, on the first outer iteration of the PIMPLE loop, so the reaction source of \protect\textcircled{4} stays fixed for the remaining outer iterations.}
\label{fig:almCouplingSchematic}
\end{figure}
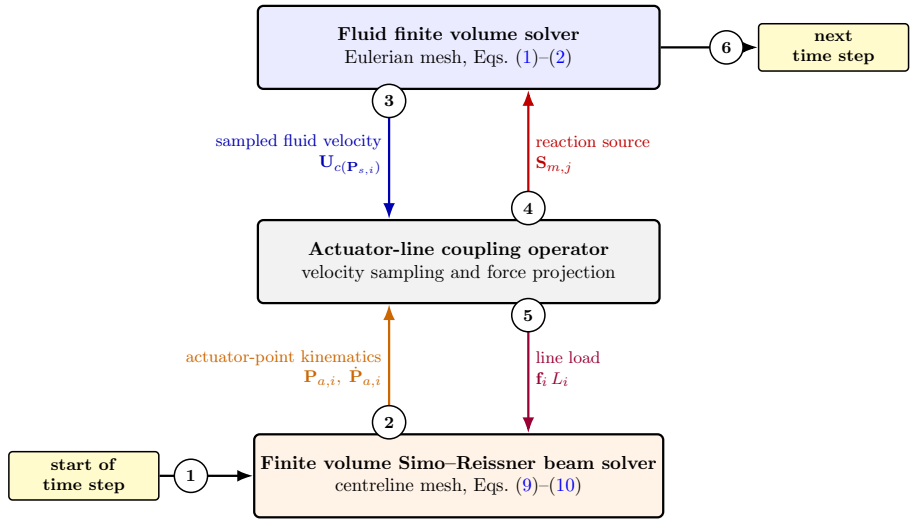

\noindent
Figure~\ref{fig:almCouplingSchematic} summarises the information exchange between the three components.

\subsubsection{Actuator points and velocity sampling}
\label{sec:almSampling}
The actuator-point position is reconstructed from three quantities stored on the beam mesh,
\begin{equation}
\mathbf{P}_{a,i}
\;=\;
\mathbf{C}^{b}_{i}
\;+\;
\mathbf{W}^{\mathrm{ref}}_{i}
\;+\;
\mathbf{W}_{i},
\label{eq:almActuatorPoint}
\end{equation}
where $\mathbf{C}^{b}_{i}$ is the initial beam-cell centre, $\mathbf{W}^{\mathrm{ref}}_{i}$ a fixed offset that places the actuator point at a chosen material location relative to the cell centre (typically zero, but retained so the point can be shifted onto, for example, the upper surface of a non-axisymmetric cross-section), and $\mathbf{W}_{i}$ the current displacement. The actuator-point velocity is the time derivative of Eq.~\eqref{eq:almActuatorPoint} and equals the beam-cell translational velocity, because the first two terms are time-independent. The local tangent $\mathbf{e}_{t,i}$ is reconstructed from a Hermite spline through the current beam-cell centres, and is needed both to decompose the relative velocity and to define the upstream sampling direction.

\begin{figure}[!ht]
\centering
\includegraphics[width=0.92\textwidth]{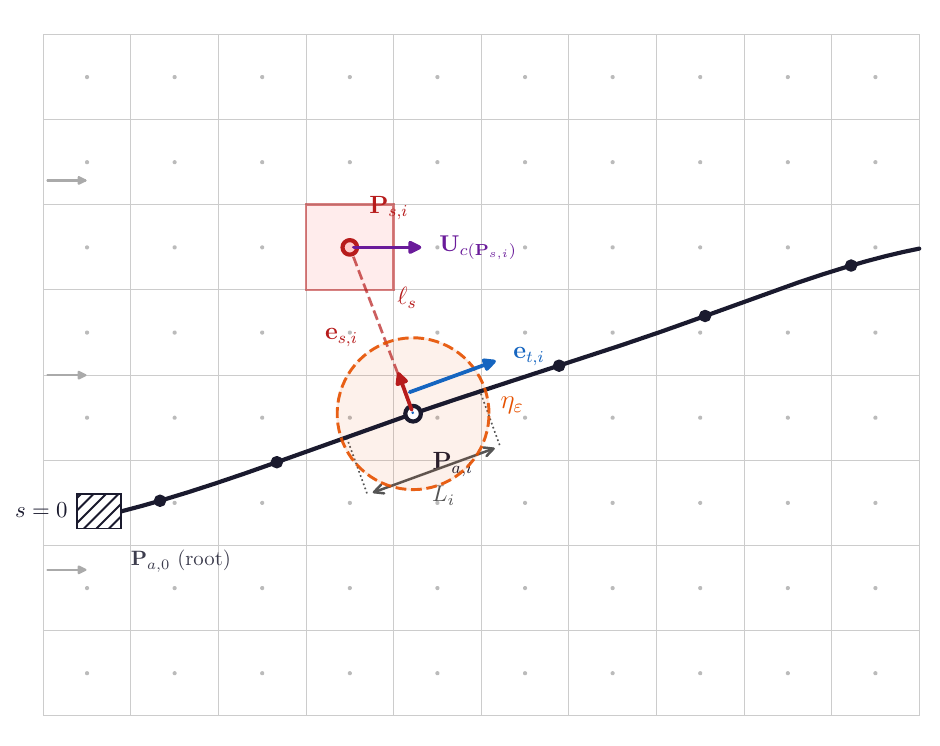}
\caption{Actuator-line velocity sampling. The beam cell centre defines the actuator point $\mathbf{P}_{a,i}$; the tangent $\mathbf{e}_{t,i}$ and upstream direction $\mathbf{e}_{s,i}$ locate the sampling point $\mathbf{P}_{s,i}$ at distance $\ell_s$. The highlighted CFD cell supplies the sampled velocity, and the dashed circle indicates the Gaussian regularisation support $\eta_{\varepsilon}$ used for the reaction force projection. The uniform square grid is used for illustration only, and the procedure applies unchanged to unstructured meshes of general polyhedral cells.}
\label{fig:almSamplingSchematic}
\end{figure}

\noindent
The sampling point is placed at a fixed distance upstream of the actuator point, in a direction reconstructed locally from the incoming flow, as Figure~\ref{fig:almSamplingSchematic} shows. A preliminary relative velocity is first evaluated at the actuator point itself,
\begin{equation}
\mathbf{u}^{0}_{r,i}
\;=\;
\mathbf{U}_{c(\mathbf{P}_{a,i})}
\;-\;
\dot{\mathbf{P}}_{a,i},
\label{eq:almPreliminaryVelocity}
\end{equation}
where $\mathbf{U}_{c(\mathbf{P}_{a,i})}$ is the velocity stored in the single fluid cell containing $\mathbf{P}_{a,i}$ and $c(\cdot)$ denotes the point-to-cell search operator. This value is used only to orient the sampling stencil, not as the line-force velocity, because the modified flow induced by the source near the beam would contaminate the force estimate. Its component normal to the beam tangent, $\mathbf{u}^{0}_{n,i} = \mathbf{u}^{0}_{r,i} - (\mathbf{u}^{0}_{r,i}\!\cdot\!\mathbf{e}_{t,i})\mathbf{e}_{t,i}$, fixes the upstream direction $\mathbf{e}_{s,i} = -\mathbf{u}^{0}_{n,i}/\lVert\mathbf{u}^{0}_{n,i}\rVert$, and the sampling point follows as
\begin{equation}
\mathbf{P}_{s,i}
\;=\;
\mathbf{P}_{a,i}
\;+\;
\bigl[\mathbf{I} - \mathbf{n}_{p}\mathbf{n}_{p}^{T}\bigr]\,\ell_{s}\,\mathbf{e}_{s,i},
\label{eq:almSamplingPoint}
\end{equation}
where $\ell_{s}$ is the sampling distance and $\mathbf{n}_{p}$ a prescribed plane normal used to flatten the stencil onto a chosen plane in three-dimensional simulations. When $\lVert\mathbf{u}^{0}_{n,i}\rVert$ falls below a small threshold, for example because the beam is nearly aligned with the flow or the flow has stagnated, the unit vector becomes ill-conditioned and the implementation switches to a fallback direction, namely the user-prescribed global upstream direction projected onto the local cross-section plane and renormalised. Because the projector in Eq.~\eqref{eq:almSamplingPoint} removes the component of $\mathbf{e}_{s,i}$ along $\mathbf{n}_{p}$ without renormalising, the realised offset is bounded above by $\ell_{s}$ and equals it only when $\mathbf{e}_{s,i}$ already lies in the sampling plane, which holds in every configuration studied here. The distance itself is set as $\ell_{s} = r_{s}^{\star}L_{\mathrm{ref}}$, with $L_{\mathrm{ref}}$ the beam radius for circular cross-sections and a representative cross-section dimension otherwise, so its dimensional value is problem-dependent rather than universal. Its influence on the response is examined directly in Section~\ref{sec:beamtunnel_sensitivity}.

\noindent
Once the sampling point has been located, the fluid cell containing it is found by the same operator, and the velocity stored in that single cell defines the relative velocity used by the line-force model,
\begin{equation}
\mathbf{U}_{r,i}
\;=\;
\mathbf{U}_{c(\mathbf{P}_{s,i})}
\;-\;
\dot{\mathbf{P}}_{a,i}.
\label{eq:almRelativeVelocity}
\end{equation}
No averaging across neighbouring cells is performed. This is a deliberate simplification compared with integral or line-averaged sampling, which removes the dependence on a sampling-volume definition and keeps the sampling local, and which avoids using the velocity at the actuator point itself, where the regularised source slows the flow and would bias the drag estimate downward.

\subsubsection{Line-force evaluation}
\label{sec:almLineForce}
The relative velocity of Eq.~\eqref{eq:almRelativeVelocity} is split into a tangential component $\mathbf{U}_{t,i} = (\mathbf{U}_{r,i}\!\cdot\!\mathbf{e}_{t,i})\mathbf{e}_{t,i}$, which controls the skin-friction contribution, and a normal component $\mathbf{U}_{n,i} = \mathbf{U}_{r,i} - \mathbf{U}_{t,i}$, which controls the cross-flow drag that dominates for slender bluff bodies. The two are combined into the quasi-steady line force per unit beam length,
\begin{equation}
\mathbf{f}^{*}_{i}
\;=\;
-\,\tfrac{1}{2}\,\rho_{f}\,d\,
\bigl\lVert\mathbf{U}_{r,i}\bigr\rVert
\bigl(C_{dn}\mathbf{U}_{n,i} + C_{dt}\mathbf{U}_{t,i}\bigr),
\label{eq:almLineForce}
\end{equation}
which is the discrete form of Eq.~\eqref{eq:fext_drag}, with the negative sign ensuring that the force opposes the relative motion of the fluid with respect to the beam. The raw value is sensitive to small fluctuations of the sampled velocity between successive coupling updates, which can excite spurious oscillations in the structural response, so a first-order relaxation against the value from the previous coupling update stabilises the coupling,
\begin{equation}
\mathbf{f}_{i}
\;=\;
\alpha_{f}\,\mathbf{f}^{*}_{i}
\;+\;
(1 - \alpha_{f})\,\mathbf{f}^{\,\mathrm{prev}}_{i},
\label{eq:almForceRelaxation}
\end{equation}
with $\alpha_{f}\in(0,1]$ set to unity to recover the un-relaxed update. Because the line force is evaluated once per time step (Section~\ref{sec:almParallelSearch}), $\mathbf{f}^{\,\mathrm{prev}}_{i}$ is the relaxed force of the previous time step. On the beam side the relaxed value is multiplied by the cell length before being added to the structural residual of Eq.~\eqref{eq:fvForceBalance}, and on the fluid side it enters the regularised source with the opposite sign.

\subsubsection{Projection to the fluid mesh}
\label{sec:almProjection}
\begin{figure}[!ht]
\centering
\includegraphics[width=0.90\textwidth]{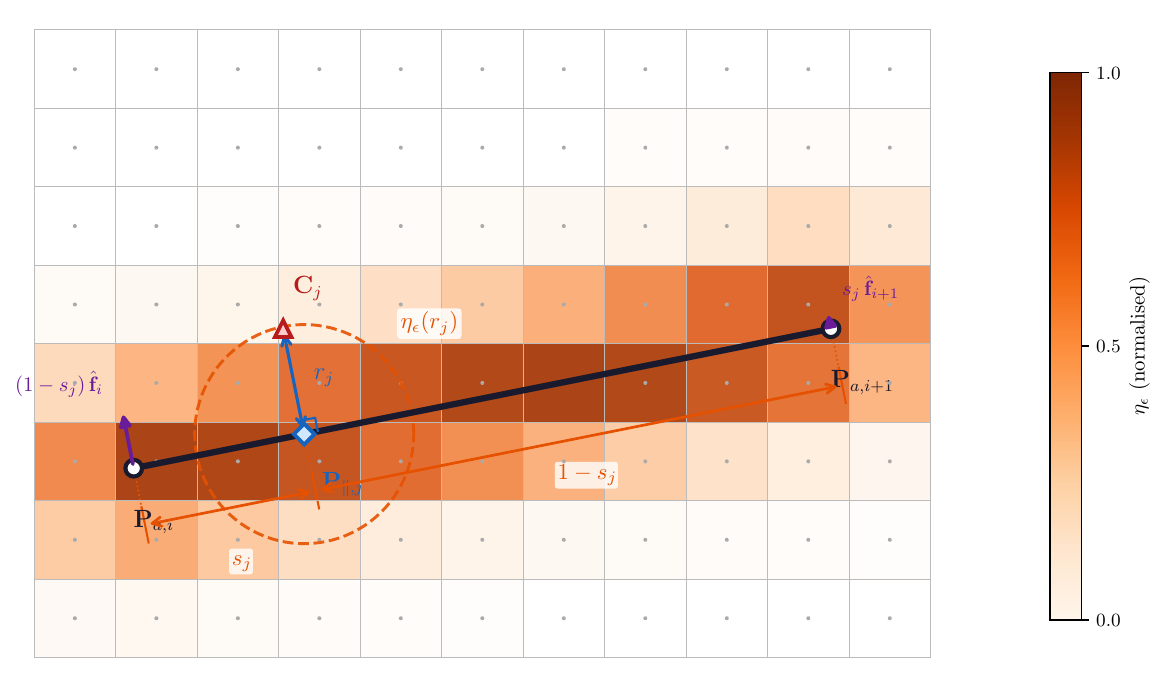}
\caption{Regularised force projection from an actuator segment to the fluid mesh. The CFD cells are coloured by the normalised Gaussian weight $\eta_{\varepsilon}$ centred on the projection point $\mathbf{P}_{\parallel,j}$. The source contribution to cell $j$ comes from linear interpolation of the normalised nodal forces along the segment, weighted by $\eta_{\varepsilon}(r_j)$. As in Figure~\ref{fig:almSamplingSchematic}, the uniform grid is for illustration only.}
\label{fig:almProjectionSchematic}
\end{figure}

Only the fluid cells near the actuator line take part in the projection. A cell whose centre $\mathbf{C}_{j}$ lies outside the bounding box of the actuator points, inflated by $3\varepsilon$ in every direction, is skipped before any distance is evaluated. For the remaining cells, the closest actuator segment is identified by evaluating the distance from $\mathbf{C}_{j}$ to each segment of the actuator line, and the cost of this step is discussed in Section~\ref{sec:almParallelSearch}. If the closest segment lies between $\mathbf{P}_{a,i}$ and $\mathbf{P}_{a,i+1}$, the foot of the perpendicular from $\mathbf{C}_{j}$ onto the segment defines the projection point
\begin{equation}
\mathbf{P}_{\parallel,j}
\;=\;
(1-s_{j})\,\mathbf{P}_{a,i} \;+\; s_{j}\,\mathbf{P}_{a,i+1},
\qquad
0\le s_{j}\le 1,
\label{eq:almProjectionPoint}
\end{equation}
and the radial distance to the segment is $r_{j} = \lVert\mathbf{C}_{j} - \mathbf{P}_{\parallel,j}\rVert$. The line force is spread over the surrounding cells with a two-dimensional Gaussian kernel,
\begin{equation}
\eta_{\varepsilon}(r_{j})
\;=\;
\frac{1}{\pi\varepsilon^{2}}\exp\!\left[-\!\left(\frac{r_{j}}{\varepsilon}\right)^{\!2}\right],
\label{eq:almGaussian}
\end{equation}
where $\varepsilon$ is the regularisation width, whose influence is examined in Section~\ref{sec:beamtunnel_sensitivity}. The kernel is truncated at $r_{j} = 3\varepsilon$, so cells farther than $3\varepsilon$ from the actuator line receive no source. The part of the kernel discarded by this truncation is a fraction $e^{-9} \approx 1.2\times10^{-4}$ of its integral, and the normalisation below returns it to the retained cells, so the truncation does not change the transferred force. The geometry of the projection is illustrated in Figure~\ref{fig:almProjectionSchematic}.

\noindent
The projected force is normalised so that the volume integral of the applied source is consistent with the force carried by the beam segment. For beam point $k$ the weight is accumulated as $w_k = \sum_{j}\omega_{jk}\eta_{\varepsilon}(r_j)V_j$, where $V_j$ is the fluid-cell volume and $\omega_{jk}$ the linear interpolation weight, with $\omega_{ji}=1-s_j$ and $\omega_{j,i+1}=s_j$. The normalised nodal force is $\widehat{\mathbf{f}}_k = \mathbf{f}_k\Delta s_k / w_k$, with $\Delta s_k$ the local actuator-segment length, and the source applied to a fluid cell is
\begin{equation}
\mathbf{S}_{m,j}^{(i)}
=
-\eta_{\varepsilon}(r_j)
\left[
(1-s_j)\widehat{\mathbf{f}}_{i}
+
s_j\widehat{\mathbf{f}}_{i+1}
\right],
\label{eq:almMomentumSource}
\end{equation}
the negative sign indicating that the source is the reaction applied to the fluid. It is integrated over the receiving cell with the cell-centred midpoint rule, consistent with the second-order accuracy of the surrounding finite volume operators.

\noindent
This normalisation conserves the resultant force by construction. Summing Eq.~\eqref{eq:almMomentumSource} over all fluid cells and regrouping by beam point gives $\sum_{j}\mathbf{S}_{m,j}V_{j} = -\sum_{k}\widehat{\mathbf{f}}_{k}\,w_{k} = -\sum_{k}\mathbf{f}_{k}\,\Delta s_{k}$, so the total force applied to the fluid equals the total actuator-line load, whatever the mesh, kernel width, or kernel truncation, provided every accumulated weight is non-zero. Without the normalisation, the discrete volume integral of the kernel differs from its continuous value whenever the kernel is under-resolved by the mesh or truncated, and the force applied to the fluid then differs from the load carried by the beam. That condition holds whenever at least one fluid cell is assigned to each actuator segment. The same guarantee does not extend to angular momentum, since the kernel redistributes the force away from the centreline.

\subsubsection{Coupled update sequence and parallel-consistent search}
\label{sec:almParallelSearch}
Algorithm~\ref{alg:almCoupling} summarises the operations performed within one time step and shows where the coupling sits inside the PIMPLE loop. Each operation is labelled with the equation it implements.

\begin{algorithm}[!ht]
\caption{Actuator-line beam--fluid coupling within one time step. The coupling operations of steps~\ref{alg:stepActuator}--\ref{alg:stepSource} run only on the first outer iteration of the PIMPLE loop, and the source $\mathbf{S}_{m,j}$ they produce is held fixed for the remaining outer iterations.}
\label{alg:almCoupling}
\begin{algorithmic}[1]
\FOR{each time step, from $t^{n}$ to $t^{n+1}$}
\FOR{each PIMPLE outer iteration $k = 1,\ldots,n_{\mathrm{out}}$}
\IF{$k = 1$}
\STATE Evaluate the actuator-point positions $\mathbf{P}_{a,i}$ and velocities $\dot{\mathbf{P}}_{a,i}$ of all beam cells $i=1,\ldots,N_{b}$ from the beam state at $t^{n}$, using Eqs.~\eqref{eq:almKinematicCoupling} and~\eqref{eq:almActuatorPoint}, and the centreline tangents $\mathbf{e}_{t,i}$. \label{alg:stepActuator}
\STATE Locate the fluid cell containing each actuator point and evaluate the preliminary relative velocity $\mathbf{u}^{0}_{r,i}$ from Eq.~\eqref{eq:almPreliminaryVelocity}. \label{alg:stepLocateA}
\STATE Construct the upstream sampling points $\mathbf{P}_{s,i}$ from Eq.~\eqref{eq:almSamplingPoint}, with the fallback direction triggered when $\lVert\mathbf{u}^{0}_{n,i}\rVert$ falls below the prescribed threshold. \label{alg:stepSamplingPts}
\STATE Locate the single fluid cell containing each $\mathbf{P}_{s,i}$ and evaluate the relative velocity $\mathbf{U}_{r,i}$ from Eq.~\eqref{eq:almRelativeVelocity}. \label{alg:stepLocateS}
\STATE Evaluate the raw line force from Eq.~\eqref{eq:almLineForce} and apply the under-relaxation of Eq.~\eqref{eq:almForceRelaxation}.
\STATE Add $\mathbf{f}_{i}\,L_{i}$ to the residual of the beam force balance~\eqref{eq:fvForceBalance} and solve the structural Newton system~\eqref{eq:fvDiscreteSystem}--\eqref{eq:beamNewtonUpdate} for the beam state at $t^{n+1}$.
\STATE At the updated actuator-point positions, identify for every fluid cell $j$ inside the bounding box of the actuator points, inflated by $3\varepsilon$, the closest actuator segment and evaluate $s_{j}$, $r_{j}$ and $\eta_{\varepsilon}(r_{j})$ from Eqs.~\eqref{eq:almProjectionPoint}--\eqref{eq:almGaussian}, with $\eta_{\varepsilon} = 0$ for $r_{j} > 3\varepsilon$. \label{alg:stepProjection}
\STATE Accumulate the nodal weights, normalise the nodal forces, and assemble the reaction source $\mathbf{S}_{m,j}$ of Eq.~\eqref{eq:almMomentumSource}. \label{alg:stepSource}
\ENDIF
\STATE Solve the flow equations of this outer iteration, namely the momentum equation~\eqref{eq:momentum} with $\mathbf{S}_{m,j}$ included and the pressure correction, preceded in the two-phase case by the phase-fraction equation~\eqref{eq:alphaTransportCompressed}.
\ENDFOR
\ENDFOR
\end{algorithmic}
\end{algorithm}

\noindent
The line force is evaluated once per time step, on the first outer iteration of the PIMPLE loop and before the structural Newton loop, and is then held fixed for the remainder of the step. The later outer iterations solve the pressure--velocity system against that frozen source and do not re-sample the fluid velocity, so the interface load is treated explicitly rather than implicitly. This choice is deliberate. The sampling point moves with the beam, so a large enough change in displacement between Newton iterations can carry that point across a fluid cell boundary. The sampled velocity then jumps between two neighbouring cells, the force jumps with it, and the iteration oscillates between two states instead of converging. Freezing the force removes that failure mode, at the cost of a staggered coupling in which the interface load lags the fluid solution by one time step. The implementation can also re-fire the sampling and the structural solve at every outer iteration, which makes the interface load implicit but returns the discontinuous sampling to the inside of the iteration and requires strong under-relaxation of the line force to stay stable, so that option is not used for any result reported here. The empirical size of the lag is bounded in Section~\ref{sec:beamtunnel_sensitivity}, where driving the pressure--velocity system to convergence within the step, rather than taking a single outer iteration, changes the steady tip displacement by $0.1\%$.

\noindent
The two cell searches in steps~\ref{alg:stepLocateA} and~\ref{alg:stepLocateS} of Algorithm~\ref{alg:almCoupling} govern the parallel behaviour of the update. Testing every fluid cell for each query point would cost $\mathcal{O}(N_{b}N_{c})$ point-in-cell tests per time step, with $N_{c}$ the number of fluid cells, which is prohibitive on the meshes used in Section~\ref{sec:flexveg}. The actuator points of adjacent beam cells lie within a beam-cell length of each other, typically much smaller than the local fluid cell size, so their containing cells are usually the same cell or two cells that share a face. The search for each actuator point therefore starts from a seed, which is the cell found for the previously processed beam cell. From that seed it walks toward the query point, each time crossing a shared face into the neighbouring cell whose centre is closest to the point, until either the containing cell is found or the walk can go no further, typically at the edge of the local subdomain. Because the seed is usually the containing cell or one of its neighbours, the walk ends after a few face crossings, independently of the mesh size. The first beam cell on each rank, which has no seed, and any walk that fails fall back to a search of the local subdomain through an octree, a hierarchical spatial index of the fluid cells that is built once at the start of the run and whose query cost grows only logarithmically with the number of cells. The sampling points of step~\ref{alg:stepLocateS} are located directly with the same octree search.

\noindent
On a domain-decomposed fluid mesh, each rank holds only a subset of the fluid cells, so a query point strictly interior to one subdomain has its containing cell on exactly one rank. The per-beam-cell sampled velocity field can therefore be assembled across ranks by a single reduction with summation, after a cheap bounding-box test rejects beam cells lying outside the local subdomain before any search is attempted. The sampling-point construction of step~\ref{alg:stepSamplingPts} requires the global preliminary relative velocity, so it is performed on the master rank and the resulting points are broadcast, after which each rank searches for them in its own subdomain. Two degenerate cases are possible. The first is a query point lying exactly on a face shared by two subdomains. The search accepts a cell as containing the point through a point-in-cell containment test, which splits the cell into tetrahedra and checks whether the point lies inside any of them, and a point on a shared face can pass this test on both sides. Its velocity would then be counted twice in the reduction. The implementation does not exclude this case explicitly, since it requires a query point to fall exactly on a processor face, and it was not observed in any simulation reported here. The second is a point found by no rank, which can happen only if it leaves the fluid domain. It contributes zero to the reduction, so the affected beam cell would silently carry no load for that step, and the case set-ups reported here keep all actuator and sampling points inside the fluid domain by construction.

\noindent
Every query is therefore answered either by a short walk or by one octree search, so the two searches together cost at most $\mathcal{O}(N_{b}\log N_{c})$ operations per rank and per time step, however often the fallback is used. The two reduction-broadcast pairs are the only collective communication the coupling requires, so the operator scales with the underlying fluid solver. The projection of step~\ref{alg:stepProjection} carries a separate cost. Every local fluid cell is first tested against the inflated bounding box of the actuator points, which is a single comparison per cell, and the distance to each of the $N_{b}$ actuator segments is evaluated only for the cells inside the box. The number of distance evaluations therefore scales with the number of fluid cells near the beam rather than with the size of the fluid mesh, and the sweep runs over local cells only, with no communication. Since the normalisation conserves the transferred force whatever the kernel truncation, the $3\varepsilon$ cutoff reduces this cost without affecting the force balance. Inside the box every cell is still tested against every segment, however, so the cost grows linearly with $N_{b}$, and the bounding box of a long, inclined, or strongly curved beam can enclose a large part of the domain. For such beams, a spatial index over the actuator segments would reduce the cost further.

\noindent
All components are implemented in a single library for the open-source finite volume toolbox OpenFOAM \citep{weller_tensorial_1998}, which its standard flow solvers load at run time. The beam solver and the coupling operator are selected through the case input files, and no source code is modified per case. The two channel-flow benchmarks solve the single-phase equations of Section~\ref{sec:fluidSubsystem}, as implemented in the \texttt{pimpleFoam} solver of OpenFOAM, and the wave-flume case solves the two-phase volume-of-fluid equations, as implemented in its \texttt{interFoam} solver. The coupling implementation is identical in both, so moving between single-phase and two-phase flow changes only the flow solver and the case input files. All simulations are run in parallel using domain decomposition of the fluid mesh, with a separate decomposition of the one-dimensional beam mesh.


\section{Verification of the structural solver}
\label{sec:cantilever}

\noindent
The three coupled cases that follow all involve a beam undergoing large rotations under a distributed hydrodynamic load, so the beam solver is first exercised in isolation on the static large-deflection cantilever of \citet{belendez_numerical_2003}, subsequently used by \citet{chen_eulerianlagrangian_2019} to validate the structural model of their flexible-vegetation framework. The case suits a combined verification and validation exercise, since the first reference provides experimental tip deflections and the second reports predictions from an independent structural model on the same loading programme.

\noindent
The beam is a slender rectangular-section steel cantilever of length $0.4~\mathrm{m}$, width $25~\mathrm{mm}$ and height $0.4~\mathrm{mm}$, giving $EI = 0.02591~\mathrm{N\,m^2}$ with $E = 194.3~\mathrm{GPa}$ and $G = 74.73~\mathrm{GPa}$. The height-to-length ratio of $10^{-3}$ places it in the large-deformation regime even at moderate loads, and at the largest load the free end displaces vertically by approximately $70\%$ of the beam length, which makes a linearised analysis inapplicable. The beam carries a uniformly distributed self-weight of $0.758~\mathrm{N\,m^{-1}}$ together with a vertical concentrated load at the free end, varied from $0$ to $0.588~\mathrm{N}$. It is solved on a uniform discretisation of $80$ beam cells with a fully clamped root and a free tip, advanced in pseudo-time to a static steady state.

\begin{figure}[!ht]
\centering
\begin{subfigure}[b]{0.49\textwidth}
  \centering
  \includegraphics[width=\linewidth]{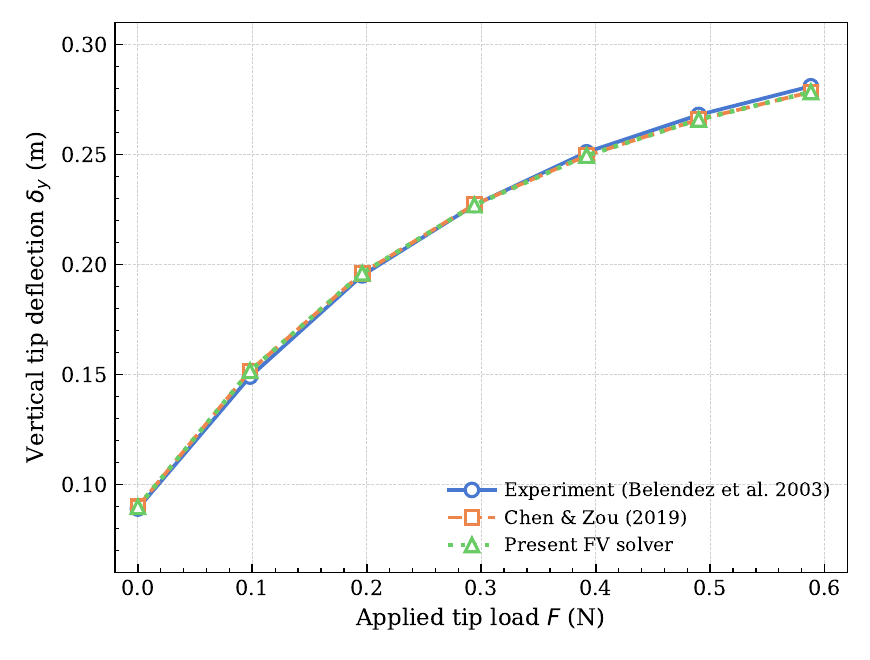}
  \caption{}
\end{subfigure}\hfill
\begin{subfigure}[b]{0.49\textwidth}
  \centering
  \includegraphics[width=\linewidth]{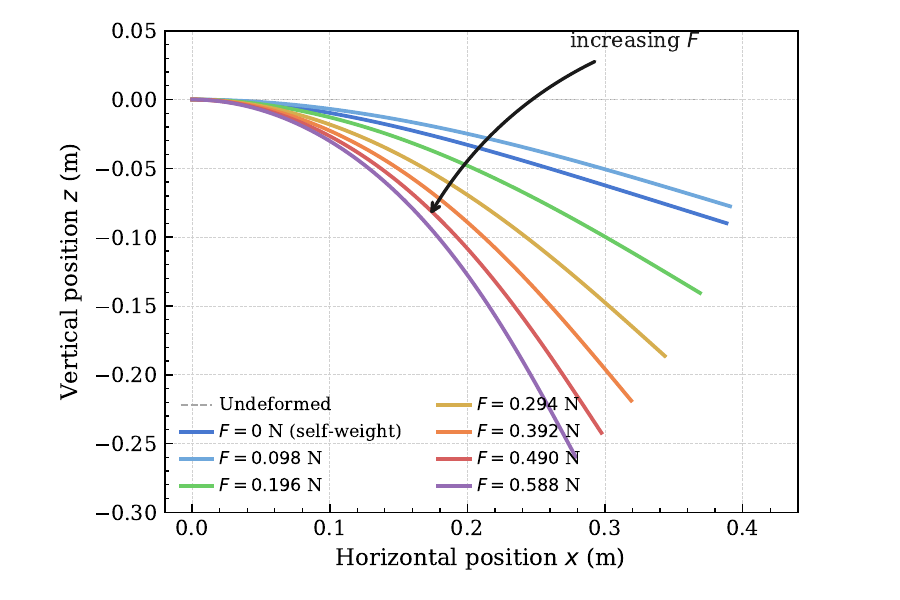}
  \caption{}
\end{subfigure}
\caption{Large-deflection cantilever benchmark. (a) Load--deflection curve, comparing the experimental measurements of \citet{belendez_numerical_2003}, the structural model of \citet{chen_eulerianlagrangian_2019}, and the present finite volume Simo--Reissner beam solver. (b) Deformed centrelines predicted by the present solver under self-weight and increasing vertical end load.}
\label{fig:cantilever_load_deflection}
\end{figure}

\noindent
Table~\ref{tab:cantilever} reports the free-end vertical displacement and Figure~\ref{fig:cantilever_load_deflection} shows the load--deflection curve together with the deformed centrelines. The incremental stiffness decreases progressively as the beam rotates away from its initial configuration and the moment arm of the applied load shortens, which the geometrically exact formulation captures without linearisation. The maximum difference from the experimental measurements is $1.74\%$, at $F = 0.098~\mathrm{N}$, where the concentrated load adds little above the self-weight equilibrium, so the percentage difference there is sensitive to small discrepancies in the applied load or material parameters. Every other case stays at or below $0.93\%$, and agreement improves steadily as the load increases. Against the independent structural model, the two agree to four decimal places in metres at every load level except $F = 0.294~\mathrm{N}$, where the present prediction of $0.2270~\mathrm{m}$ lands on the experimental measurement while the reference value differs by $0.09\%$. The structural solver is therefore both validated against the measurements and verified against an independent model before it is embedded in the coupled cases.

\begin{table}[!ht]
\centering
\small
\caption{Large-deflection cantilever benchmark. Free-end vertical displacement $\delta_y$ under applied tip load $F$, comparing the experimental measurements of \citet{belendez_numerical_2003}, the values reported by \citet{chen_eulerianlagrangian_2019}, and the present finite volume beam solver.}
\label{tab:cantilever}
\setlength{\tabcolsep}{4pt}
\renewcommand{\arraystretch}{1.1}
\begin{tabularx}{\linewidth}{c c >{\centering\arraybackslash}X c c c}
\toprule
& \multicolumn{3}{c}{Free-end vertical displacement $\delta_y$ (m)} & \multicolumn{2}{c}{Difference vs.\ experiment (\%)} \\
\cmidrule(lr){2-4}\cmidrule(lr){5-6}
$F$ (N) & Experiment & Chen and Zou (structural model) & This work & Chen and Zou & This work \\
\midrule
0.000 & 0.089 & 0.0898 & 0.0898 & 0.90 & 0.90 \\
0.098 & 0.149 & 0.1516 & 0.1516 & 1.74 & 1.74 \\
0.196 & 0.195 & 0.1960 & 0.1960 & 0.51 & 0.51 \\
0.294 & 0.227 & 0.2272 & 0.2270 & 0.09 & 0.00 \\
0.392 & 0.251 & 0.2495 & 0.2495 & 0.60 & 0.60 \\
0.490 & 0.268 & 0.2659 & 0.2659 & 0.78 & 0.78 \\
0.588 & 0.281 & 0.2784 & 0.2784 & 0.93 & 0.93 \\
\bottomrule
\end{tabularx}
\end{table}

\section{Flexible beam in a confined channel flow}
\label{sec:beamtunnel}

\noindent
The first coupled verification is the soft-beam case of \citet{zhang_immersed_2012} (Section~6.4, Case~1), in which a slender flexible cantilever clamped to the no-slip wall of a narrow channel is bent into a steady deformed shape by a steady, pressure-driven viscous flow. It is an established benchmark for immersed and meshless FSI formulations \citep{long_coupling_esfem_sph_2021,wang_immersed_esfem_2022,yang_peridynamics_iblbm_2022,yan_explicit_velocity_iblbm_2023}. Because the flow is steady and the cantilever settles into a fixed shape, the problem reduces to a static equilibrium in which the drag distributed along the deflected beam is balanced everywhere by the elastic restoring moment. That steadiness is what makes it a convenient first test, since there is no moving interface to capture, no time variation of the inflow, and no vortex shedding.

\noindent
Every reference study resolves the cantilever as a two-dimensional continuum, and they differ from one another only in the fluid solver and the interface coupling, as Table~\ref{tab:beamtunnel_refstudies} shows. Here the square cross-section enters the formulation only through its area, its second moment of area, and the drag reference length of Section~\ref{sec:almLineForce}. The benchmark is two-dimensional, and it is solved on a three-dimensional fluid mesh that is one cell thick in the spanwise direction $z$, with symmetry conditions on the two faces normal to $z$, so the flow does not vary across the span. The beam is the same Simo--Reissner model used in the other cases, a one-dimensional centreline carrying a three-dimensional cross-section, so its physical thickness still enters the drag law and the projection. What the benchmark tests is therefore whether a one-dimensional beam loaded by a distributed normal drag can reproduce the deflection that a resolved two-dimensional continuum gives, in a confined flow at low Reynolds number. Collapsing the cross-section onto a line discards the resolved pressure and viscous tractions on the beam faces and replaces them with a single drag coefficient, so the reduction can be trusted only once the tip deflection it predicts falls within the spread of the reference solutions.

\begin{table}[!ht]
\centering
\small
\caption{Published studies of the flexible-beam-in-channel benchmark of \citet{zhang_immersed_2012} (Section~6.4, Case~1).}
\label{tab:beamtunnel_refstudies}
\setlength{\tabcolsep}{5pt}
\renewcommand{\arraystretch}{1.30}
\begin{tabularx}{\linewidth}{l X X X}
\toprule
Study & Structural model & Fluid solver & Coupling \\
\midrule
\citet{zhang_immersed_2012}          & 2D ES-FEM-T3 continuum          & Characteristic Galerkin FEM            & Immersed smoothed FEM \\
\citet{long_coupling_esfem_sph_2021} & 2D ES-FEM continuum             & Smoothed particle hydrodynamics        & Fictitious boundary particles \\
\citet{yan_explicit_velocity_iblbm_2023} & 2D smoothed point interpolation & Hybrid lattice-Boltzmann flux solver & Immersed boundary \\
\citet{yang_peridynamics_iblbm_2022} & 2D bond-based peridynamics      & Lattice-Boltzmann                      & Multi-direct-forcing IB \\
\citet{wang_immersed_esfem_2022}     & 2D ES-FEM continuum             & Edge-based smoothed FEM fluid          & Immersed ES-FEM/ES-FEM \\
\midrule
Present work                          & 1D Simo--Reissner FV beam       & Finite volume Navier--Stokes (OpenFOAM) & Actuator-line distributed forcing \\
\bottomrule
\end{tabularx}
\end{table}

\subsection{Problem definition}
\label{sec:beamtunnel_geometry}
\begin{figure}[!ht]
\centering
\includegraphics[width=0.95\textwidth]{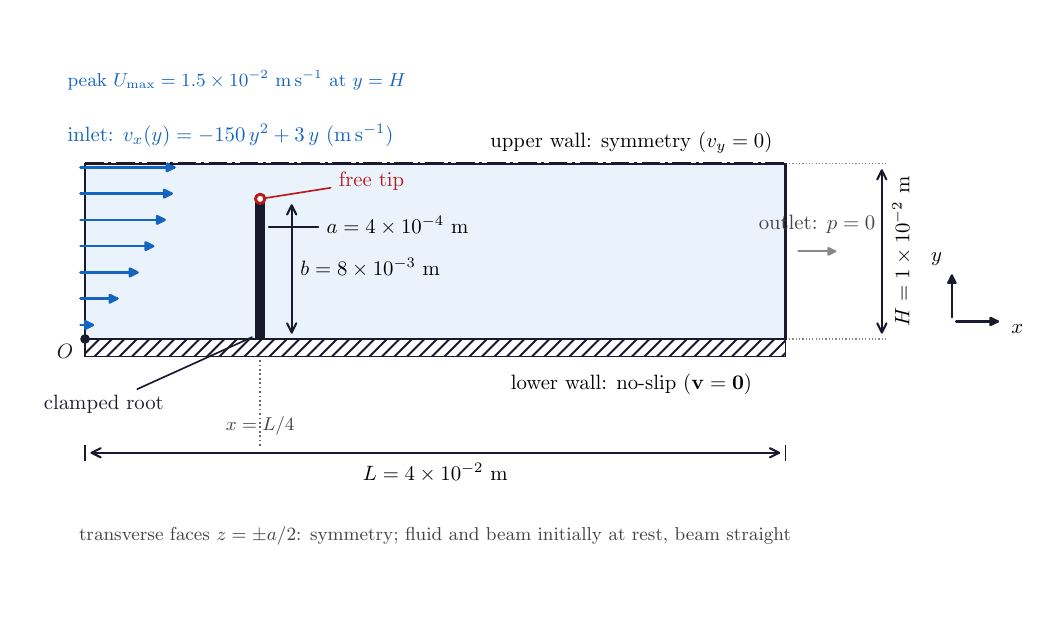}
\caption{The flexible-beam-in-channel verification problem. The origin $O$ lies where the inlet plane meets the lower wall, with $x$ measured downstream and $y$ upward. The cantilever of length $b$ and square cross-section of side $a$ is clamped at $x = L/4$ on the lower no-slip wall and free at its tip. A steady parabolic profile is imposed at the inlet, a zero-pressure outflow on the right, the lower wall is no-slip, and the upper wall is a symmetry plane. Dimensions reproduce \citet{zhang_immersed_2012}, Section~6.4, Case~1, converted to SI units.}
\label{fig:beamInTunnel_schematic}
\end{figure}

The geometry is presented in Figure~\ref{fig:beamInTunnel_schematic}. The fluid domain is a rectangular channel of streamwise length $L = 4\times10^{-2}~\mathrm{m}$ and transverse height $H = 1\times10^{-2}~\mathrm{m}$, with the origin where the inlet plane meets the lower no-slip wall, $x$ measured downstream and $y$ upward. A slender flexible cantilever of length $b = 8\times10^{-3}~\mathrm{m}$ and uniform square cross-section of side $a = 4\times10^{-4}~\mathrm{m}$ is clamped at $x = L/4$ on the lower wall and extends upward into the flow, occupying $80\%$ of the channel height when undeformed. The thickness-to-length ratio $a/b = 1/20$ places the beam in the slender-cantilever regime.

\noindent
A steady parabolic horizontal velocity profile is prescribed at the inlet,
\begin{equation}
v_x(y) = -150\,y^2 + 3\,y~~~\mathrm{m\,s^{-1}}, \qquad v_y(y) = 0,
\label{eq:beamtunnel_inlet}
\end{equation}
with $y$ in metres, which is the SI form of the profile $v_x = 1.5(-y^2 + 2y)~\mathrm{cm\,s^{-1}}$ reported in the original paper. It vanishes at the lower wall and attains $U_{\max} = 1.5\times10^{-2}~\mathrm{m\,s^{-1}}$ at the upper symmetry plane. The downstream boundary is an outflow with a fixed reference pressure and zero normal gradient of velocity, the lower boundary a stationary no-slip wall on which the cantilever root is clamped, and the upper boundary a symmetry plane, consistent with the half-domain interpretation of the reference paper in which the mirror across $y = H$ recovers an infinite array of identical cantilevers. The front and back faces of the one-cell-thick domain, normal to $z$, are also symmetry planes. Both the fluid and the beam start at rest, with the beam undeformed, and the simulation is advanced with $\Delta t = 10^{-4}~\mathrm{s}$ to $t = 2.5~\mathrm{s}$, by which point the tip velocity has vanished.

\noindent
Properties are listed in Table~\ref{tab:beamtunnel_properties} and are taken without modification from the reference. The beam resultants are linear in the strain measures of Section~\ref{sec:beamSection}, with Young's modulus $E$ and Poisson ratio $\nu_s$, from which the shear modulus follows as $G = E/[2(1+\nu_s)]$. The flow regime is characterised by a thickness-based Reynolds number $\mathrm{Re}_a = \rho\,U_{\max}\,a/\mu \approx 0.6$, in the viscous-dominated Stokes-like regime, with a height-based value of $15$. Gravity is neglected, consistent with the reference.

\begin{table}[!ht]
\centering
\small
\caption{Fluid and beam properties for the flexible-beam-in-channel verification problem, following \citet{zhang_immersed_2012}, Section~6.4, Case~1.}
\label{tab:beamtunnel_properties}
\setlength{\tabcolsep}{6pt}
\renewcommand{\arraystretch}{1.15}
\begin{tabularx}{\linewidth}{l c X}
\toprule
Quantity & Symbol & Value \\
\midrule
\multicolumn{3}{l}{\textbf{Fluid domain}} \\
\cmidrule(lr){1-3}
Channel length (streamwise)      & $L$    & $4.0\times10^{-2}~\mathrm{m}$ \\
Channel height                   & $H$    & $1.0\times10^{-2}~\mathrm{m}$ \\
Fluid density                    & $\rho$ & $1.0\times10^{3}~\mathrm{kg\,m^{-3}}$ \\
Dynamic viscosity                & $\mu$  & $1.0\times10^{-2}~\mathrm{Pa\,s}$ \\
Peak inlet velocity              & $U_{\max}$ & $1.5\times10^{-2}~\mathrm{m\,s^{-1}}$ \\
\midrule
\multicolumn{3}{l}{\textbf{Beam}} \\
\cmidrule(lr){1-3}
Length                           & $b$        & $8.0\times10^{-3}~\mathrm{m}$ \\
Cross-section side               & $a$        & $4.0\times10^{-4}~\mathrm{m}$ \\
Solid density                    & $\rho_s$   & $7.8\times10^{3}~\mathrm{kg\,m^{-3}}$ \\
Young's modulus                  & $E$        & $1.0\times10^{4}~\mathrm{Pa}$ \\
Poisson ratio                    & $\nu_s$    & $0.3$ \\
\bottomrule
\end{tabularx}
\end{table}

\subsection{Steady-CFD calibration of the normal drag coefficient}
\label{sec:beamtunnel_cd_calibration}

\noindent
The normal drag coefficient in Eq.~\eqref{eq:almLineForce} has to be fixed before the coupled result can be discussed. Because the formulation reduces the cross-section to a centreline, the near-field tractions on it are not resolved and are absorbed instead into a single coefficient multiplying the upstream-sampled velocity squared, so that coefficient has to reflect the drag on a rigid cross-section of the same geometry in the same confined channel.

\begin{figure}[!ht]
\centering
\begin{subfigure}[b]{0.55\textwidth}
  \centering
  \includegraphics[width=\linewidth]{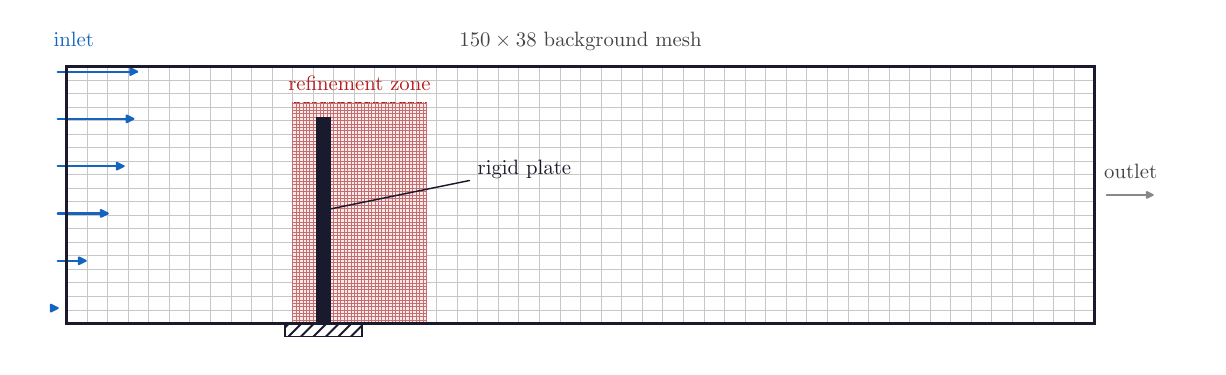}
  \caption{}
  \label{fig:beamInTunnel_dragcalib_mesh}
\end{subfigure}\hfill
\begin{subfigure}[b]{0.43\textwidth}
  \centering
  \includegraphics[width=\linewidth]{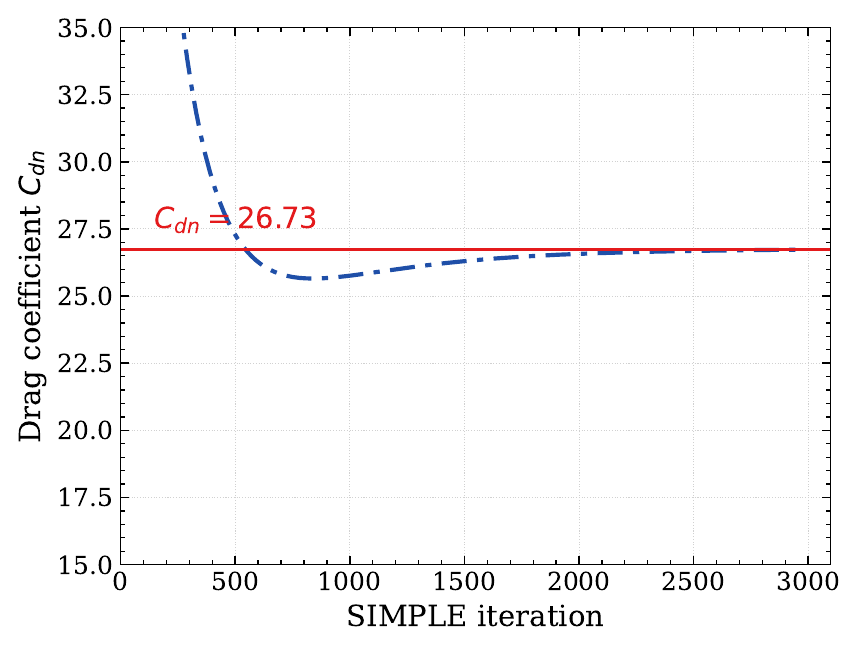}
  \caption{}
  \label{fig:beamInTunnel_cd_calibration}
\end{subfigure}
\caption{Steady-CFD calibration of the normal drag coefficient. (a) Mesh topology: a uniform $150 \times 38$ background mesh with one level of local refinement around the rigid plate (red zone), which occupies the same root location and cross-section as the flexible cantilever. (b) Convergence of $C_{dn}$ over the SIMPLE outer iterations (blue dash-dotted), with the converged value $C_{dn} = 26.73$ marked in red.}
\label{fig:beamInTunnel_dragcalib}
\end{figure}

\noindent
A separate steady incompressible CFD simulation was run on the same channel domain, with the flexible cantilever replaced by an undeformed rigid plate of the same square cross-section clamped at the same root location. The geometry, boundary conditions, and fluid properties are identical to the coupled case. It is solved as a steady flow with the SIMPLE algorithm \citep{patankar_calculation_1972}, as implemented in the \texttt{simpleFoam} solver of OpenFOAM, on a uniform $150 \times 38$ background hexahedral mesh with one level of local refinement around the plate (Figure~\ref{fig:beamInTunnel_dragcalib_mesh}), which captures the upstream stagnation point, the accelerated jet between the plate tip and the upper symmetry boundary, and the leeward recirculation. The reference area is $A_{\mathrm{ref}} = a b = 3.2\times10^{-6}~\mathrm{m^2}$, so the reported coefficient corresponds directly to the normal drag coefficient of the closure. The plate is held straight while the coupled beam bends downstream and blocks less of the channel, which matters less than it appears, because the actuator line multiplies $C_{dn}$ by the velocity sampled in the deformed configuration, and that velocity already reflects the wider gap above the deflected tip.

\noindent
Figure~\ref{fig:beamInTunnel_cd_calibration} reports the convergence history. A rapid initial transient brings $C_{dn}$ down to approximately $25.5$ at iteration $700$, after which a slow drift, reflecting the residual pressure adjustment in the highly viscous flow, carries it toward an asymptotic value of $26.7$ at iteration $2950$, the change between iterations $2000$ and $2950$ being below $0.5\%$. The value $C_{dn} = 26.73$ is read directly off the converged plateau, with no rounding, no empirical adjustment, and no tuning against the reference tip displacement, so the calibration is a single one-way handover from the rigid-plate CFD to the coupled run. The tangential coefficient is fixed at $C_{dt} = 0.01$, the size of the skin-friction coefficient of a smooth surface at these Reynolds numbers and one to two orders of magnitude below the drag coefficient of a bluff cross-section \citep{hoerner_fluid_1965}. Its value matters little here, because the beam stays close to normal to the flow.

\noindent
This calibration is a one-time, geometry-specific step rather than a free parameter of the method, and it is not a standing requirement of the coupling either. What makes it necessary here is confinement. The undeformed cantilever spans $80\%$ of the channel height, so the drag on it is set by the proximity of the walls as much as by the cross-section itself, whereas the generic correlations one might reach for instead, such as the flat-plate values of \citet{hoerner_fluid_1965} or the low-Reynolds cylinder estimate from Oseen's linearisation \citep{lamb_hydrodynamics_1932}, are derived for a body in an unbounded stream. Reading $C_{dn}$ off a steady CFD run on the confined geometry sidesteps that mismatch, and the same procedure can be repeated for any new blockage ratio. In an unbounded or weakly confined flow no such run is needed, and the wave-driven stem of Section~\ref{sec:flexveg} takes its coefficient from drag fits published alongside the reference experiment. Confinement changes the value of the coefficient, not the form of the closure.

\subsection{Verification against the reference solutions}
\label{sec:beamtunnel_verification}

\begin{figure}[!ht]
\centering
\includegraphics[width=0.90\textwidth]{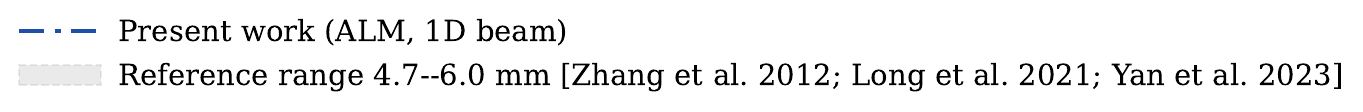}\\[0.4em]
\includegraphics[width=0.68\textwidth]{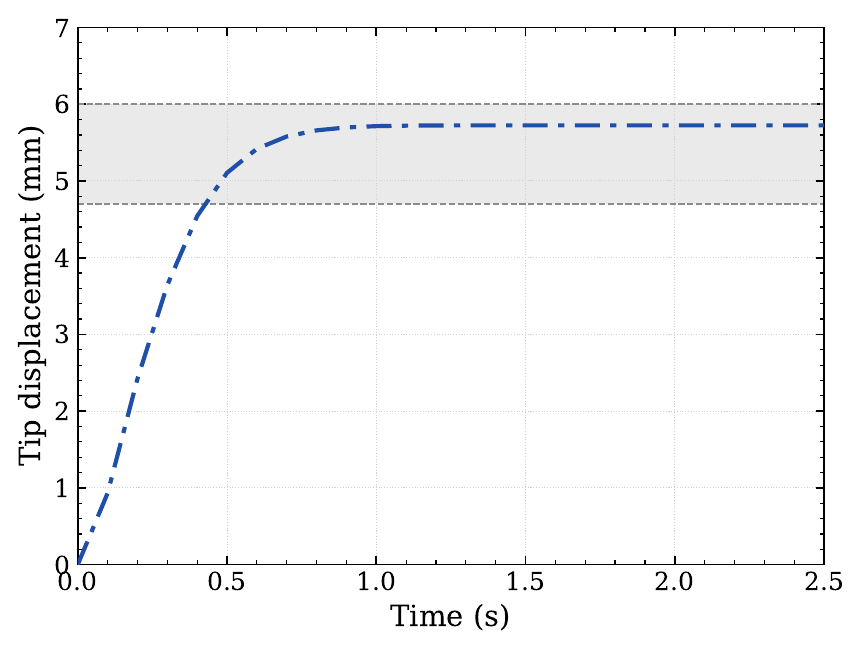}
\caption{Steady tip horizontal displacement of the flexible cantilever in the channel benchmark. The blue dash-dotted curve is the present result; the grey band spans the range $4.7$ to $6.0~\mathrm{mm}$ covered by three of the reference studies \citep{zhang_immersed_2012,long_coupling_esfem_sph_2021,yan_explicit_velocity_iblbm_2023}. The other two reference studies \citep{yang_peridynamics_iblbm_2022,wang_immersed_esfem_2022} report lower values, below the band.}
\label{fig:beamInTunnel_verification}
\end{figure}

\noindent
With $C_{dn}$ fixed, the coupled simulation is run to steady state. Figure~\ref{fig:beamInTunnel_verification} compares the predicted steady tip horizontal displacement against the values reported in the literature. The shaded band spans $4.7$ to $6.0~\mathrm{mm}$ and brackets the steady tip displacements digitised from the published curves of \citet{zhang_immersed_2012}, \citet{long_coupling_esfem_sph_2021}, and \citet{yan_explicit_velocity_iblbm_2023}. The other two studies of Table~\ref{tab:beamtunnel_refstudies}, \citet{yang_peridynamics_iblbm_2022} and \citet{wang_immersed_esfem_2022}, report lower steady displacements, below $4.7~\mathrm{mm}$, so the five published solutions together span a wider range that extends below the band.

\noindent
The present transient reaches a stationary plateau at approximately $t = 1.4~\mathrm{s}$, and the converged tip displacement is $5.73~\mathrm{mm}$, inside the band and therefore also within the range of all five studies. The present result and the reference solutions are not expected to converge to the same value under mesh refinement, because the reference methods resolve the solid and the tractions on its faces while the present formulation replaces them with a drag law on a line, so the underlying mathematical models differ. Nor is there a single agreed value within the family of two-dimensional continuum methods, the spread across the five studies reflecting sensitivities to the structural discretisation, the smoothing kernel adopted in the immersed coupling, and the treatment of the singular pressure field at the sharp upstream corners of the beam.

\begin{figure}[!ht]
\centering
\begin{subfigure}[b]{0.90\textwidth}
  \centering
  \includegraphics[width=\linewidth,trim={0 270 0 20},clip]{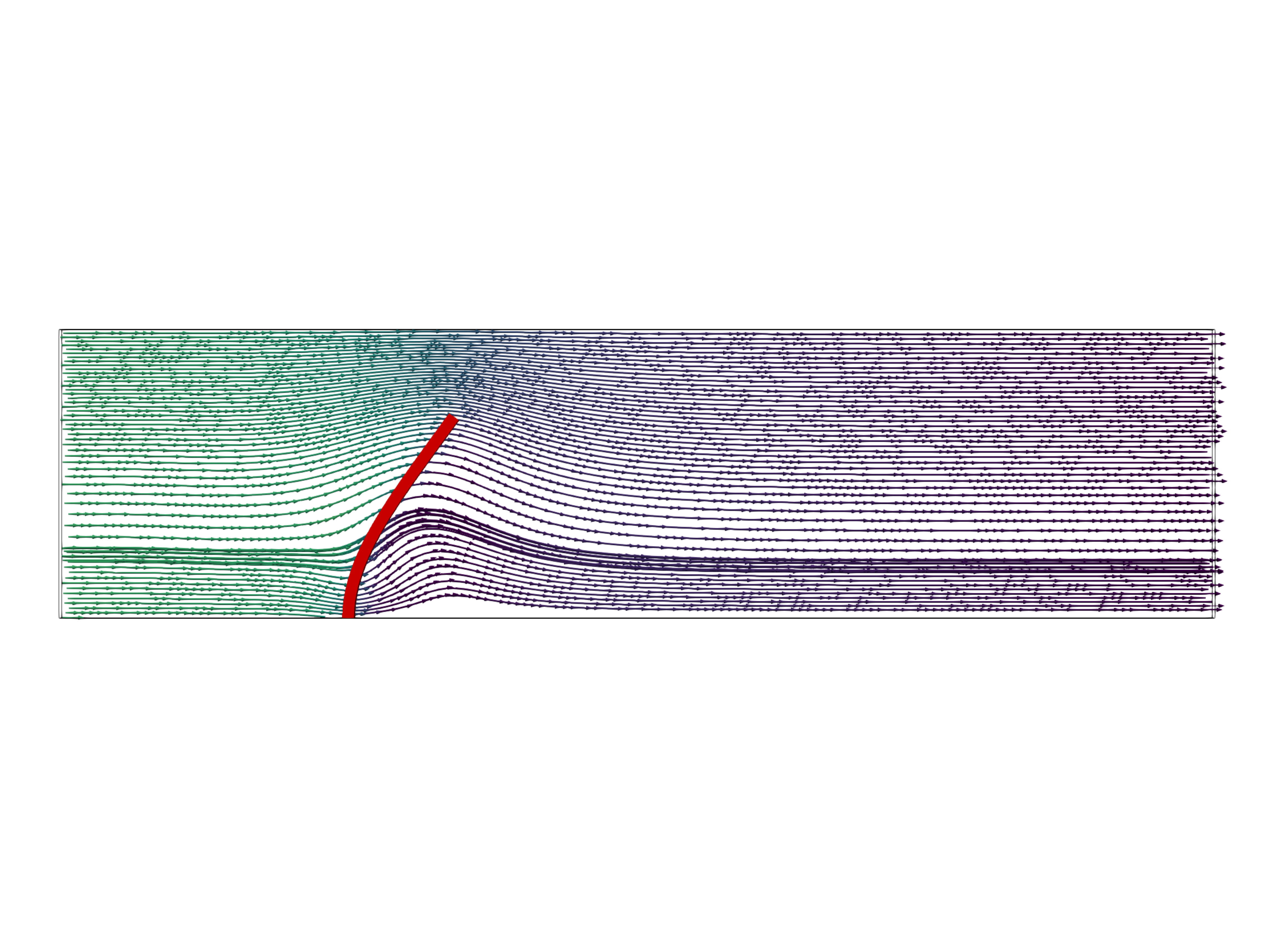}
  \caption{$t = 0.3~\mathrm{s}$}
\end{subfigure}\\[0.6em]
\begin{subfigure}[b]{0.90\textwidth}
  \centering
  \includegraphics[width=\linewidth,trim={0 270 0 20},clip]{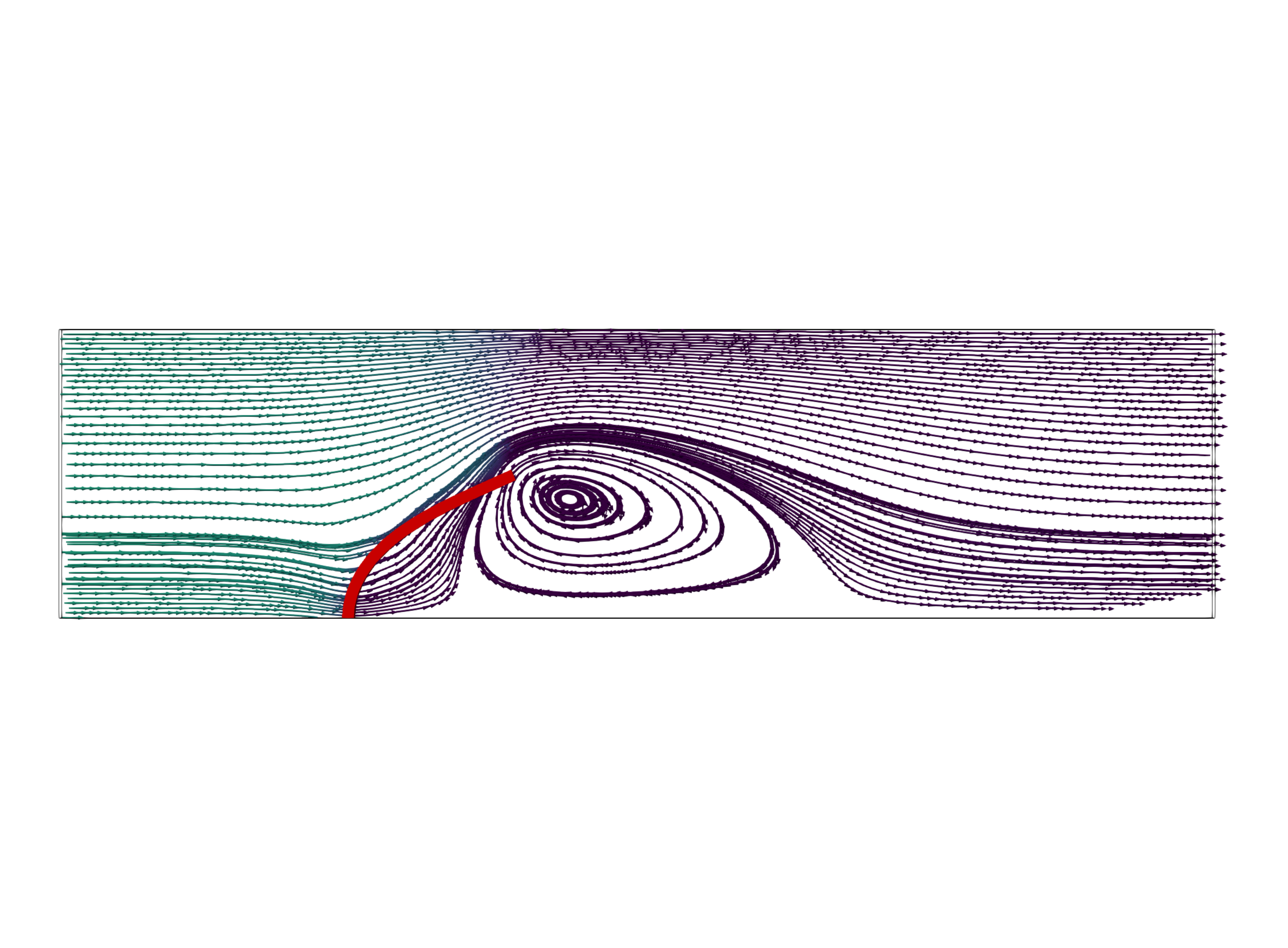}
  \caption{$t = 1.5~\mathrm{s}$}
\end{subfigure}
\caption{Flow field during the early transient and at the steady asymptote. Streamlines are coloured by velocity magnitude (in $\mathrm{m\,s^{-1}}$, common scale) and the deflected cantilever is rendered in red. At $t = 0.3~\mathrm{s}$ the leeward recirculation cell has not yet formed; at $t = 1.5~\mathrm{s}$ the accelerated jet between the bent tip and the upper symmetry boundary and the leeward eddy are both fully developed.}
\label{fig:beamInTunnel_snapshots}
\end{figure}

\noindent
Figure~\ref{fig:beamInTunnel_snapshots} shows the flow field early in the transient and at the steady asymptote. At $t = 0.3~\mathrm{s}$ the tip has swung downstream by roughly half of its final displacement and the streamlines bend smoothly around the bent cantilever, with no closed recirculation cell yet in the leeward region. The cell emerges between $t = 0.3$ and $0.6~\mathrm{s}$ together with an accelerated jet above the tip, and by $t = 1.5~\mathrm{s}$ the field is indistinguishable from its steady asymptote, the jet reaching $2.9\times 10^{-2}~\mathrm{m\,s^{-1}}$, almost twice the inlet peak velocity.

\noindent
This is the mechanism behind the steady balance. As the cantilever deflects, it forces the incoming flow into the narrowing upper portion of the channel, raising the local velocity above the tip and therefore the drag transferred to the beam, until the elastic restoring moment matches it. Those are the same flow features the steady-CFD calibration sees on the equivalent rigid plate, which is what justifies transferring $C_{dn}$ into the coupled run. The stagnation region and the accelerated jet also explain why the steady tip displacement is sensitive to the upstream sampling distance, which must be taken beyond the stagnation region and short of both the jet and the inlet plane.

\subsection{Numerical sensitivity studies}
\label{sec:beamtunnel_sensitivity}

\noindent
Six sensitivity studies establish the converged status of the result. Each perturbs a single parameter around a baseline of $150 \times 38$ fluid cells, $N_b = 20$ beam cells, $\Delta t = 10^{-4}~\mathrm{s}$, $\varepsilon = 6\times10^{-4}~\mathrm{m}$, $r_s = 3.18~\mathrm{mm}$, and the baseline solver tolerance set.

\begin{figure}[!ht]
\centering
\begin{subfigure}[t]{0.32\textwidth}
  \centering
  \includegraphics[width=\linewidth]{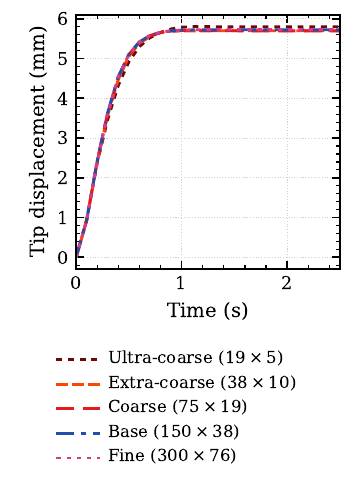}
  \caption{fluid mesh}
\end{subfigure}\hfill
\begin{subfigure}[t]{0.32\textwidth}
  \centering
  \includegraphics[width=\linewidth]{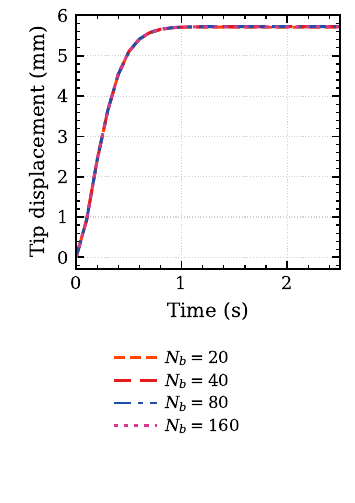}
  \caption{beam mesh}
\end{subfigure}\hfill
\begin{subfigure}[t]{0.32\textwidth}
  \centering
  \includegraphics[width=\linewidth]{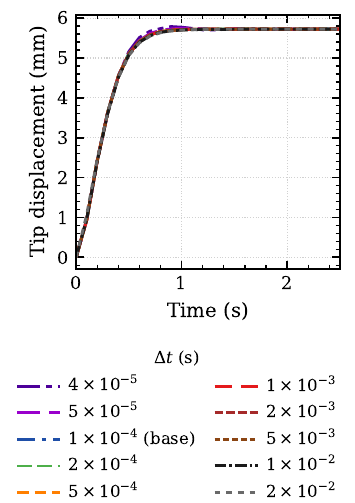}
  \caption{time step}
\end{subfigure}
\caption{Discretisation studies for the flexible-beam-in-channel problem, showing the transient tip horizontal displacement for (a) five successively refined fluid meshes, (b) four beam discretisations spanning $N_b = 20$ to $160$ cells, and (c) time steps spanning a factor of $500$. Each panel carries its own legend beneath its axes. Within each study the converged traces are visually indistinguishable in the steady regime.}
\label{fig:beamInTunnel_discretisation}
\end{figure}

\paragraph{Fluid mesh, beam mesh, and time step.}
Figure~\ref{fig:beamInTunnel_discretisation} collects the three discretisation studies. Five fluid meshes are compared, each successive mesh halving the cell size in both directions, from ultra-coarse ($19 \times 5$) up to fine ($300 \times 76$). The coarse, base, and fine meshes give $5.699$, $5.725$ and $5.728~\mathrm{mm}$, within $0.5\%$ of one another, and the corresponding trajectories are visually indistinguishable. Only the two coarsest meshes are under-resolved, the extra-coarse mesh giving $5.701~\mathrm{mm}$ with only ten cells across the channel and the ultra-coarse mesh $5.800~\mathrm{mm}$, where the additional offset reflects an inability to resolve the velocity gradient between the deflected beam and the upper boundary. The insensitivity follows from the regularised forcing, since the beam force is smeared over several fluid cells rather than acting on a no-slip interface, so the velocity field never develops the sharp boundary-layer profile that would require fine wall-normal grading.

\noindent
Four beam discretisations spanning $N_b = 20$ to $160$ give $5.71$, $5.73$, $5.73$, and $5.72~\mathrm{mm}$, within $0.3\%$ of one another and visually indistinguishable throughout. The deformed shape is close to the first bending mode and carries no sharp curvature, and the extent of the fluid momentum source is set by the kernel width rather than by the beam cell spacing. The closure also samples the upstream velocity once per beam cell, so once adjacent sampling points fall within a single fluid cell the refinement repeats the same sampled value. The coarsest discretisation is therefore adopted as the baseline.

\noindent
The time step is varied over a factor of $500$, from $2\times10^{-2}$ down to $4\times10^{-5}~\mathrm{s}$, and across the entire stable range the steady plateau is recovered within $\pm 0.1\%$ of $5.725~\mathrm{mm}$. The response is monotonic with a single slow timescale, so even a coarse temporal discretisation captures the asymptotic equilibrium, and steps larger than $2\times10^{-2}~\mathrm{s}$ trigger instability in the partitioned coupling. The base step of $10^{-4}~\mathrm{s}$ is therefore chosen to resolve the early transient accurately enough for the trace in Figure~\ref{fig:beamInTunnel_verification} to be comparable with the reference solutions, not to converge the steady displacement, which tolerates a much larger step.

\begin{figure}[!ht]
\centering
\begin{subfigure}[t]{0.48\textwidth}
  \centering
  \includegraphics[width=\linewidth]{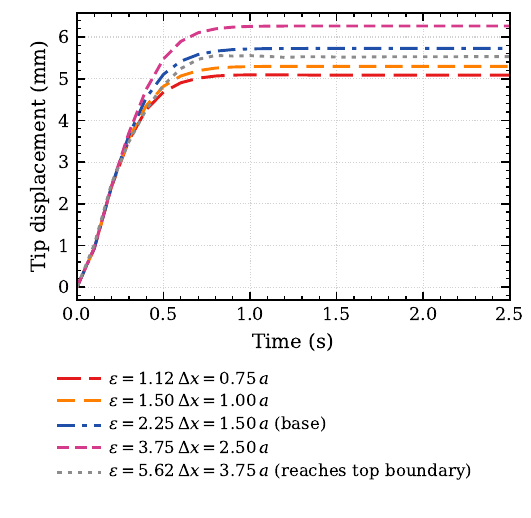}
  \caption{kernel width $\varepsilon$}
\end{subfigure}\hfill
\begin{subfigure}[t]{0.48\textwidth}
  \centering
  \includegraphics[width=\linewidth]{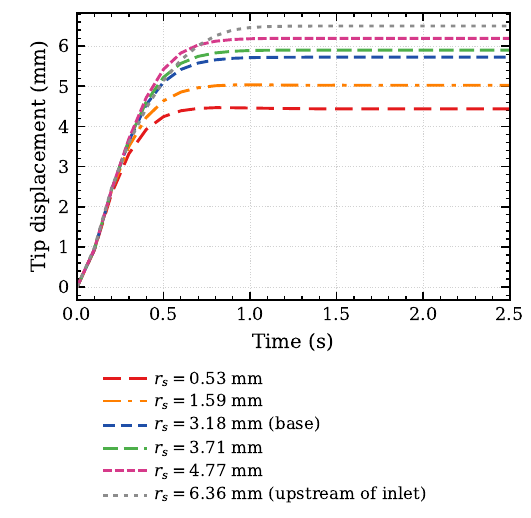}
  \caption{sampling offset $r_s$}
\end{subfigure}
\caption{Sensitivity of the steady tip displacement to the two coupling-operator settings, with each panel carrying its own legend. (a) Gaussian kernel width, each case labelled both in units of the base-mesh cell size $\Delta x = 2.67\times10^{-4}~\mathrm{m}$ and the cross-section side $a$. The case $\varepsilon = 5.62\,\Delta x$, shown in grey, reaches the upper symmetry boundary and lies outside the admissible range. (b) Upstream sampling offset. The smallest offset samples inside the stagnation zone, and the offset of $6.36~\mathrm{mm}$, shown in grey, places the sampling point upstream of the inlet for the lower part of the beam. The individual values are listed in Table~\ref{tab:beamtunnel_operator_cases}. Unlike the discretisation studies of Figure~\ref{fig:beamInTunnel_discretisation}, both parameters change the response by tens of percent.}
\label{fig:beamInTunnel_operator}
\end{figure}

\paragraph{Regularisation kernel width.}
Figure~\ref{fig:beamInTunnel_operator}(a) examines the sensitivity to the Gaussian kernel width of Eq.~\eqref{eq:almGaussian}. Following the dual normalisation of \citet{bral_actuator_2025}, each value is expressed both as a multiple of the base-mesh cell size $\Delta x = 2.67\times10^{-4}~\mathrm{m}$ and of the cross-section side $a$. Because the projection is normalised so that the total force on the fluid is independent of the kernel width (Section~\ref{sec:almProjection}), what is measured here is only the indirect effect of the kernel on the surrounding velocity field, which modifies the upstream-sampled velocity.

\noindent
Within the admissible range $1.12\,\Delta x \le \varepsilon \le 3.75\,\Delta x$, equivalently $0.75\,a$ to $2.5\,a$, the steady tip displacement varies monotonically from $5.09$ to $6.26~\mathrm{mm}$, a spread of approximately $20\%$ (Table~\ref{tab:beamtunnel_operator_cases}). Smaller $\varepsilon$ concentrates the reaction closer to the centreline, producing a more localised wake and a lower sampled velocity, hence smaller drag and smaller deflection, while larger $\varepsilon$ spreads the reaction into the freestream and raises the sampled velocity. The baseline $\varepsilon = 2.25\,\Delta x = 1.5\,a$ sits above the minimum required for a resolved projection and clear of the upper boundary. The case at $5.62\,\Delta x$ lies outside this range, because its kernel is so wide that its tail reaches the upper symmetry boundary. The nodal normalisation then redistributes the truncated mass along the whole node support instead of preserving the local profile, so the momentum injected near the beam drops and its displacement of $5.53~\mathrm{mm}$ falls back below the monotonic trend. The increasing relation between $\varepsilon$ and tip displacement holds only while the kernel stays clear of the domain boundaries.

\paragraph{Upstream sampling distance.}
Figure~\ref{fig:beamInTunnel_operator}(b) reports the sensitivity to the offset $r_s$. In an unbounded flow this offset is chosen large enough to sample the undisturbed freestream rather than the stagnation-affected velocity adjacent to the beam, but in a confined channel it must also remain inside the domain and below the upper symmetry plane. Over the admissible range the steady displacement varies monotonically from $4.44~\mathrm{mm}$ at $r_s = 0.53~\mathrm{mm}$ to $6.20~\mathrm{mm}$ at $4.77~\mathrm{mm}$, a spread of approximately $31\%$. The smallest offset samples within the beam's own stagnation zone and the largest samples the accelerated jet between the deflected beam and the upper boundary. The case $r_s = 6.36~\mathrm{mm}$ pushes the sampling point upstream of the inlet for the lower part of the beam and lies outside the admissible range.

\begin{table}[!ht]
\centering
\small
\caption{Kernel-width and sampling-distance studies. Steady tip displacement for each Gaussian regularisation width $\varepsilon$, listed as a multiple of the base-mesh cell size $\Delta x = 2.67\times10^{-4}~\mathrm{m}$ and of the cross-section side $a = 4\times10^{-4}~\mathrm{m}$, and for each upstream sampling offset $r_s$.}
\label{tab:beamtunnel_operator_cases}
\setlength{\tabcolsep}{4pt}
\renewcommand{\arraystretch}{1.15}
\begin{tabular}{c c c c l @{\qquad} c c l}
\toprule
\multicolumn{5}{l}{\textbf{Kernel width}} & \multicolumn{3}{l}{\textbf{Sampling offset}} \\
$\varepsilon$ (m) & $\varepsilon/\Delta x$ & $\varepsilon/a$ & Tip (mm) & Note & $r_s$ (mm) & Tip (mm) & Note \\
\midrule
$3.0\times10^{-4}$ & $1.12$ & $0.75$ & $5.09$ & lower bound        & $0.53$ & $4.44$ & stagnation zone \\
$4.0\times10^{-4}$ & $1.50$ & $1.00$ & $5.30$ &                    & $1.59$ & $5.03$ & \\
$6.0\times10^{-4}$ & $2.25$ & $1.50$ & $5.73$ & baseline           & $3.18$ & $5.73$ & baseline \\
$1.0\times10^{-3}$ & $3.75$ & $2.50$ & $6.26$ & upper bound        & $3.71$ & $5.90$ & \\
$1.5\times10^{-3}$ & $5.62$ & $3.75$ & $5.53$ & reaches top boundary & $4.77$ & $6.20$ & upper bound \\
                   &        &        &        &                    & $6.36$ & $6.50$ & upstream of inlet \\
\bottomrule
\end{tabular}
\end{table}

\noindent
The baseline $r_s = 3.18~\mathrm{mm}$ lands in the middle of the admissible range and yields, together with the kernel-width baseline, the result closest to the literature range. Both settings therefore act as effective calibration parameters rather than discretisation-only convergence controls, and in this confined channel they calibrate the quasi-steady closure to the local blockage. Only $C_{dn}$ was fixed independently of the reference, whereas the pair $(\varepsilon, r_s)$ was selected within its admissible plateau with the reference result in view, so the agreement is an assessment against other numerical solutions with two calibrated operator settings rather than a fully blind prediction. The selected pair is carried \emph{unchanged} into the benchmark of Section~\ref{sec:leaflet}, and the wave-flume case of Section~\ref{sec:flexveg} sizes the kernel by the same rule relative to the local cell size, so what those later cases test is the transferred settings rather than the settings themselves.

\paragraph{Solver tolerance hierarchy.}
The remaining numerical artefact is the iterative tolerance of the structural Newton solver and the PIMPLE outer loop. Table~\ref{tab:beamtunnel_tol_cases} reports three successively tightened configurations. The loose-structural configuration over-predicts the steady displacement by $3.3\%$ and drifts slowly away from a transient overshoot, which is the characteristic signature of a partitioned coupling whose structural side has not converged within each time step. Tightening the Newton tolerance alone reduces both the overshoot and the drift to under $1\%$ at negligible cost, because the structural solve is cheap relative to the fluid solve, so the dominant iterative error is on the structural side. The third configuration keeps the tight Newton tolerance and allows up to twenty PIMPLE outer iterations per time step instead of one. The outer loop stops earlier once the initial residuals of the velocity and pressure equations fall below $10^{-7}$, or below $10^{-4}$ of their values at the first outer iteration of the step. This brings the plateau within $0.1\%$ of its asymptotic value, at roughly five to ten times the per-step fluid cost, and the third configuration is used for Figure~\ref{fig:beamInTunnel_verification} and for all the studies above. The last step also bounds the error introduced by freezing the interface force for the duration of the time step (Section~\ref{sec:almParallelSearch}), because driving the pressure--velocity system to convergence against that frozen force changes the answer by $0.1\%$.

\begin{table}[!ht]
\centering
\small
\caption{Solver-tolerance hierarchy. The structural Newton tolerance and the maximum number of PIMPLE outer iterations $n_{\mathrm{out}}$ are tightened in turn. The last configuration is adopted for the verification figure and all sensitivity studies.}
\label{tab:beamtunnel_tol_cases}
\setlength{\tabcolsep}{5pt}
\renewcommand{\arraystretch}{1.15}
\begin{tabular}{l c c c l}
\toprule
Configuration & Newton tol. & $n_{\mathrm{out}}$ & Tip (mm) & Note \\
\midrule
Loose structural          & $10^{-6}$ & $1$  & $5.92$ & overshoot, slow drift \\
Tight structural          & $10^{-9}$ & $1$  & $5.75$ & drift below $1\%$ \\
Tight structural + PIMPLE & $10^{-9}$ & $20$ & $5.73$ & reference (within $0.1\%$) \\
\bottomrule
\end{tabular}
\end{table}

\noindent
Taken together, the six studies separate two kinds of sensitivity. The fluid mesh, the beam mesh, the time step, and the solver tolerances have small effects that can be controlled, so the steady tip displacement of $5.73~\mathrm{mm}$ is converged in the discretisation sense, and it stays inside the literature band of $4.7$ to $6.0~\mathrm{mm}$. The two coupling-operator settings are the dominant sensitivities of the method in this benchmark. Across their admissible ranges the kernel width changes the tip displacement by about $20\%$ and the sampling offset by about $31\%$. The dependence on the sampling offset is a property of the closure rather than a discretisation error, since the offset selects the velocity that drives the drag law, so it does not diminish under mesh refinement or tighter tolerances. The kernel is sized relative to the local cell size in practice, so part of its influence behaves like a discretisation effect, but whether it vanishes when the kernel and the mesh are refined together was not tested here. A confined channel in which the beam spans most of the cross-section is also the least favourable case for both parameters, because the velocity upstream of the beam varies strongly with position, from the stagnation region near the root to the accelerated jet above the tip. The coupling is therefore verified against reference solutions obtained with other methods for steady, viscous-dominated flow in a confined channel, which is a check against other numerical solutions rather than a validation against physical measurements, and it motivates moving to the time-periodic benchmark of the next section.


\section{Leaflet driven by a sinusoidal channel flow}
\label{sec:leaflet}

\noindent
The second benchmark exercises the coupling under strong time-dependent forcing. A thin flexible leaflet, clamped along one edge, oscillates in response to a periodic channel flow whose direction reverses every half-period. The configuration is taken from the immersed finite element method (IFEM) study of \citet{zhang_immersed_2007}, where it serves as the canonical FSI verification case, and it has since been used by several groups developing related immersed and fictitious-domain formulations \citep{baaijens_fictitiousdomain_2001,de_hart_threedimensional_2003,zhang_immersed_2012,roy_benchmarking_2015}. The geometry is idealised, but the combination of strong unsteady forcing, large structural rotation, and bidirectional loading makes it a stringent test of whether an FSI scheme transfers momentum and kinematic information consistently across the interface.

\noindent
In the original study the leaflet is a two-dimensional nearly incompressible hyperelastic continuum discretised by quadrilateral finite elements and immersed in a non-conforming Eulerian finite element fluid mesh. The two are coupled through a smoothed Dirac-delta-like interpolation operator that both spreads the interaction forces to the fluid and interpolates the fluid velocity onto the solid, which generalises Peskin's immersed boundary method \citep{peskin_immersed_2002} by removing the assumption of a fibre-like solid and the requirement of a uniform background mesh \citep{de_hart_threedimensional_2003}. The leaflet response is fully resolved as a two-dimensional continuum, including through-thickness stretching and shearing and the surrounding tractions on its faces. The present approach instead reduces the structure to a slender Simo--Reissner beam whose centreline coincides with the leaflet mid-line, with the rectangular cross-section entering only through its inertia and bending stiffness, and couples it through the actuator-line operators of Section~\ref{sec:almCouplingTheory}. The leaflet therefore exerts a force on the fluid without occupying a discrete volume in the CFD mesh, and the channel mesh remains static. The framework is lighter, but it assumes the leaflet is loaded by quasi-steady normal drag rather than by resolved surface tractions, and whether that assumption holds in the oscillatory low-Reynolds regime is what the benchmark assesses.

\subsection{Problem definition}
\label{sec:leaflet_geometry}
\begin{figure}[!ht]
\centering
\includegraphics[width=0.88\textwidth]{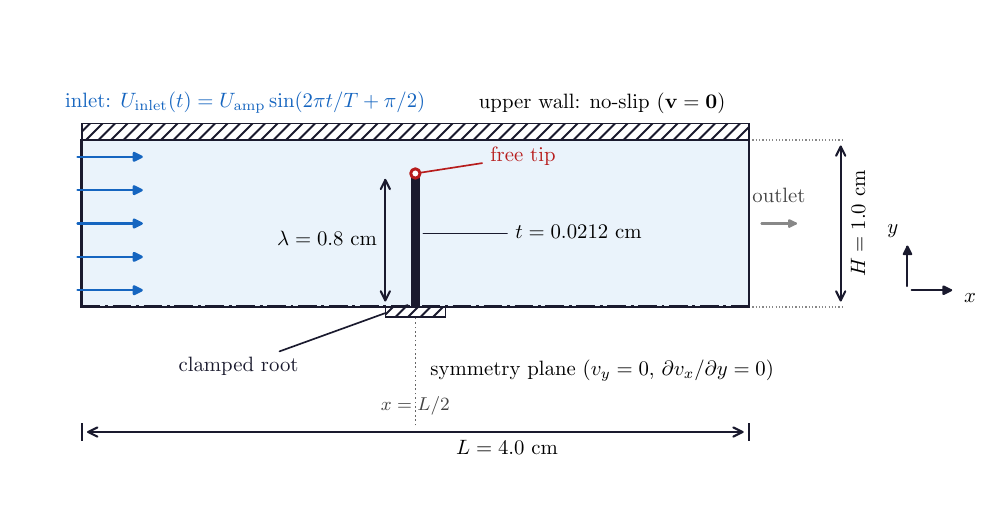}
\caption{The leaflet-in-channel verification problem, drawn as the simulated half-domain. The leaflet, of length $\lambda$ and thickness $t_{\ell}$, is clamped at the channel mid-plane at $x = L/2$ and free at its tip. A sinusoidal channel flow is imposed at the inlet, the opposite wall is no-slip, and the lower edge is a symmetry boundary. Dimensions reproduce \citet{zhang_immersed_2007}, Section~3.2.}
\label{fig:leaflet_schematic}
\end{figure}

The geometry is presented in Figure~\ref{fig:leaflet_schematic} and follows \citet{zhang_immersed_2007}, Section~3.2. The fluid domain is a rectangular channel of length $L = 4~\mathrm{cm}$ and full width $2H = 2~\mathrm{cm}$, with the origin where the inlet plane meets the channel mid-plane. The leaflet is a thin flexible body of length $\lambda = 0.8~\mathrm{cm}$ and thickness $t_{\ell} = 0.0212~\mathrm{cm}$, where the subscript distinguishes it from the time $t$, clamped at the mid-length of the channel and free at the other end. The clamped end is anchored on the channel mid-plane, so only one half of the original domain is simulated, with the mid-plane treated as a symmetry boundary carrying the leaflet root and a simulated transverse extent of $H = 1~\mathrm{cm}$. As in Section~\ref{sec:beamtunnel}, the problem is solved on a three-dimensional fluid mesh one cell thick in the spanwise direction $z$, here of depth $t_{\ell}$, with symmetry conditions on the two faces normal to $z$, so the flow does not vary across the span while the beam keeps its physical cross-section.

\noindent
A spatially uniform, time-periodic horizontal velocity is prescribed at the inlet,
\begin{equation}
U_{\mathrm{inlet}}(t) = U_{\mathrm{amp}} \sin\!\left(\frac{2\pi t}{T} + \frac{\pi}{2}\right),
\label{eq:leaflet_inlet}
\end{equation}
so the flow begins at its peak amplitude, the phase shift being chosen for consistency with the original paper in which the leaflet is released from rest into a fully developed inflow. The downstream boundary is an outflow with a fixed reference pressure and zero normal gradient of velocity, the upper boundary a stationary no-slip wall, and the lower boundary a symmetry plane that simultaneously models the channel mid-plane and anchors the leaflet root. Both the fluid and the leaflet start at rest with the leaflet undeformed.

\noindent
The fluid is a Newtonian incompressible liquid of density $10^{3}~\mathrm{kg\,m^{-3}}$, and the leaflet has density $\rho_{s} = 6\times10^{3}~\mathrm{kg\,m^{-3}}$, Young's modulus $E = 10^{6}~\mathrm{Pa}$, and Poisson ratio $\nu = 0.5$. In the present beam formulation the Poisson ratio enters only through the shear modulus $G = E/[2(1+\nu)]$, evaluated at the reference value for consistency with the original constitutive description. The regime is characterised by a Reynolds number $\mathrm{Re} = \rho\,U_{\mathrm{amp}}\,H / \mu$ and a Strouhal number $\mathrm{St} = H / (U_{\mathrm{amp}}\,T)$, defined as in the reference paper. The four combinations used here are listed in Table~\ref{tab:leaflet_cases}.

\begin{table}[!ht]
\centering
\small
\caption{Forcing parameters for the four (Re, St) cases of \citet{zhang_immersed_2007}, Section~3.2, with the cross-section Reynolds and Keulegan--Carpenter numbers $\mathrm{Re}_{t} = \rho\,U_{\mathrm{amp}}\,t_{\ell}/\mu$ and $\mathrm{KC} = U_{\mathrm{amp}}\,T/t_{\ell}$. All cases share $\rho = 10^{3}~\mathrm{kg\,m^{-3}}$, $H = 10^{-2}~\mathrm{m}$, and $t_{\ell} = 2.12\times10^{-4}~\mathrm{m}$. The last column is the normal drag coefficient passed to the closure, taken from the rigid-plate channel calibration of Section~\ref{sec:leaflet_cd_calibration} at each case's own Reynolds number.}
\label{tab:leaflet_cases}
\setlength{\tabcolsep}{6pt}
\renewcommand{\arraystretch}{1.15}
\begin{tabular}{c c c c c c c c}
\toprule
Case & $U_{\mathrm{amp}}$ (m\,s$^{-1}$) & $T$ (s) & $\mu$ (Pa\,s) & $\mathrm{Re}$ & $\mathrm{St}$ & $\mathrm{KC}$ & $C_{dn}$ \\
\midrule
C1 & $0.01$ & $1$ & $10^{-2}$ & $10$ & $1.0$ & $47$ & $108$ \\
C2 & $0.01$ & $1$ & $10^{-1}$ & $1$  & $1.0$ & $47$ & $332$ \\
C3 & $0.01$ & $2$ & $10^{-1}$ & $1$  & $0.5$ & $94$ & $332$ \\
C4 & $0.01$ & $2$ & $10^{-2}$ & $10$ & $0.5$ & $94$ & $108$ \\
\bottomrule
\end{tabular}
\end{table}

\subsection{Steady-CFD calibration of the normal drag coefficient}
\label{sec:leaflet_cd_calibration}

\begin{figure}[!ht]
\centering
\includegraphics[width=0.88\textwidth]{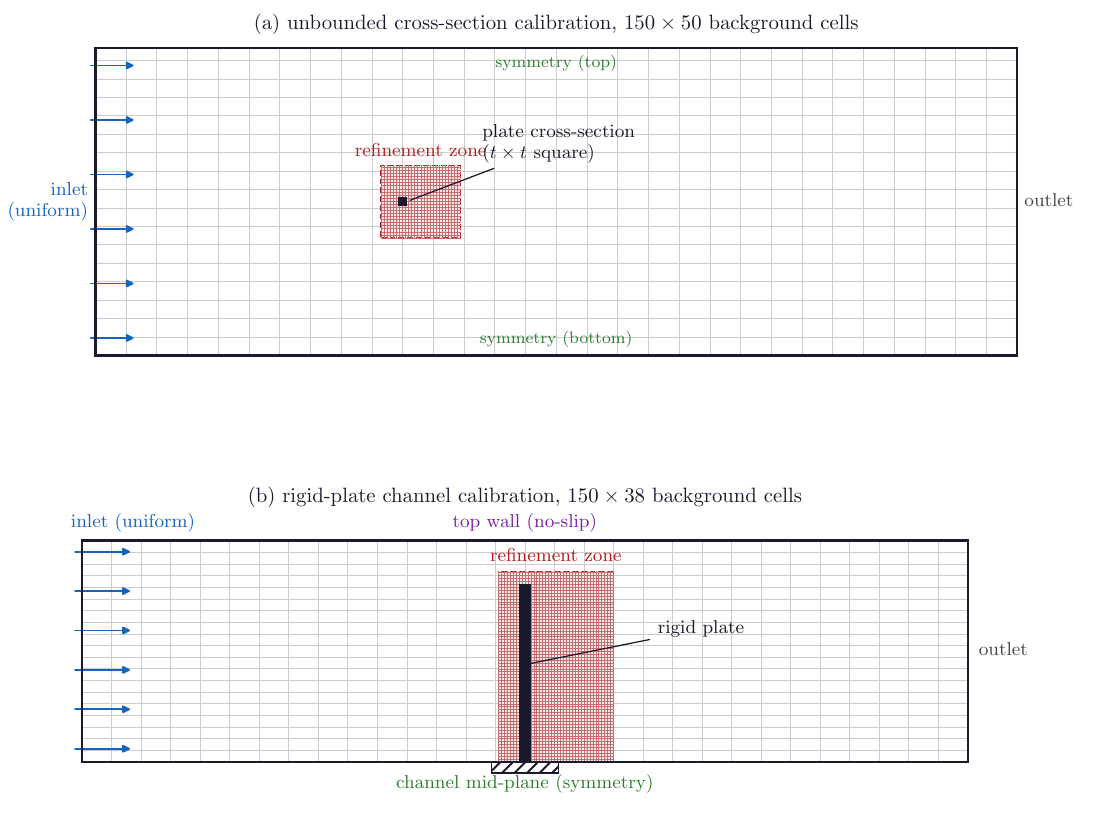}
\caption{The two steady-CFD drag-calibration setups. (a) Unbounded cross-section in plan view, with the square cross-section of side $t_{\ell}$ in an unbounded stream, symmetry on the two lateral boundaries and on the two faces that bound the thin domain along the plate axis, returning the free-stream drag ($A_{\mathrm{ref}} = t_{\ell}^{2}$). (b) Rigid-plate channel, with a rigid plate of length $\lambda$ and thickness $t_{\ell}$ in the coupled channel, the mid-plane treated as a symmetry plane and the opposite wall as no-slip, returning the in-channel drag ($A_{\mathrm{ref}} = t_{\ell}\,\lambda$). A local refinement zone is shown in red.}
\label{fig:leaflet_dragCalib_mesh}
\end{figure}

\noindent
As in Section~\ref{sec:beamtunnel_cd_calibration}, $C_{dn}$ must reflect the drag that would act on a rigid leaflet of the same cross-section in the same channel at the same Reynolds number. Two steady-CFD setups are used, illustrated in Figure~\ref{fig:leaflet_dragCalib_mesh}. The first is an unbounded cross-section calibration, which measures the free-stream drag $C_{d}^{\infty}$ of the leaflet cross-section in isolation. The second is a rigid-plate channel calibration, which measures the in-channel drag $C_{d}$ of the full leaflet in the coupled domain. Both are solved as steady flows with the SIMPLE algorithm, as implemented in the \texttt{simpleFoam} solver of OpenFOAM. The meshes are generated by octree-based splitting of a hexahedral background mesh around the plate surface, as implemented in the \texttt{snappyHexMesh} utility of OpenFOAM, where each refinement level halves the cell size. Three levels are applied in a zone enveloping the plate and four to five on the plate surface, giving a smallest cell at the plate of $1.3\times10^{-5}~\mathrm{m}$ and $8\times10^{-6}~\mathrm{m}$ respectively. The channel calibration uses the same $150\times38$ background mesh as the coupled simulation, which removes any mesh-related component of the discrepancy between the calibration result and the value seen by the closure at run time.

\begin{figure}[!ht]
\centering
\begin{subfigure}[b]{0.48\textwidth}
  \centering
  \includegraphics[width=\linewidth]{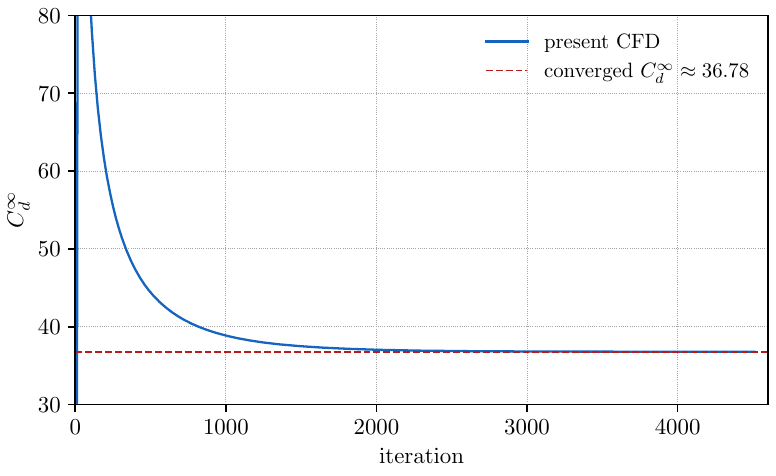}
  \caption{}
\end{subfigure}\hfill
\begin{subfigure}[b]{0.48\textwidth}
  \centering
  \includegraphics[width=\linewidth]{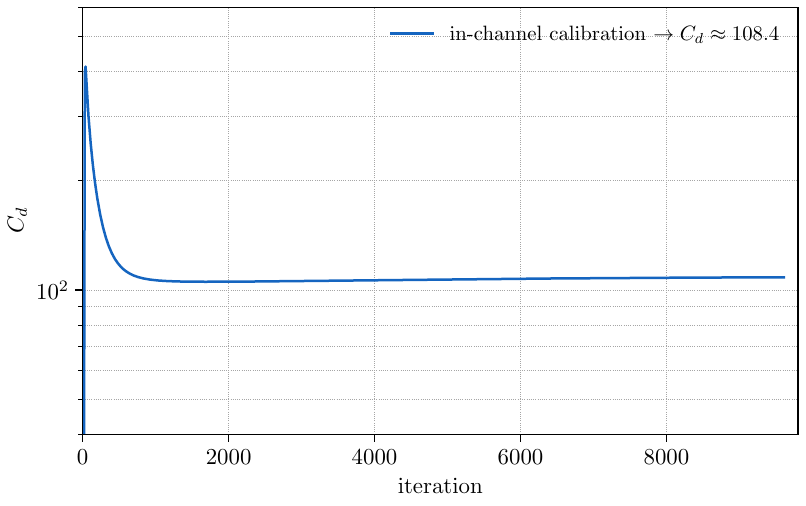}
  \caption{}
\end{subfigure}
\caption{Convergence of the two steady-CFD drag calibrations. (a) Free-stream drag coefficient for the unbounded cross-section at $\mathrm{Re}_{t} = 0.212$, converging to $36.8$. (b) In-channel drag coefficient at $\mathrm{Re} = 10$, converging to $C_{d} \approx 108$.}
\label{fig:leaflet_dragCalib}
\end{figure}

\noindent
The unbounded configuration has four boundaries parallel to the flow, namely the two lateral boundaries drawn at the top and bottom of Figure~\ref{fig:leaflet_dragCalib_mesh}(a) and the two faces normal to the plate axis, which bound the thin domain above and below the plane of that drawing. All four are symmetry planes. The symmetry on the two faces normal to the plate axis makes the cross-section behave as an infinitely long cylinder, so the case is equivalent to a two-dimensional one. Figure~\ref{fig:leaflet_dragCalib}(a) shows the trace decreasing monotonically to $C_{d}^{\infty} \approx 36.8$ at the C1 cross-section Reynolds number $\mathrm{Re}_{t} = 0.212$, between the Oseen estimate \citep{lamb_hydrodynamics_1932} for a low-Reynolds two-dimensional cylinder ($\approx 25$) and the empirical estimate of \citet{hoerner_fluid_1965} for a flat-plate cross-section ($\approx 50$). This is not the coefficient passed to the closure, because the leaflet sits in a channel of half-width $H = 10\,t_{\ell}$ where the blockage and the walls modify the drag. Its role is a literature-comparable check on the cross-section drag and a baseline for the confinement amplification.

\noindent
In the channel configuration the lower boundary is the channel mid-plane and is treated as a symmetry plane, because reflecting the half-domain across it recovers a plate floating freely at the centreline of a full channel of width $2H$, which is the reference geometry. Figure~\ref{fig:leaflet_dragCalib}(b) shows the in-channel coefficient converging to $C_{d} \approx 108$ at $\mathrm{Re} = 10$, almost three times the unbounded value. That ratio of $2.9$ quantifies the cumulative effect of blockage, aspect ratio, and viscous coupling along the full plate length, and decomposing the force shows that pressure accounts for approximately $98\%$ of the total, so the in-channel drag is dominated by upstream stagnation and leeward suction. At $\mathrm{Re} = 1$ the same calibration gives $C_{d} \approx 332$, about three times larger, reflecting the thickening of the viscous boundary layer around the plate, which effectively enlarges the bluff cross-section and raises the upstream-face pressure rise. A single calibration cannot therefore be transferred between the two Reynolds numbers, and each coupled case uses the value calibrated at its own. This is a real limitation of the closure, since a coefficient that has to be re-established whenever the Reynolds number or the blockage changes restricts where the method can be applied without a preparatory run. The value passed to the closure is the in-channel coefficient at the matching Reynolds number, identified with the closure input through $C_{dn} = C_{d}$ as checked in Section~\ref{sec:leaflet_probe}, and the tangential coefficient is fixed at $C_{dt} = 0.01$ as before.

\subsection{Check of the identification $C_{dn} = C_{d}$}
\label{sec:leaflet_probe}

\noindent
Identifying $C_{dn}$ with $C_{d}$ is not automatic, because the two are normalised by different velocities. This subsection checks whether the coefficient calibrated in Section~\ref{sec:leaflet_cd_calibration} can be passed to the closure unchanged. The check samples the converged channel calibration and uses no new configuration, so it verifies the calibration rather than adding a second one. At each beam node the closure samples the local fluid velocity $U_{s}$ at the fixed offset $r_{s} = 3.18\times10^{-3}~\mathrm{m}$ upstream of the node and applies a normal load per unit length of $\tfrac{1}{2}\rho\,C_{dn}\,t_{\ell}\,U_{s}^{2}$, whereas the steady CFD normalises the integrated rigid-plate drag by the channel inlet velocity, giving an equivalent force per unit length of $\tfrac{1}{2}\rho\,C_{d}\,t_{\ell}\,U_{\mathrm{in}}^{2}$ averaged over the plate length. Requiring the two to apply the same force per unit length on average along the beam, so that the calibration carries unchanged into the coupled run, gives
\begin{equation}
\label{eq:cdn_correction}
C_{dn} \;=\; C_{d}\,
\frac{U_{\mathrm{in}}^{2}}{\langle U_{s}^{2}\rangle_{\mathrm{beam}}},
\qquad
\langle U_{s}^{2}\rangle_{\mathrm{beam}} \;\equiv\; \frac{1}{\lambda}\int_{0}^{\lambda} U_{s}^{2}(y)\,\mathrm{d}y,
\end{equation}
so the identification is exact if and only if the beam-averaged squared sampled velocity equals the inlet squared velocity.

\begin{figure}[!ht]
\centering
\includegraphics[width=0.72\textwidth]{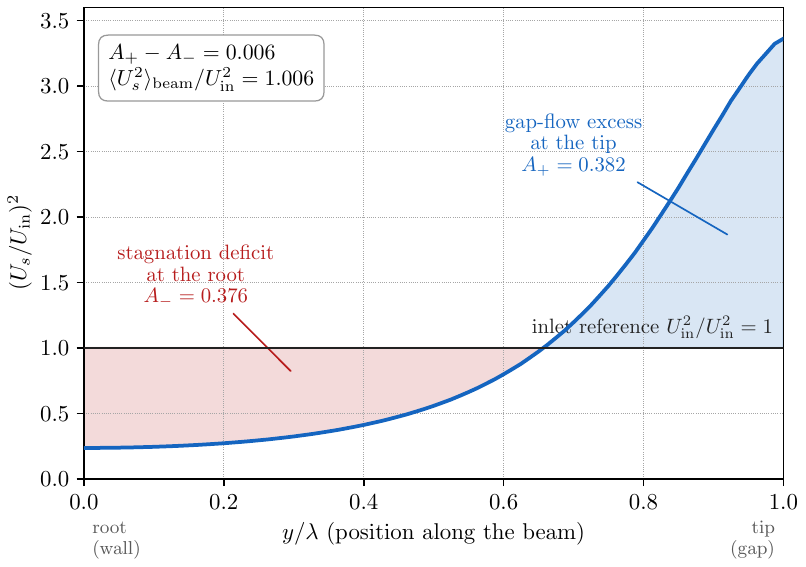}
\caption{Squared sampled velocity $(U_{s}/U_{\mathrm{in}})^{2}$ along the beam at the upstream-sampling line, in the converged $\mathrm{Re} = 10$ channel calibration. For $C_{dn} = C_{d}$ to hold exactly, the beam average must equal the inlet reference (horizontal line). The deficit area below the reference near the root ($A_- = 0.376$) is cancelled almost exactly by the excess above it near the tip ($A_+ = 0.382$), leaving the beam average at $1.006$.}
\label{fig:leaflet_probe_profile}
\end{figure}

\noindent
To check the condition, the converged $\mathrm{Re} = 10$ channel calibration is sampled along a line at $x = L/2 - r_{s}$, with the streamwise velocity extracted at $81$ uniformly spaced points spanning the beam. Squaring the profile and applying Eq.~\eqref{eq:cdn_correction} gives $\langle U_{s}^{2}\rangle_{\mathrm{beam}}/U_{\mathrm{in}}^{2} = 1.006$. Figure~\ref{fig:leaflet_probe_profile} shows why. The curve varies strongly along the beam, reduced to $0.24$ at the root where the plate stagnation slows the flow and rising to $3.36$ at the tip where the fluid accelerates through the narrow gap to the opposite wall, but the two departures balance almost exactly, so the average sits within $0.6\%$ of the reference and the correction can be dropped.

\noindent
That the two nearly cancel is specific to this geometry. Without the plate, the sampling line would lie in the undisturbed fully developed channel flow, whose Poiseuille profile peaks at $1.5\,U_{\mathrm{in}}$ on the symmetry plane and integrates to $1.474$, so a leaflet in undisturbed flow would see a $47\%$ enhancement of the squared driving velocity and the correction could not be dropped. The actual plate blocks $80\%$ of the channel half-width, so most of the upstream flow either stagnates near the root or accelerates through the tip gap, which redistributes the squared profile sharply away from the parabolic shape. In a less confined channel the same probe-line analysis would return a value distinctly above unity and the explicit correction would have to be retained.

\subsection{Time step and flow snapshots}
\label{sec:leaflet_snapshots}

\begin{figure}[!ht]
\centering
\includegraphics[width=0.90\textwidth]{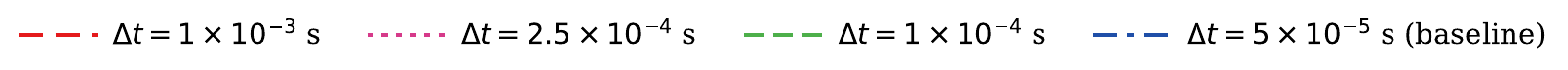}\\[0.3em]
\includegraphics[width=0.68\textwidth]{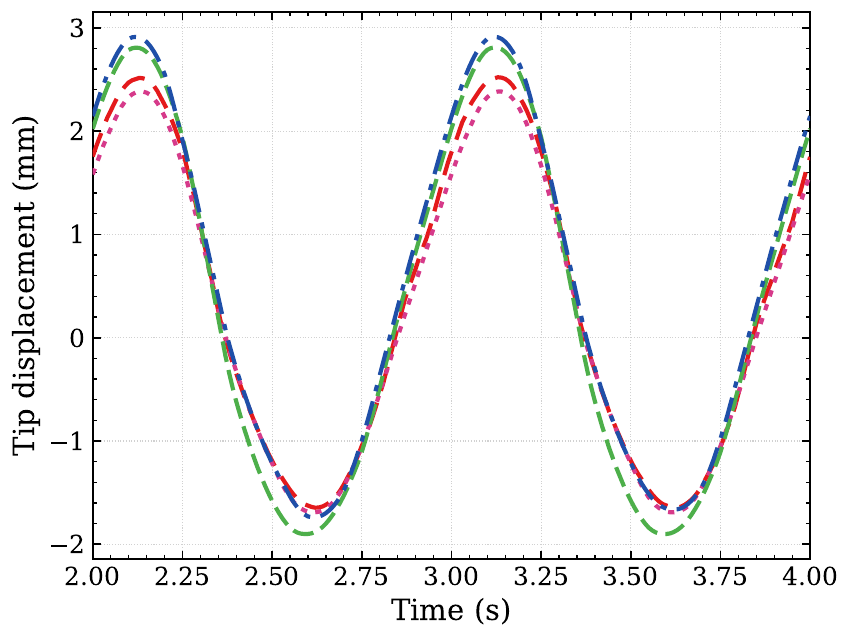}
\caption{Time-step convergence of the leaflet tip horizontal displacement for case C1 ($\mathrm{Re} = 10$, $\mathrm{St} = 1$), over the converged window $t \in [2, 4]~\mathrm{s}$, with $\Delta t = 5\times10^{-5}~\mathrm{s}$ taken as the converged reference.}
\label{fig:leaflet_dt_study}
\end{figure}

\noindent
The temporal resolution is selected from a study on case C1, in which four time steps spanning a factor of twenty are run to statistical convergence with all other inputs held fixed (Figure~\ref{fig:leaflet_dt_study}). The two finest steps collapse onto a common trace, with half-amplitudes of $2.324~\mathrm{mm}$ at $\Delta t = 5\times10^{-5}~\mathrm{s}$ and $2.355~\mathrm{mm}$ at $10^{-4}~\mathrm{s}$, a difference of $1.3\%$ and no visible phase drift over the two reported periods. The two coarser steps under-predict the response, giving $2.036~\mathrm{mm}$ at $2.5\times10^{-4}~\mathrm{s}$ and $2.084~\mathrm{mm}$ at $10^{-3}~\mathrm{s}$, so the coarsest step gives the larger of the two values. The four steps are not related by a constant refinement ratio, the successive reductions being by factors of $4$, $2.5$, and $2$, so they are not used to estimate an order of convergence. The non-monotonic ordering of the two coarser values indicates that at time steps of $2.5\times10^{-4}~\mathrm{s}$ and larger the explicit sampling and quasi-steady closure are no longer integrated with sufficient resolution to track the half-cycle build-up of tip excursion. The finest setting is adopted for all four cases, so that the same step is shared by the higher-$\mathrm{KC}$ cases C3 and C4 without further verification.

\begin{figure}[!ht]
\centering
\begin{subfigure}[b]{0.48\textwidth}
\centering
\includegraphics[width=\textwidth]{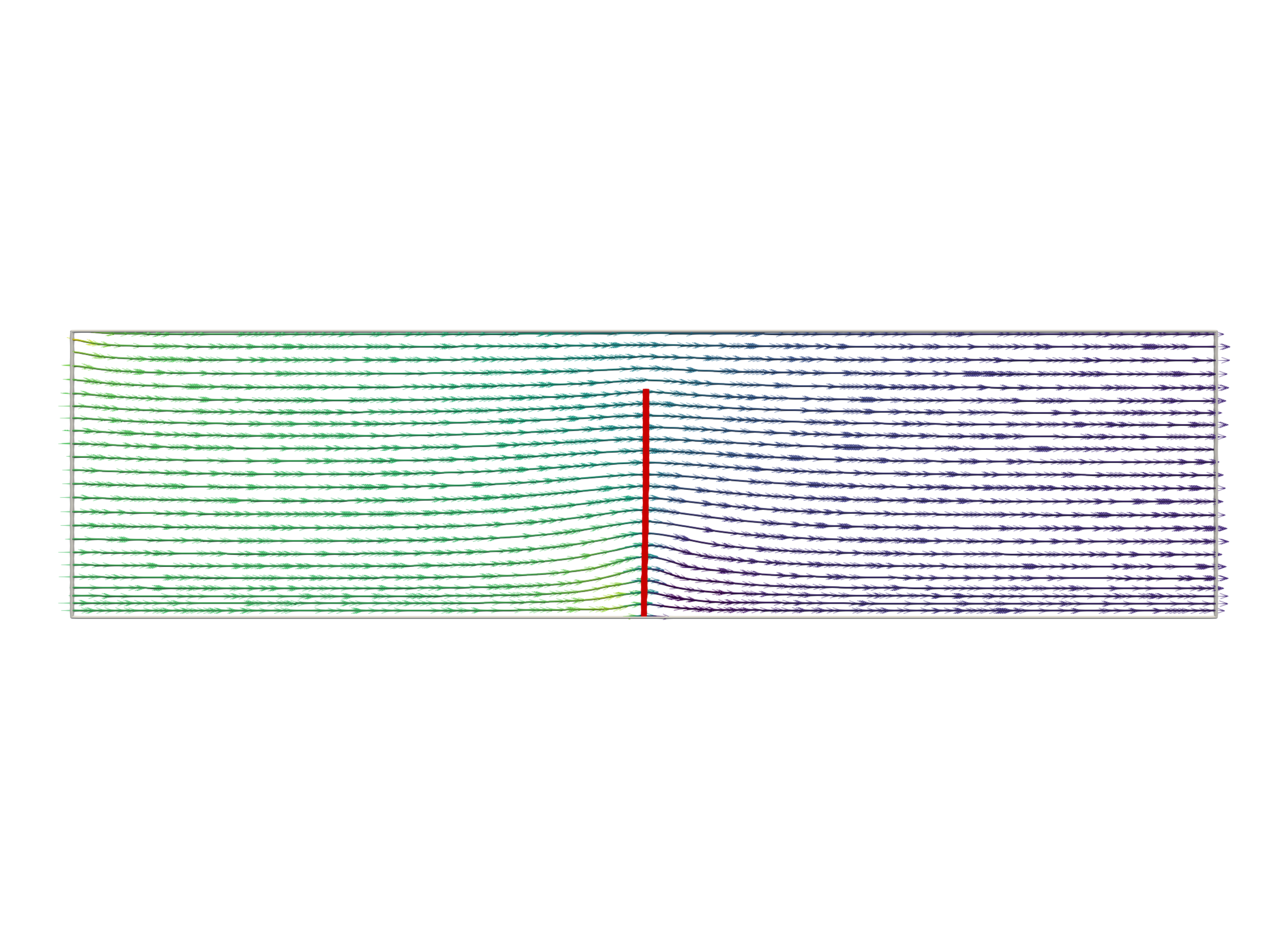}
\caption{$t = 0$}
\end{subfigure}\hfill
\begin{subfigure}[b]{0.48\textwidth}
\centering
\includegraphics[width=\textwidth]{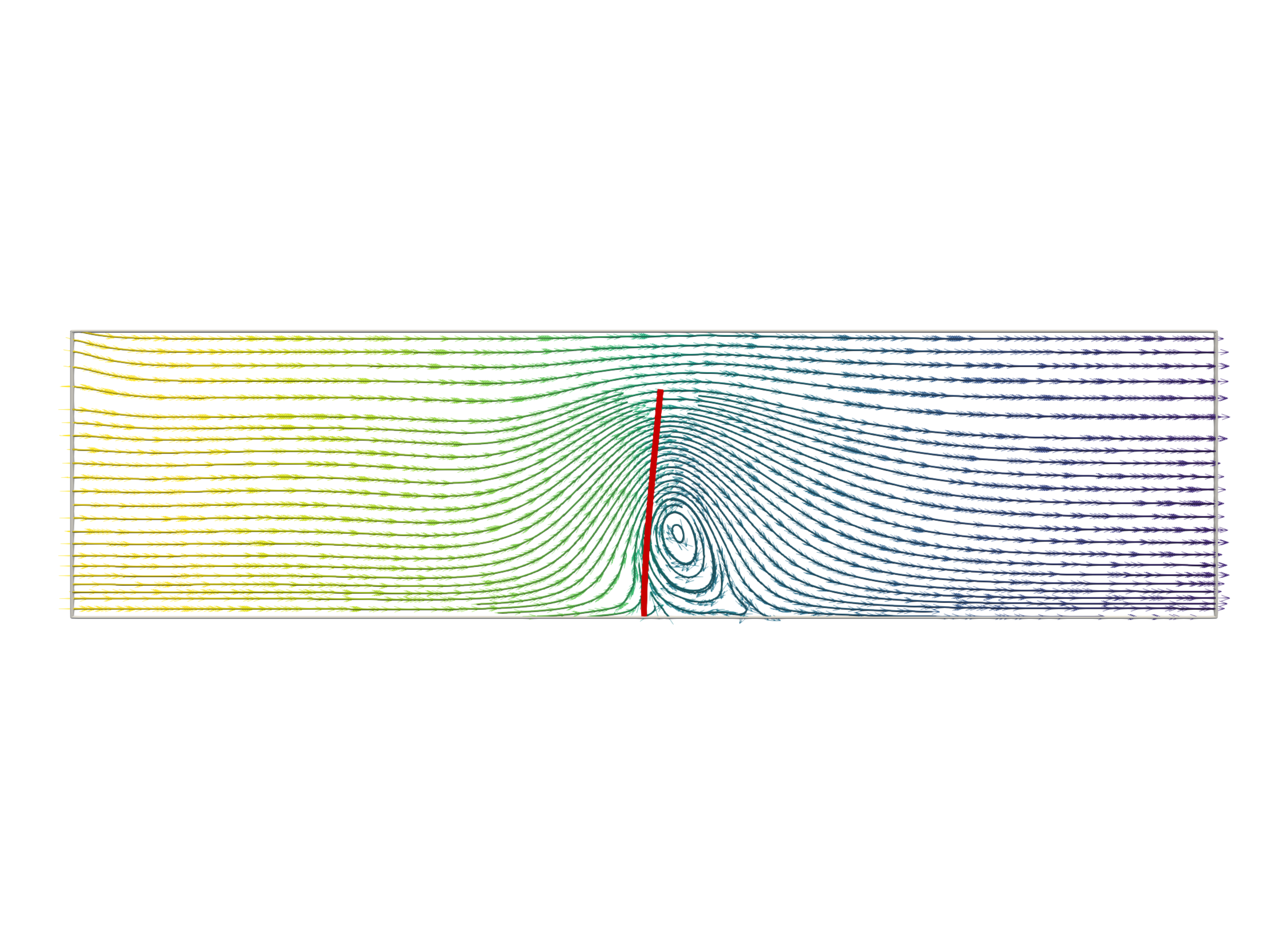}
\caption{$t = T/8$}
\end{subfigure}\\[0.4em]
\begin{subfigure}[b]{0.48\textwidth}
\centering
\includegraphics[width=\textwidth]{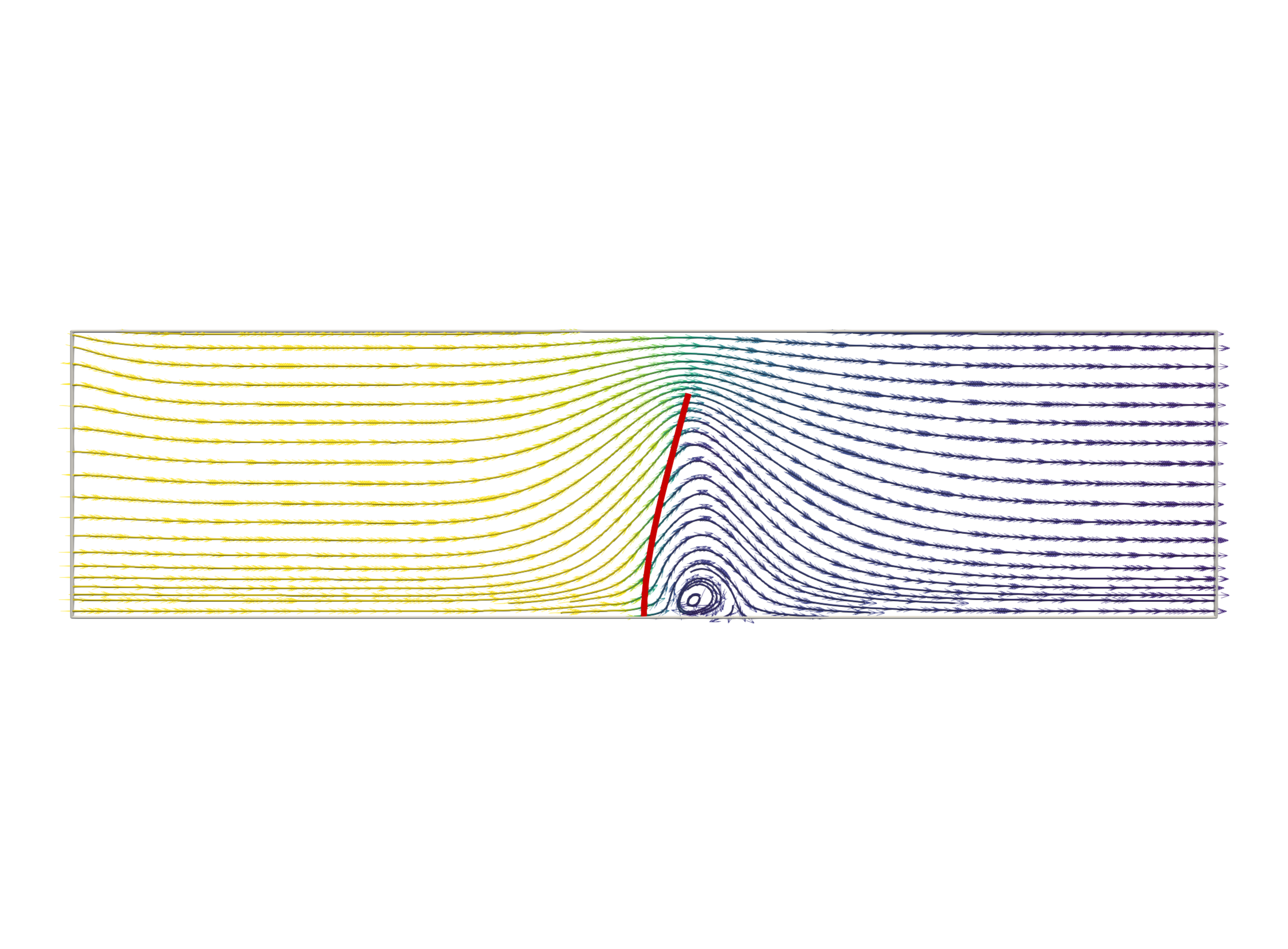}
\caption{$t = T/4$}
\end{subfigure}\hfill
\begin{subfigure}[b]{0.48\textwidth}
\centering
\includegraphics[width=\textwidth]{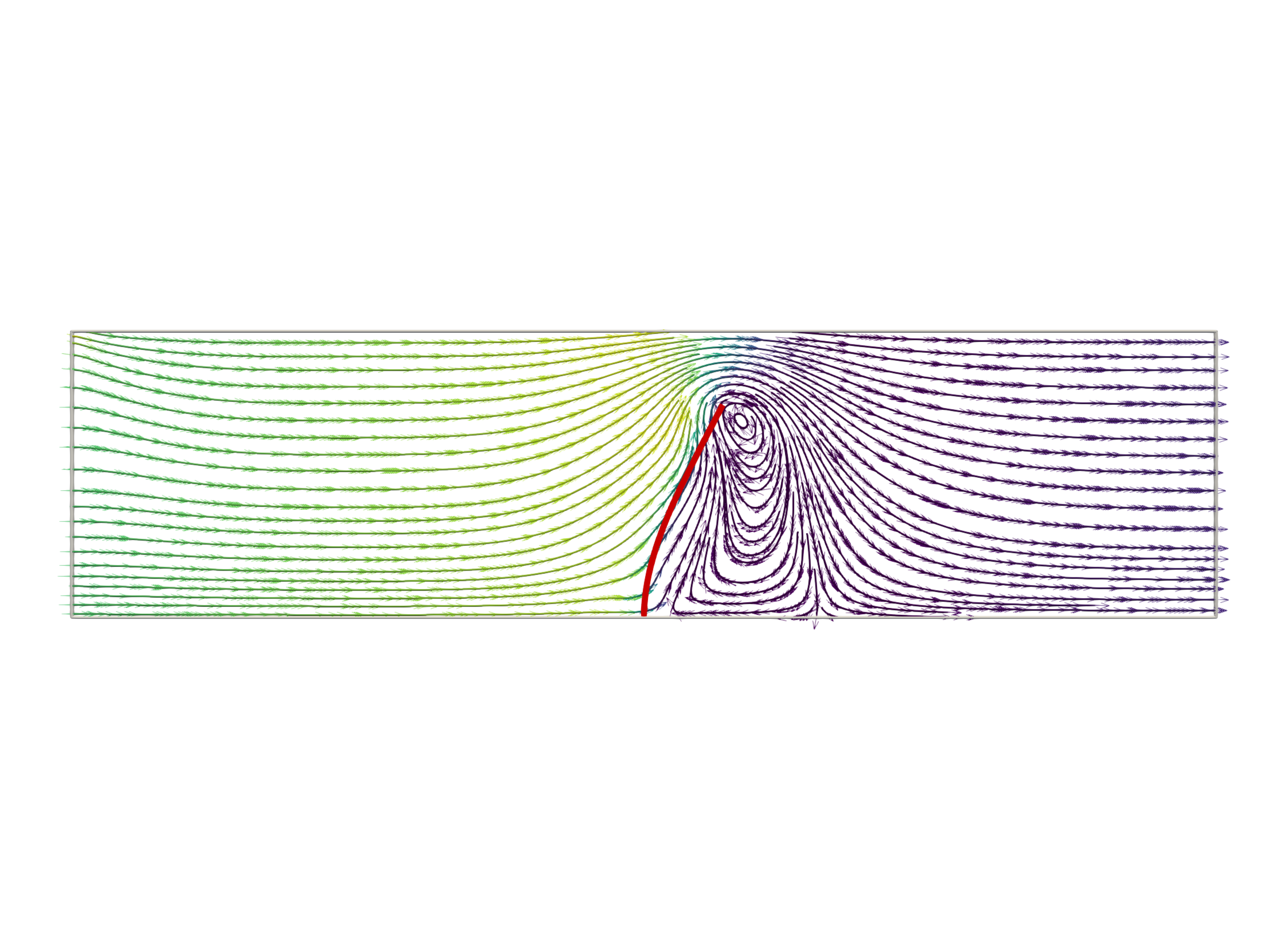}
\caption{$t = 3T/8$}
\end{subfigure}\\[0.4em]
\includegraphics[width=0.80\textwidth]{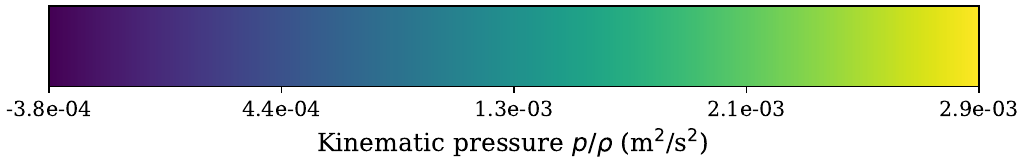}
\caption{Mid-plane streamlines around the leaflet, coloured by kinematic pressure $p/\rho$ (shared colour bar below), over the first half of a converged forcing period of case C1 ($\mathrm{Re} = 10$, $\mathrm{St} = 1$, $T = 1~\mathrm{s}$), with the deflected actuator line overlaid in red. The second half of the period mirrors the first.}
\label{fig:leaflet_snapshots_panel}
\end{figure}

\noindent
Figure~\ref{fig:leaflet_snapshots_panel} shows the converged field over the first half of a forcing period of case C1, at intervals of $T/8$, with the in-plane streamlines coloured by kinematic pressure and the deflected actuator line overlaid in red. The second half of the period is the mirror image of the first and is not reproduced. Because the imposed inlet signal of Eq.~\eqref{eq:leaflet_inlet} carries a $+\pi/2$ phase shift, the velocity extrema fall at $t = 0$ and $T/2$ and the zero crossings at $T/4$ and $3T/4$. At $t = 0$ the leaflet attains its instantaneous maximum tip excursion and a clear pressure dipole develops across the line, with a stagnation patch on the upstream face and a suction patch on the downstream face that switch sides at the opposite extremum. At $T/4$ the streamlines straighten out as the bulk flow reverses and the pressure field is comparatively uniform. The small recirculation pair that lingers near the deflected leaflet at that phase is the residual of the previous half-cycle wake, which confirms that the quasi-steady closure generates no wake of its own, only the redistribution of the resolved field around the moving line of momentum sinks.

\noindent
Three features of these snapshots bear on the comparison with the reference in Section~\ref{sec:leaflet_results}. The absence of any small-scale vortex shedding throughout the period is consistent with the low cross-section Reynolds number of case C1 and supports the use of a $\mathrm{KC}$-independent drag coefficient. The tip excursion in the peak frame is visually consistent with the $\pm 2.324~\mathrm{mm}$ half-amplitude of the time-step study, which checks the integral trace at field level. The pressure dipole is generated entirely by the projected body force back-reacting on the fluid, so although the leaflet is geometrically absent from the mesh, its presence in the momentum equation produces the upstream-stagnation and downstream-suction structure that a body-fitted simulation would resolve at the plate surface, with the redistribution length set by the kernel width rather than by the plate thickness.

\subsection{Verification across the four (Re, St) cases}
\label{sec:leaflet_results}

\noindent
The coupled simulation is run at the four combinations of Table~\ref{tab:leaflet_cases}. All four runs share the same beam discretisation, fluid mesh, kernel width, and time step, and only the inlet amplitude, the forcing period, and the matching $C_{dn}$ vary. The reference traces are digitised from Figs.~13 and 14 of the reference paper.

\begin{figure}[!ht]
\centering
\includegraphics[width=0.90\textwidth]{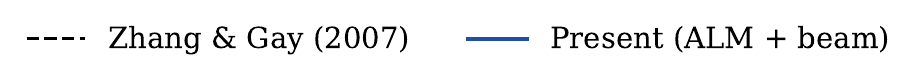}\\[0.3em]
\includegraphics[width=\textwidth]{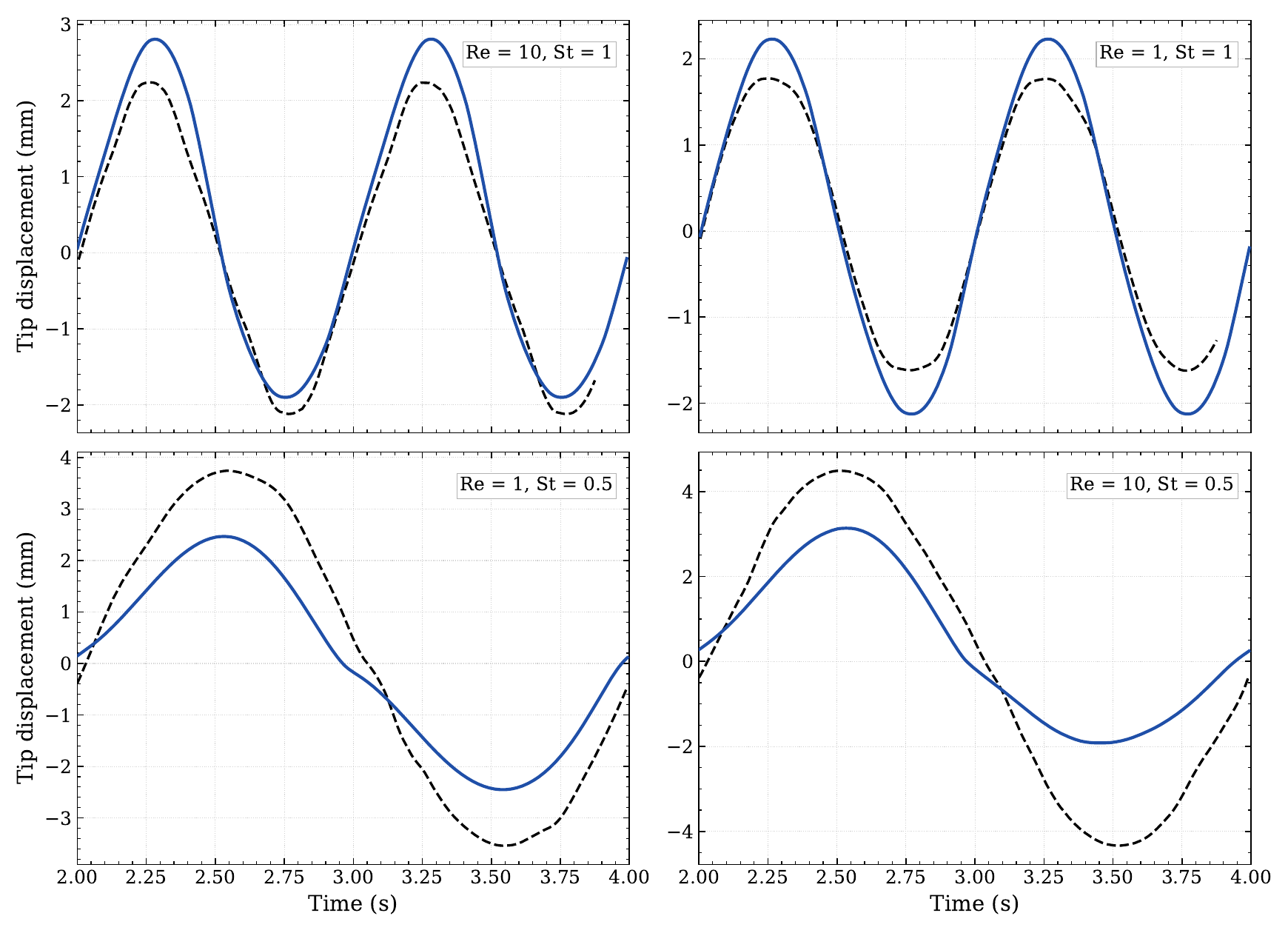}
\caption{Tip horizontal displacement traces for the four (Re, St) cases of \citet{zhang_immersed_2007}, Section~3.2. Dashed black lines are the digitised reference traces and solid blue lines the present result. Each trace has been phase-aligned to the reference by cross-correlation and windowed to the converged interval $t \in [2, 4]~\mathrm{s}$.}
\label{fig:leaflet_results_4cases}
\end{figure}

\noindent
Figure~\ref{fig:leaflet_results_4cases} compares the simulated tip displacement against the reference over the converged window. The agreement falls into two families. The cases at $\mathrm{St} = 1$ (C1 and C2) reproduce the reference period exactly and overshoot its peak excursion, the present values being $\pm 2.324~\mathrm{mm}$ against $\pm 2.2~\mathrm{mm}$ at C1 and $\pm 2.2~\mathrm{mm}$ against $\pm 1.7~\mathrm{mm}$ at C2, with the phase coinciding within plotting accuracy and the cycle-to-cycle shape closely matched. The cases at $\mathrm{St} = 0.5$ (C3 and C4) behave differently. The simulated peak excursion is essentially fixed near $\pm 2.5~\mathrm{mm}$, whereas the reference reaches $\pm 3.7~\mathrm{mm}$ at C3 and $\pm 4.5~\mathrm{mm}$ at C4, so the simulation under-predicts the reference peak excursion by $30\%$ in both cases. The period is again reproduced exactly, so the discrepancy is purely in amplitude.

\paragraph{Origin of the Strouhal-dependent discrepancy.}
The quasi-steady closure of Eq.~\eqref{eq:almLineForce} depends only on the instantaneous sampled velocity and carries no fluid-inertia term proportional to the acceleration, whereas in the IFEM reference the leaflet is immersed in a fully resolved Navier--Stokes field, so its force balance includes an added-mass term scaling with the local fluid acceleration and a history term depending on the past evolution of the relative motion. Halving the Strouhal number doubles the forcing period and halves the peak fluid acceleration at the inlet, and for a stationary member in an oscillating stream the classical Morison scaling would then predict the \emph{opposite} of the observed trend, since the ratio of inertia to drag scales as $1/\mathrm{KC}$ \citep{sarpkaya_mechanics_1981}. The growing under-prediction at $\mathrm{St} = 0.5$ must therefore originate in the unsteady forces associated with the leaflet's own motion, which the fixed-body scaling does not describe. At the longer period the reference excursion roughly doubles, so the relative acceleration between the leaflet and the surrounding fluid, and with it the added-mass reaction on the moving structure, remains dynamically significant even as the inlet acceleration halves, while at $\mathrm{Re}_{t} \approx 0.2$ the viscous history contribution depends on the past relative motion in a way that no instantaneous coefficient can represent. Apportioning the deficit among these terms would require a force decomposition of the reference solution, which the published data do not provide, so what the present results establish is the operational boundary of the closure rather than the split between the missing terms.

\noindent
One further observation supports this picture. The drag coefficient of the rigid plate is essentially independent of the Keulegan--Carpenter number in this regime, an unsteady re-run of the rigid-plate calibration at $\mathrm{KC} = 47$ and $94$ giving $C_{d} \approx 107$ in both. The textbook $C_{dn}(\mathrm{KC})$ trend that operates at higher Reynolds numbers does not apply at $\mathrm{Re}_{t} \approx 0.2$, where the wake remains attached and vortex shedding is suppressed. The Strouhal-dependent discrepancy is therefore not driven by a Strouhal-dependent drag coefficient on the rigid plate, but by the unsteady inertia of the surrounding fluid acting on the moving leaflet, which is the term that the classical Morison equation separates from the quadratic drag and that the present closure does not carry.

\noindent
The deterioration is documented quantitatively rather than removed by case-specific recalibration, since replacing the steady-CFD $C_{dn}$ with a per-case empirical value would mask the modelling limitation behind a parameter sweep. The framework is therefore verified over a clearly defined range of conditions, namely the drag-dominated regime represented here by the Strouhal number of $1$ rather than $0.5$, and extending it further would require an unsteady closure that adds explicit added-mass and history terms to Eq.~\eqref{eq:almLineForce}.


\section{Wave-induced motion of a flexible vegetation stem}
\label{sec:flexveg}

\noindent
The two preceding benchmarks verified the coupling against reference numerical solutions in single-phase channel flows. The final case applies it to a vegetation stem swaying under free-surface waves, which is the first validation against a physical experiment and the first use of the coupling in a two-phase setting, where the beam is driven by the orbital velocities of a propagating wave rather than by a unidirectional or uniformly oscillating channel flow. It is also the first case in which the structure barely confines the flow. The blade, $8~\mathrm{mm}$ wide in a flume $0.8~\mathrm{m}$ wide, blocks a small part of the flume cross-section, whereas the beam and the leaflet of the two channel benchmarks spanned $80\%$ of their simulated channel heights. Without that confinement the velocity approaching the stem stays close to the undisturbed wave kinematics, so the magnitude of the sampled velocity depends little on the sampling offset (Section~\ref{sec:flexveg_setup}). Flexible vegetation attenuates incident wave energy, modifies the mean flow and sediment transport within canopies, and is increasingly considered as a component of nature-based coastal defences, and predictive models for these processes rest on the ability to describe the motion of, and the force distribution along, an individual stem \citep{luhar_waveinduced_2016,jacobsen_waveinduced_2019}.

\noindent
The configuration reproduces the single-stem experiment of \citet{jacobsen_waveinduced_2019}, in which the motion of, and force on, individual plastic vegetation mimics under regular waves were measured in a laboratory wave flume. The data set suits the present purpose for three reasons. It provides the intra-wave kinematics along the entire stem rather than only at the tip, together with the base force and an estimate of the distributed loading, so the validation can interrogate the spatial structure of the coupled response. Its non-dimensional analysis classifies each test into quasi-static and dynamic response regimes, which allows the validation case to be chosen deliberately inside the range where the quasi-steady closure was shown to be accurate in Section~\ref{sec:leaflet_results}. And the authors derive measured drag coefficients from the experiment itself, so the coefficient required by the closure can be fixed a priori from published data. That transfer is possible only because of the weak confinement noted above, which is why the two channel benchmarks needed their own steady-CFD calibrations instead.

\subsection{Reference experiment and selection of the validation case}
\label{sec:flexveg_case_selection}
The experiment was performed in a wave flume at Delft University of Technology, $41.55~\mathrm{m}$ long, $0.8~\mathrm{m}$ wide and $0.9~\mathrm{m}$ deep, with a false bottom in the central test section accommodating a force transducer, so that the water depth was $0.65~\mathrm{m}$ at the wave paddle and $0.45~\mathrm{m}$ at the measuring location. A single stem was mounted on the transducer for each test, so the measurements characterise an isolated stem rather than a canopy. Surface elevation was recorded by seven wave gauges, the velocity with an electromagnetic flow meter traversed over the stem height, and the stem motion with a video camera at $25~\mathrm{Hz}$, giving the displacement along the stem at increments of $2.5~\mathrm{mm}$. Four plastic mimics were tested at two lengths under eight regular wave conditions, covering Cauchy numbers over five orders of magnitude.

\noindent
The present validation targets the blade labelled ``Mimic~3'' at the full length $l = 0.30~\mathrm{m}$, whose properties are listed in Table~\ref{tab:flexveg_properties}, under the wave condition designated H040T20. The choice is deliberate and follows directly from the modelling limitation quantified in the leaflet benchmark. \citet{jacobsen_waveinduced_2019} characterise the stem response by the Cauchy number, the Keulegan--Carpenter number and the relative stem length,
\begin{equation}
\label{eq:flexveg_nondim}
Ca = \frac{\rho\,\delta_y\,u_w^{2}\,l^{3}}{EI},
\qquad
KC = \frac{2\pi\,u_w}{\omega\,\delta_y},
\qquad
L = \frac{l\,\omega}{u_w},
\end{equation}
where $u_w$ is the characteristic depth-averaged root-mean-square orbital velocity and $\omega = 2\pi/T$ the cyclic wave frequency. The combination $CaL$ measures the ratio of hydrodynamic drag to elastic restoring stiffness, while $CaL/KC = (2/\pi^{2})(T_n\omega)^{2}$ controls whether the stem responds quasi-statically or dynamically. Following \citet{leclercq_reconfiguration_2018}, confirmed experimentally by \citet{jacobsen_waveinduced_2019}, the response is quasi-static, with the shape following the instantaneous force balance in a single spatial mode, for $CaL/KC < 0.20$, whereas larger values excite a dynamic response in which the added mass of the surrounding water amplifies the motion and introduces a second mode shape.

\noindent
This classification matches the range in which the present framework is accurate, since the closure of Eq.~\eqref{eq:almLineForce} carries no added-mass or history force and the leaflet benchmark showed that the framework under-predicts once fluid-inertia effects become dynamically significant. For Mimic~3 under H040T20 the measured characteristic velocity is $u_w = 7.7\times10^{-2}~\mathrm{m\,s^{-1}}$ and the orbital excursion is $a_w = u_w/\omega = 2.5\times10^{-2}~\mathrm{m}$, the natural length scale for the stem displacement throughout this section. These give $CaL/KC = 0.135$, inside the quasi-static regime, and the proper orthogonal decomposition reported in the reference confirms the placement, the motion being described entirely by the mean and a single mode. The inertia force reconstructed by the reference contributes less than $10\%$ of the maximum distributed force, so neglecting fluid inertia is justified a priori rather than assumed. By contrast the geometrically similar Mimic~2, a thinner blade of thickness $0.50~\mathrm{mm}$, gives $CaL/KC = 3.69$ under the same wave and would demand exactly the added-mass modelling that the present closure omits.

\noindent
One experimental artefact must be carried into the comparison. The reference reports a small negative mean deflection of this blade in stagnant water, attributed to anisotropic pre-tension introduced during manufacturing, which biases the mean position of the measured motion by about $-0.015\,l$ at the tip without affecting the oscillatory content. The comparison below therefore concentrates on the oscillation range and shape, and treats the mean offset as an acknowledged experimental bias.

\noindent
Three further wave conditions of the campaign are simulated with the same stem, summarised in Table~\ref{tab:flexveg_conditions}, chosen to span the widest range of forcing that remains defensible for the quasi-steady closure. H070T25 and H110T20 are quasi-static points at twice and three times the Keulegan--Carpenter number of the primary case, and H110T20 also drives the tip excursion to approximately one third of the stem length, which exercises the geometrically exact formulation at large deflection. H070T15 lies deliberately just \emph{beyond} the quasi-static limit at $CaL/KC = 0.241$ and probes where the drag-only closure stops being accurate. The two larger-amplitude conditions have Ursell numbers near $20$ and are generated with fifth-order Stokes kinematics \citep{fenton_fifth_1985}.

\begin{table}[!ht]
\centering
\small
\caption{The four simulated wave conditions (Mimic~3, $l = 0.30~\mathrm{m}$, $d = 0.45~\mathrm{m}$). Wave parameters and non-dimensional numbers follow Tables~1 and 3 of \citet{jacobsen_waveinduced_2019}, where $a_w$ and $u_w$ are the measured orbital excursion and characteristic velocity. $C_D$ is the drag-coefficient range fitted by the reference from the measured base force at each condition's $KC$, and $C_{dn}$ the value or values adopted for the actuator line.}
\label{tab:flexveg_conditions}
\footnotesize
\setlength{\tabcolsep}{3pt}
\renewcommand{\arraystretch}{1.15}
\begin{tabular}{l c c c c c c c c c c}
\toprule
Condition & $H$ (m) & $T$ (s) & Wave theory & $a_w$ (cm) & $u_w$ (cm/s) & $KC$ & $Ca$ & $CaL/KC$ & $C_D$ (meas.) & $C_{dn}$ \\
\midrule
H040T20 & $0.04$ & $2.0$ & Stokes II & $2.5$ & $7.7$  & $19.3$ & $0.21$ & $0.135$ & $3.5$--$4.0$ & $3.0$, $4.0$ \\
H070T15 & $0.07$ & $1.5$ & Stokes II & $2.7$ & $11.2$ & $20.9$ & $0.45$ & $0.241$ & $3.1$--$3.7$ & $4.0$ \\
H070T25 & $0.07$ & $2.5$ & Stokes V  & $5.0$ & $12.7$ & $39.7$ & $0.58$ & $0.088$ & $2.9$       & $3.0$ \\
H110T20 & $0.11$ & $2.0$ & Stokes V  & $7.0$ & $22.0$ & $55.1$ & $1.75$ & $0.137$ & $2.2$--$2.4$ & $2.4$ \\
\bottomrule
\end{tabular}
\end{table}

\subsection{Problem definition}
\label{sec:flexveg_geometry}
\begin{figure}[!htb]
\centering
\includegraphics[width=0.95\textwidth]{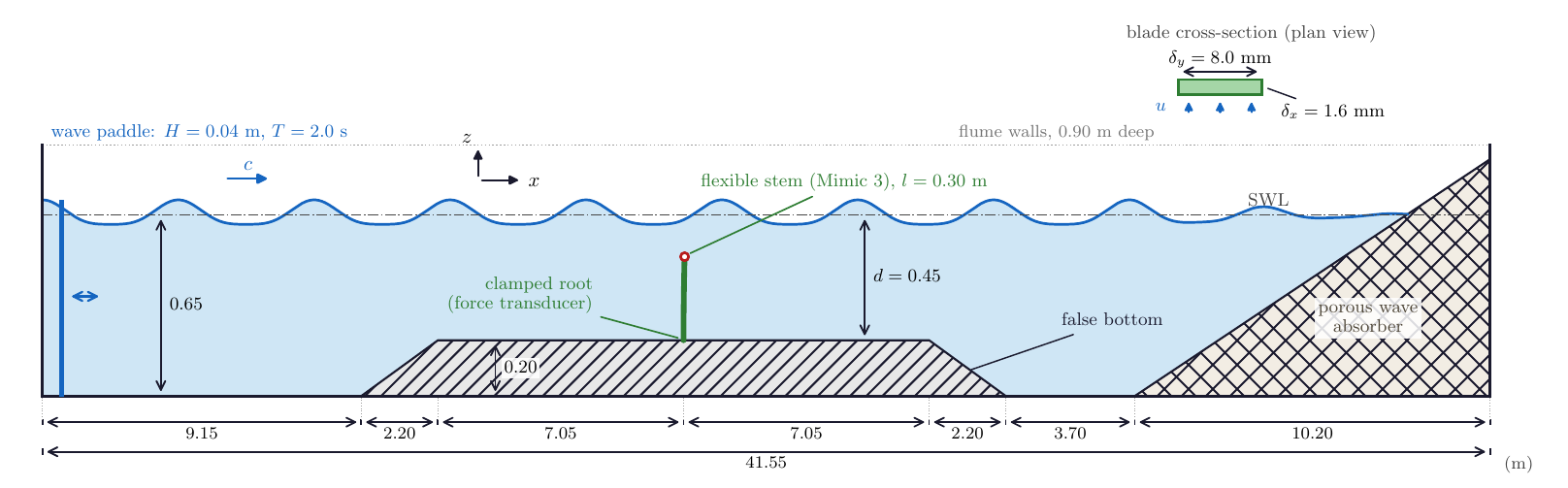}
\caption{The flexible-vegetation validation problem in side view, with the vertical scale exaggerated and dimensions in metres, following Figure~1 of \citet{jacobsen_waveinduced_2019}. The $0.20~\mathrm{m}$ false bottom, accessed by $2.20~\mathrm{m}$ ramps, raises the bed in the central test section so that the still water depth is $0.65~\mathrm{m}$ at the paddle and $d = 0.45~\mathrm{m}$ over the $14.1~\mathrm{m}$ plateau. The Mimic~3 blade is clamped at the plateau centre. The computational domain reproduces the whole flume and is three-dimensional. The inset shows the blade cross-section in plan view, with the wide face $\delta_y$ normal to the wave direction.}
\label{fig:flexVeg_schematic}
\end{figure}

Figure~\ref{fig:flexVeg_schematic} presents the flume layout together with the computational domain. The bed is raised in the central test section by a $0.20~\mathrm{m}$ false bottom, flat at the paddle depth over the first $9.15~\mathrm{m}$, rising along a $2.20~\mathrm{m}$ ramp onto a $14.1~\mathrm{m}$ elevated plateau, and descending along a second ramp to a flat section, with a gently sloped porous absorber occupying the final $10.20~\mathrm{m}$. The computational domain reproduces the whole flume, spanning $41.55~\mathrm{m} \times 0.8~\mathrm{m} \times 1.1~\mathrm{m}$, with air above the still water level. The paddle section, both ramps, the plateau, and the absorber section are all meshed, so the wave shoals onto the plateau in the simulation as it does in the experiment, and the physical absorber is replaced by the numerical outlet absorption of Section~\ref{sec:theory_waves}. The simulation is three-dimensional throughout (Figure~\ref{fig:flexveg_domain}), and the schematic is drawn in side view because the motion is confined to the vertical plane of wave propagation. The stem is clamped at the mid-length of the plateau, $18.4~\mathrm{m}$ from the inlet and on the flume centreline, extending vertically to its free tip $0.15~\mathrm{m}$ below the still water level. The upstream fetch corresponds to approximately $4.7$ wavelengths of the target wave. The blade is represented by its centreline, with the rectangular cross-section entering through its area, second moment of area, and drag reference length as in the two preceding benchmarks, and the wide face normal to the wave direction, so the blade bends about its weak axis. The sensitivity of the coupled solution on this domain to the discretisation is examined in Appendix~\ref{app:flexveg_mesh}.

\noindent
The case is solved with the two-phase formulation of Section~\ref{sec:fluidSubsystem}, using the volume-of-fluid phase fraction to capture the free surface and the modified pressure $p_{\mathrm{rgh}}$ as the pressure variable, as implemented in the \texttt{interFoam} solver of OpenFOAM. At the inlet, a regular wave of the appropriate order is imposed through the boundary conditions of Section~\ref{sec:theory_waves}, with active absorption so that reflected waves are not re-injected \citep{higuera_realistic_2013} and the amplitude ramped up smoothly over the first two wave periods, and at the outlet a shallow-water absorption condition dissipates the outgoing wave. The bed is a stationary no-slip wall on which the stem root is clamped, the top boundary is open to the atmosphere at fixed total pressure, and the two lateral boundaries represent the flume side walls. The fluid is initialised at rest with a hydrostatic pressure distribution and a flat interface at the still water level, and the stem starts undeformed, so the flow at the stem is not yet periodic when the run begins and all quantitative comparisons are taken from the end of the run. The blade is close to neutrally buoyant, so gravity and buoyancy nearly cancel along the stem and the response is dominated by the balance between hydrodynamic drag and elastic restoring moment.

\begin{table}[!ht]
\centering
\small
\caption{Physical parameters of the flexible-vegetation validation case, following \citet{jacobsen_waveinduced_2019} (Mimic~3, $l = 0.30~\mathrm{m}$, wave condition H040T20).}
\label{tab:flexveg_properties}
\setlength{\tabcolsep}{6pt}
\renewcommand{\arraystretch}{1.15}
\begin{tabular}{l c l}
\toprule
Quantity & Symbol & Value \\
\midrule
\multicolumn{3}{l}{\textbf{Wave and flume}} \\
\cmidrule(lr){1-3}
Wave height (target)             & $H$    & $0.04~\mathrm{m}$ \\
Wave period                      & $T$    & $2.0~\mathrm{s}$ \\
Still water depth (test section) & $d$    & $0.45~\mathrm{m}$ \\
Water density                    & $\rho$ & $1.0\times10^{3}~\mathrm{kg\,m^{-3}}$ \\
Kinematic viscosity              & $\nu$  & $1.0\times10^{-6}~\mathrm{m^{2}\,s^{-1}}$ \\
Characteristic orbital velocity  & $u_w$  & $7.7\times10^{-2}~\mathrm{m\,s^{-1}}$ \\
Orbital excursion                & $a_w$  & $2.5\times10^{-2}~\mathrm{m}$ \\
\midrule
\multicolumn{3}{l}{\textbf{Stem (Mimic~3)}} \\
\cmidrule(lr){1-3}
Length                           & $l$        & $0.30~\mathrm{m}$ \\
Blade width (facing the wave)    & $\delta_y$ & $8.0\times10^{-3}~\mathrm{m}$ \\
Blade thickness                  & $\delta_x$ & $1.6\times10^{-3}~\mathrm{m}$ \\
Young's modulus                  & $E$        & $2.2\times10^{9}~\mathrm{Pa}$ \\
Solid density                    & $\rho_s$   & $1.09\times10^{3}~\mathrm{kg\,m^{-3}}$ \\
Bending stiffness                & $EI$       & $6.0\times10^{-3}~\mathrm{N\,m^{2}}$ \\
Natural period (submerged)       & $T_n$      & $0.26~\mathrm{s}$ \\
\bottomrule
\end{tabular}
\end{table}

\subsection{Choice of the drag coefficient}
\label{sec:flexveg_cd}
The steady-CFD calibration used in the two channel benchmarks is not transferable here, because the flow past the blade is oscillatory and depth-varying and a steady rigid-plate simulation in a wave flume has no meaningful steady state. The reference experiment supplies the coefficient instead. \citet{jacobsen_waveinduced_2019} derive the average drag coefficient for each mimic by least-squares fitting of the measured base force to the relative-velocity Morison formulation integrated over the instantaneous stem shape, and report it against the Keulegan--Carpenter number, the fitted forces reproducing the measured forces generally within $10$ to $20\%$. At $KC = 19.3$, the two repeat tests of the present blade yield $C_D = 3.5$ and $4.0$, with the neighbouring cluster for the shorter $l = 0.15~\mathrm{m}$ specimen lying slightly lower at $3.2$ to $3.3$.

\noindent
Rather than committing to a single reading, the primary case is simulated twice, at $C_{dn} = 3.0$ and $4.0$. The upper value is the upper repeat fit of the present blade, whereas $3.0$ sits slightly \emph{below} the blade's own fitted cluster and is retained as a deliberately conservative lower bracket, so the pair brackets the family of published fitted values and this propagated range appears explicitly in every comparison that follows. The drag reference length is the blade width. For the three additional conditions the coefficient is read from the same figure at each condition's $KC$ (Table~\ref{tab:flexveg_conditions}). For H070T25 and H110T20 the adopted values match the fitted clusters, at $3.0$ against $2.85$ to $2.94$, marginally above within reading accuracy, and $2.4$ inside $2.2$ to $2.4$. For the boundary probe H070T15 the upper reading of the primary case is retained even though it lies marginally above that condition's own cluster of $3.1$ to $3.7$, a deliberately generous choice made so that any shortfall observed at the boundary of the quasi-static regime cannot be attributed to a conservative drag input.

\noindent
The tangential coefficient is fixed at $C_{dt} = 0.01$ as before, which matters little because the stem stays close to vertical while the wave-induced velocity is predominantly horizontal, so the relative velocity is nearly normal to the blade axis throughout the cycle. No recalibration of either coefficient against the measured stem motion is performed. The drag coefficients are fixed from the published fits and the coupling-operator settings are carried over from the channel benchmarks, so the comparison is blind with respect to the validation data, with the operator settings inherited rather than independently derived. One additional simulation at $C_{dn} = 5.0$, deliberately \emph{outside} the measured band, is reported below purely as a sensitivity diagnostic, to test whether a coefficient above the measured band could make up a shortfall in the response amplitude, and it is never used as a calibration.

\subsection{Numerical set-up}
\label{sec:flexveg_setup}

\begin{figure}[!htb]
\centering
\includegraphics[width=0.95\textwidth,trim={20 5 40 310},clip]{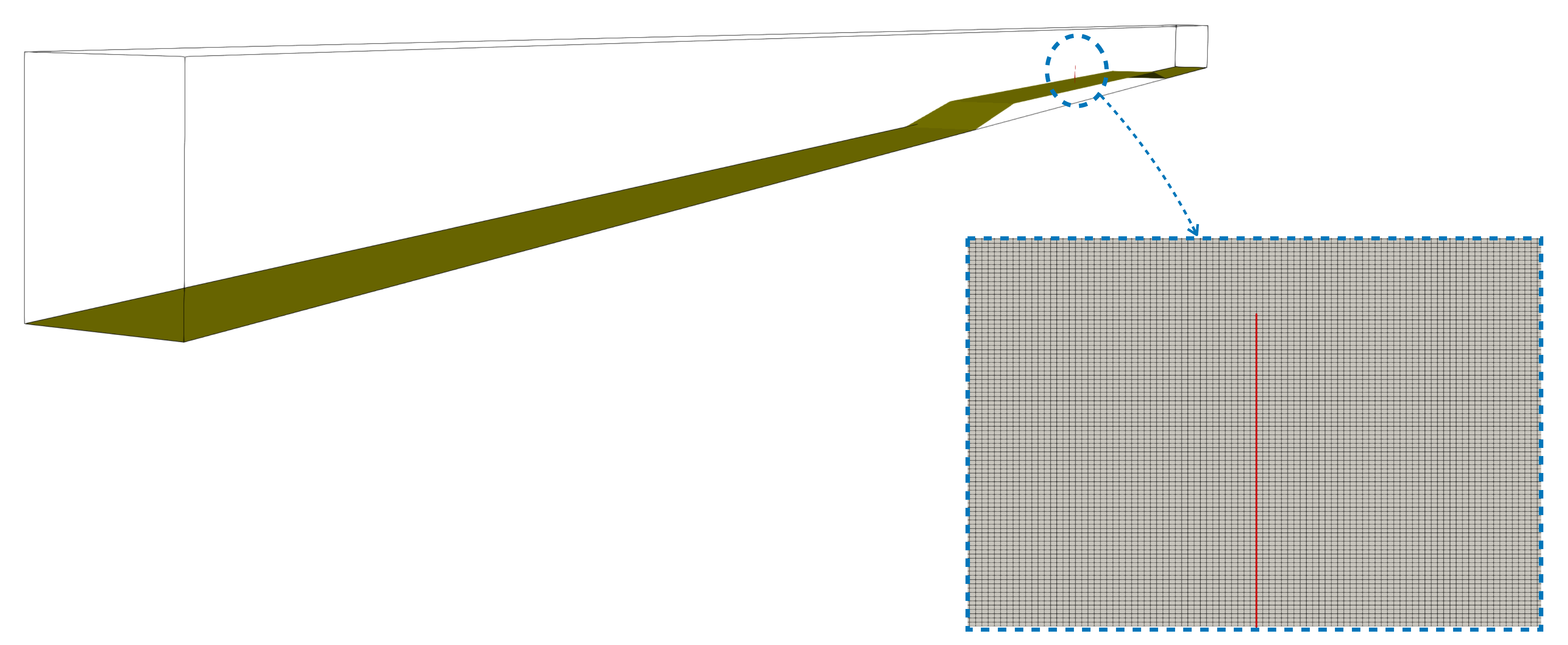}
\caption{Snapshot of the three-dimensional computational domain of the wave flume, with the bed and the false bottom shaded and the stem shown in red. The dashed circle marks the stem region, and the inset is a close-up of the mesh near the stem in a vertical plane through the stem.}
\label{fig:flexveg_domain}
\end{figure}

\noindent
The fluid domain is discretised with a structured hexahedral mesh of approximately $1.35\times10^{7}$ cells, built in six blocks spanning the whole flume with the \texttt{blockMesh} utility of OpenFOAM, with $20$ cells across the width and $220$ in the vertical. Over the plateau the streamwise spacing is uniform at $\Delta x = 6~\mathrm{mm}$, with $\Delta y = 40~\mathrm{mm}$ across the flume and $\Delta z = 4.1~\mathrm{mm}$ in the vertical. Away from the plateau the streamwise spacing is graded from $6~\mathrm{mm}$ at the plateau edges to $32~\mathrm{mm}$ over the paddle section and about $200~\mathrm{mm}$ at the outlet, so the wave is resolved finely where it is measured and the mesh coarsens where it is not. Figure~\ref{fig:flexveg_domain} shows the three-dimensional domain together with a close-up of the mesh near the stem. The vertical resolution corresponds to about $10$ cells per wave height for the primary condition, rising to $27$ for the largest wave, and the streamwise spacing over the plateau to about $650$ cells per wavelength. \citet{larsen_performance_2019} propagated a nonlinear regular wave with the same solver family and found that $12.5$ cells per wave height carried it over about five wavelengths without appreciable decay provided the Courant number was kept near $0.05$, whereas $6.25$ cells per wave height smeared the interface. The primary condition lies between these values, the larger waves are resolved more finely than either, the celerity-based Courant number here is about $0.016$, and the upstream fetch is close to the propagation distance over which that assessment was made. The adequacy of the resolution is shown directly in Section~\ref{sec:flexveg_results_wave} by comparing the achieved elevation and orbital velocities at the stem against the measured values, which is the sharper test because it checks the quantity that actually drives the stem, and Appendix~\ref{app:flexveg_mesh} repeats the largest-deflection condition on meshes a factor of $\sqrt{2}$ coarser and finer in every direction, where the load and the response move by about $2\%$.

\noindent
The beam is discretised into $60$ finite volume segments, consistent with the beam-mesh study of Section~\ref{sec:beamtunnel_sensitivity}, and the adequacy of that resolution for the present load distribution is confirmed independently by the quasi-static test of Section~\ref{sec:flexveg_conditions}. The kernel width is $\varepsilon = 2.5\,\Delta x = 0.015~\mathrm{m}$, sized by the same rule adopted in the channel benchmarks, and the upstream sampling is performed $0.375~\mathrm{m}$ upstream of each beam segment in the wave propagation direction, projected onto the vertical plane through the stem. That offset places the sampled velocity outside the momentum-sink region, but it is a nonlocal input in a propagating wave, corresponding to about a tenth of the local wavelength, so the sampled kinematics lead the at-stem wave by a known phase $k \times 0.375~\mathrm{m}$, approximately $35^{\circ}$ for the primary condition and evaluated per condition. Two properties keep this admissible. The offset affects phase but not magnitude, since the wave propagates over a flat plateau without decay and the sampled orbital-velocity amplitude was verified to match the at-stem amplitude to within a few percent. The phase lead is also deterministic and is corrected for wherever phase enters a comparison. The absolute phase of the response relative to the local wave is therefore not claimed as an independently validated quantity, and the internal phase consistency of the coupled solution is certified instead by the in-phase quasi-static reconstruction of Section~\ref{sec:flexveg_conditions}. A dedicated sampling-distance sweep was not repeated in the wave flume, the sensitivity to that distance having been quantified in Section~\ref{sec:beamtunnel_sensitivity}, and this transfer is acknowledged as a limitation.

\noindent
The simulation is advanced with a fixed time step of $\Delta t = 5\times10^{-5}~\mathrm{s}$, which is $4\times10^{4}$ steps per wave period and well below the interface-Courant limit. The small step is dictated by the robustness of the beam Newton solve within the partitioned coupling rather than by the fluid, for the reason given in Section~\ref{sec:almParallelSearch}, so the run time is set by the structural solve. Each case simulates $32~\mathrm{s}$ of physical time, sixteen wave periods at the primary condition, and all results below are extracted from the final three periods, over which the per-period tip-displacement extrema vary by less than $1\%$ in every run. That repeatability is what establishes that the start-up is over.

\subsection{Wave-field verification at the stem}
\label{sec:flexveg_results_wave}
\begin{figure}[!htb]
\centering
\includegraphics[width=0.78\textwidth]{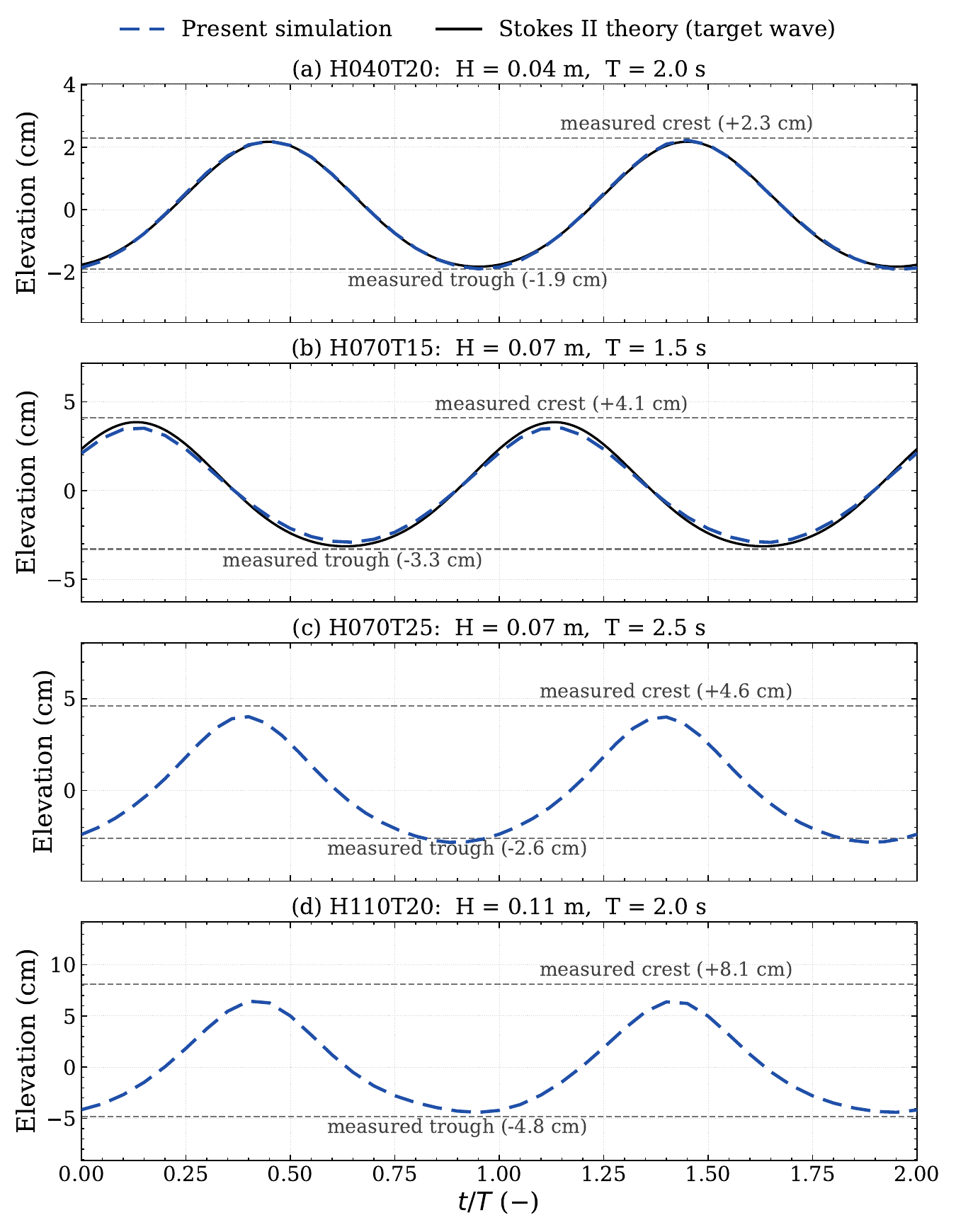}
\caption{Verification of the wave field at the stem location for the four simulated conditions. Surface elevation over two developed wave periods, about the local mean water level, against the crest and trough elevations measured at the same location \citep[Table~1]{jacobsen_waveinduced_2019}, and, for the Stokes~II cases (a, b), against second-order theory at the target wave height. The measured values describe the \emph{achieved} experimental wave while the simulation generates the target wave, and both heights are listed in Table~\ref{tab:flexveg_wave_check}.}
\label{fig:flexveg_wave}
\end{figure}

Any error in the local wave kinematics propagates quadratically into the drag force, so the generated wave is verified at the stem location first. Figure~\ref{fig:flexveg_wave} shows the simulated surface elevation over two developed wave periods for all four conditions, against the crest and trough elevations measured at the same location, and Table~\ref{tab:flexveg_wave_check} complements it with the vertical-profile quantities that drive the stem, namely the crest and trough horizontal orbital velocities averaged over the stem height and the characteristic velocity $u_w$.

\begin{table}[!ht]
\centering
\small
\caption{Wave kinematics at the stem location, simulation against the measurements reported in Tables~1 and 3 of \citet{jacobsen_waveinduced_2019}. The first pair of columns gives the wave height each simulation was driven at and the height the flume achieved, because the two differ and the remaining columns therefore describe slightly different waves. $\eta_c$ and $\eta_t$ are the crest and trough elevations, $\langle \tilde{u}_c \rangle_z$ and $\langle \tilde{u}_t \rangle_z$ the crest and trough horizontal orbital velocities averaged over the stem height, and $u_w$ the characteristic orbital velocity.}
\label{tab:flexveg_wave_check}
\footnotesize
\setlength{\tabcolsep}{3.5pt}
\renewcommand{\arraystretch}{1.15}
\begin{tabular}{l cc cc cc cc cc cc}
\toprule
 & \multicolumn{2}{c}{$H$ (cm)} & \multicolumn{2}{c}{$\eta_c$ (cm)} & \multicolumn{2}{c}{$\eta_t$ (cm)} & \multicolumn{2}{c}{$\langle \tilde{u}_c \rangle_z$ (cm/s)} & \multicolumn{2}{c}{$\langle \tilde{u}_t \rangle_z$ (cm/s)} & \multicolumn{2}{c}{$u_w$ (cm/s)} \\
\cmidrule(lr){2-3}\cmidrule(lr){4-5}\cmidrule(lr){6-7}\cmidrule(lr){8-9}\cmidrule(lr){10-11}\cmidrule(lr){12-13}
Condition & target & achieved & sim & meas & sim & meas & sim & meas & sim & meas & sim & meas \\
\midrule
H040T20 & $4.0$  & $4.1$  & $+2.3$ & $+2.3$ & $-1.9$ & $-1.9$ & $+8.0$  & $+8.0$  & $-7.3$  & $-7.3$  & $7.7$  & $7.7$ \\
H070T15 & $7.0$  & $7.4$  & $+3.5$ & $+4.1$ & $-2.9$ & $-3.3$ & $+13.4$ & $+11.5$ & $-12.8$ & $-10.8$ & $13.1$ & $11.2$ \\
H070T25 & $7.0$  & $7.2$  & $+4.0$ & $+4.6$ & $-2.8$ & $-2.6$ & $+15.7$ & $+15.6$ & $-12.1$ & $-9.9$  & $14.0$ & $12.7$ \\
H110T20 & $11.0$ & $12.9$ & $+6.5$ & $+8.1$ & $-4.4$ & $-4.8$ & $+21.4$ & $+25.4$ & $-16.9$ & $-19.1$ & $19.3$ & $22.0$ \\
\bottomrule
\end{tabular}
\end{table}

\noindent
For the primary condition the agreement is essentially exact. The simulated elevation follows the second-order theory curve, the crest and trough match the measured values to within $0.05~\mathrm{cm}$, and every velocity measure agrees to within $1\%$, including the Stokes-II asymmetry between the stronger, shorter crest phase and the weaker, longer trough phase and the small negative Eulerian mean at the stem. The wave that drives the primary validation is therefore the experimental wave, and any discrepancy in the stem response cannot be attributed to the wave field. The tabulated values are those of the $C_{dn} = 3.0$ run; in the $C_{dn} = 4.0$ run the crest velocity at the stem is $7.68$ instead of $8.0~\mathrm{cm\,s^{-1}}$, about $4\%$ lower, because the stronger momentum sink slows the flow locally. That difference is the two-way coupling acting on the wave it sees, not a defect in the wave generation. The simulated mean water level sits $0.7$ to $0.9~\mathrm{cm}$ above the still water level in all runs, a constant set-up associated with the mass injected during the inlet ramp that the outlet absorption maintains but does not drain. It is steady over the entire analysis window and elevations are reported about the local mean, exactly as a wave gauge reports them, and a depth increase of $0.8~\mathrm{cm}$ on $0.45~\mathrm{m}$ shifts the wavenumber and the orbital-velocity amplitude by of the order of $1\%$, an order of magnitude below the response deficits under discussion.

\noindent
For the three additional conditions the comparison must separate model error from a mismatch of inputs, because the simulation generates the \emph{target} wave of each condition whereas the published kinematics describe the \emph{achieved} flume wave. For H110T20 the flume overshot its target substantially, at $12.9~\mathrm{cm}$ against $11~\mathrm{cm}$, so the simulated crest and orbital velocities sit systematically below the measured ones by essentially that margin, the velocity ratio matching the height ratio almost exactly. The opposite arises for H070T15, where the simulated velocities \emph{exceed} the measurement by about $17\%$ even though the simulated crest is low. At this condition the relative depth is $kd \approx 1.04$, at the edge of validity of the shallow-water outlet absorption, and the partial reflection creates a weak standing-wave pattern that raises the velocity locally at the stem while lowering the elevation. The contamination can be estimated from its own signature, the crest sitting about $14\%$ below the measured value while the velocity sits about $17\%$ above it. That is the pattern of a partial standing wave sampled near an elevation node and velocity antinode, and it implies an amplitude reflection coefficient of the order of $0.15$, an order of magnitude larger than the few-percent pattern detectable at the other conditions. This contaminates the local wave rather than the wave generation, and it is the second of two grounds on which this condition is excluded from the quantitative closure-accuracy claim below. H070T25 sits in between, with crest velocity matching the measurement and a trough somewhat too strong.

\subsection{Stem motion along the blade}
\label{sec:flexveg_results_motion}
\begin{figure}[!htb]
\centering
\includegraphics[width=0.95\textwidth]{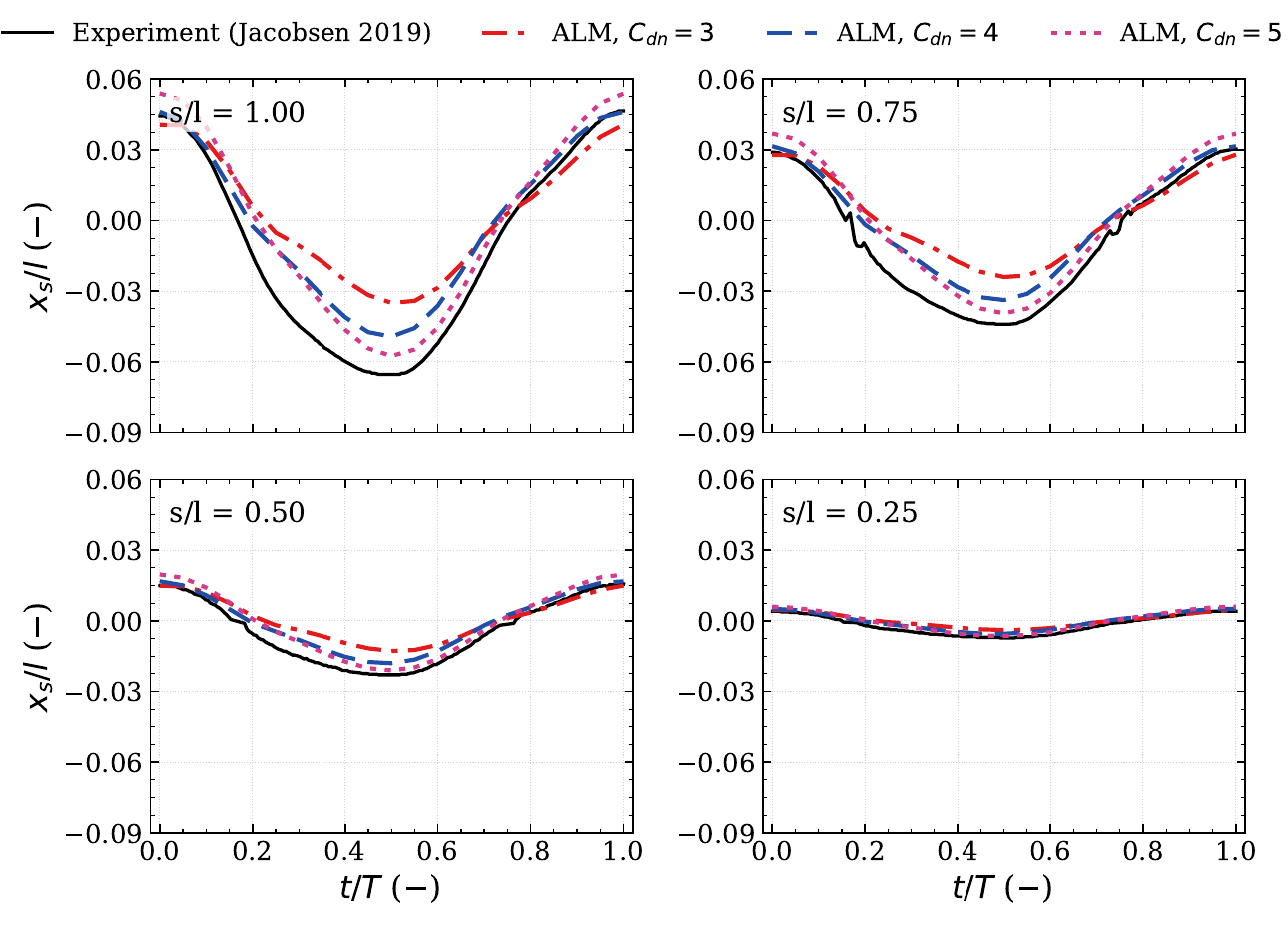}
\caption{Horizontal stem displacement over one wave period at four stations along the blade. Present two-way simulations at $C_{dn} = 3.0$ and $4.0$, the propagated coefficient range bracketing the published fitted values, and at the out-of-band sensitivity value $C_{dn} = 5.0$, against the video-tracked measurement of \citet[Figure~14d]{jacobsen_waveinduced_2019}. Each run is aligned in time by its tip peak, with one shift per run applied at all stations. The experimental curves carry the $-0.015\,l$ pre-tension mean offset, which deepens their troughs.}
\label{fig:flexveg_alongblade}
\end{figure}

Figure~\ref{fig:flexveg_alongblade} compares the simulated stem motion over one wave period against the video-tracked displacement of the experiment, for both ends of the propagated coefficient range and for the out-of-band sensitivity value $C_{dn} = 5.0$. The compared quantity is the horizontal displacement $x_s$ of the centreline, normalised by the stem length, at the four stations $s/l = 0.25$, $0.50$, $0.75$ and $1.00$ tracked in the experiment, where $s$ is the arc-length from the clamped root. Because the upstream sampling shifts the phase of the whole response relative to the local wave, each run is aligned in time by its tip peak, with one shift per run applied at all four stations, so the relative phasing along the blade is preserved and remains a genuine prediction.

\noindent
At every station the simulated motion reproduces the shape of the measured signal, including the sawtooth asymmetry between the fast forward sweep under the crest and the slower backward relaxation under the trough, the simultaneous motion of all four stations that characterises the single-mode quasi-static response, and the monotonic growth of amplitude from root to tip. A proper orthogonal decomposition of the simulated motion attributes $100.0\%$ of the fluctuation energy to a single spatial mode, exactly as the reference reports for the measurement. At $C_{dn} = 4.0$ the positive tip peak matches the measurement to within $1\%$, at $+13.9~\mathrm{mm}$ against $+14.0~\mathrm{mm}$. The measured troughs are deeper than the simulated ones at all stations, most of that difference being the documented pre-tension offset, which shifts the whole measured signal by about $-4.5~\mathrm{mm}$, and the remainder the amplitude shortfall quantified next.

\begin{figure}[!htb]
\centering
\begin{subfigure}[b]{0.53\textwidth}
  \centering
  \includegraphics[width=\linewidth]{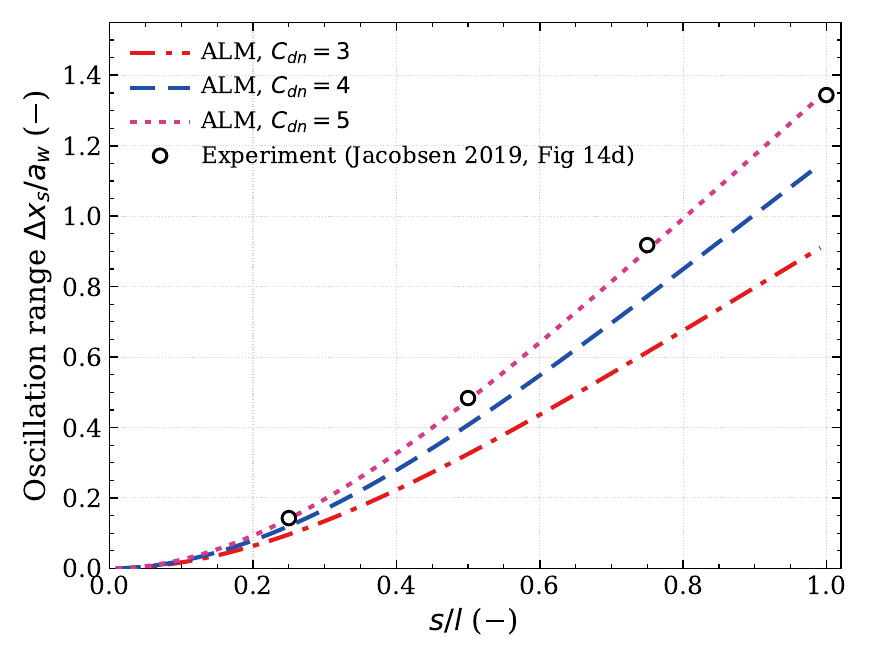}
  \caption{}
\end{subfigure}\hfill
\begin{subfigure}[b]{0.45\textwidth}
  \centering
  \includegraphics[width=\linewidth]{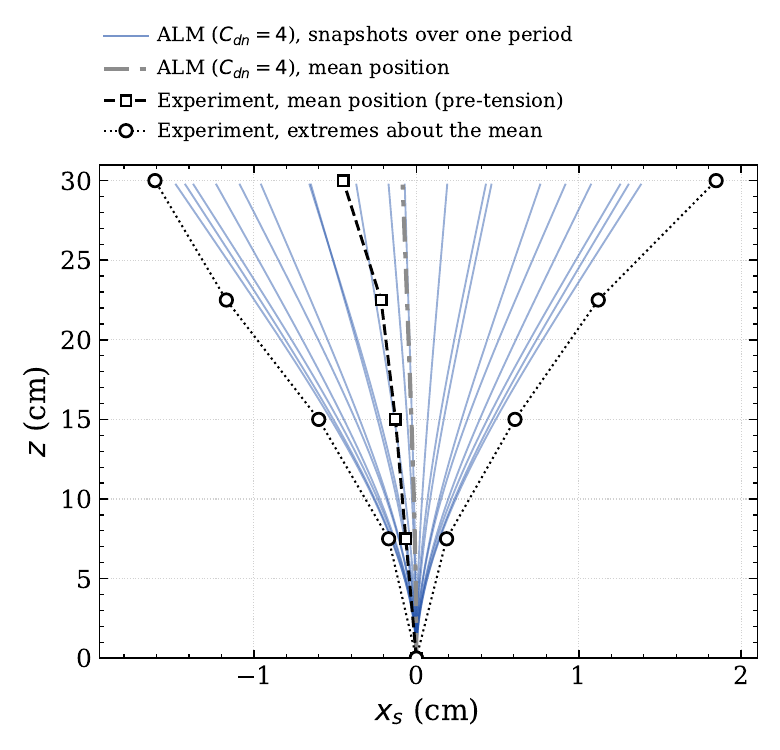}
  \caption{}
\end{subfigure}
\caption{(a) Peak-to-trough oscillation range of the horizontal displacement along the blade, normalised by the wave excursion $a_w$, for the three simulations against the four digitised experimental stations. The captured fraction is nearly uniform along the blade, so the simulated mode shape is exact and the residual deficit is a single amplitude factor. (b) Swaying envelope at $C_{dn} = 4.0$, showing stem centrelines at twenty equispaced instants of one wave period and the time-mean position, compared with the experiment reconstructed from the four digitised stations of \citet[Figure~14d]{jacobsen_waveinduced_2019}: the measured mean position (squares), which carries the $-0.015\,l$ static pre-tension offset of the physical blade, and the measured extreme-phase positions referenced about that mean (circles), which is the like-for-like comparison with an unbiased model.}
\label{fig:flexveg_range_envelope}
\end{figure}

\noindent
The peak-to-trough oscillation range, which is unaffected by the mean offset and is therefore the cleanest scalar measure, is shown in Figure~\ref{fig:flexveg_range_envelope}(a) as a profile along the blade normalised by $a_w$. The two runs of the propagated range bracket the response, capturing $68$ to $72\%$ of the measured range at $C_{dn} = 3.0$ and $85$ to $89\%$ at $4.0$, depending on station. These station-level fractions are referenced to the time-resolved series of Figure~14d of the reference, whose tip range of $1.34\,a_w$ is slightly smaller than the $1.41\,a_w$ of the compilation across conditions in their Figure~5, which gives tip fractions of $65\%$ and $81\%$ instead. Both normalisations appear in this section and are always identified, and the summary statement of $65$ to $89\%$ used elsewhere spans the two coefficients, the four stations, and the two experimental references. Two features of the profile carry more information than the percentages. The captured fraction is nearly constant along the blade, so the simulated \emph{mode shape} is correct at both coefficients and the entire deficit is a single amplitude factor. The response also grows visibly slower than linearly with the coefficient, a $33\%$ increase of $C_{dn}$ yielding a $26\%$ increase of range, because the stem follows the water more closely as the drag grows and the stronger momentum sink weakens the local flow, so no coefficient inside the measured band can close the residual gap.

\noindent
The envelope of Figure~\ref{fig:flexveg_range_envelope}(b) completes the kinematic comparison, with the experiment reconstructed in two separate parts. The first is the experimental \emph{mean} position, which leans backward with a tip offset of $-4.5~\mathrm{mm}$, the static pre-tension of the physical blade that no unbiased model reproduces. The simulated mean position is essentially vertical, which is the expected quasi-static result, since a symmetric oscillatory drag produces no rectified mean deflection at this order. The second part is the experimental extreme-phase configurations referenced \emph{about that mean}, with no measured value altered, and on this basis the envelopes agree closely. The simulated fan is forward-backward asymmetric in the same sense as the experiment, because the longer trough phase of the second-order wave acts on the stem for a larger fraction of the period, its extremes track the mean-referenced experimental extremes along the entire blade, and the residual difference is the $85$ to $89\%$ oscillatory amplitude fraction already quantified.

\subsection{Distributed hydrodynamic force}
\label{sec:flexveg_force}
\begin{figure}[!htb]
\centering
\begin{subfigure}[b]{0.49\textwidth}
  \centering
  \includegraphics[width=\linewidth]{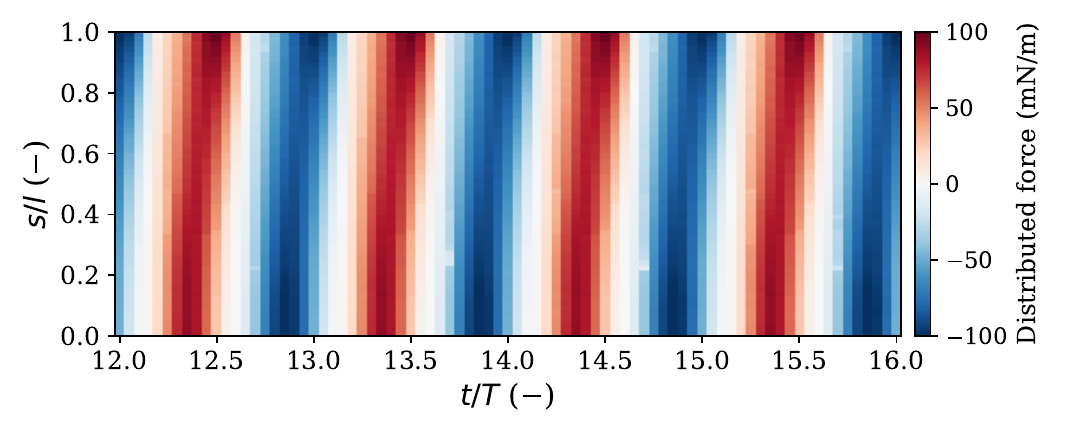}
  \caption{}
\end{subfigure}\hfill
\begin{subfigure}[b]{0.49\textwidth}
  \centering
  \includegraphics[width=\linewidth]{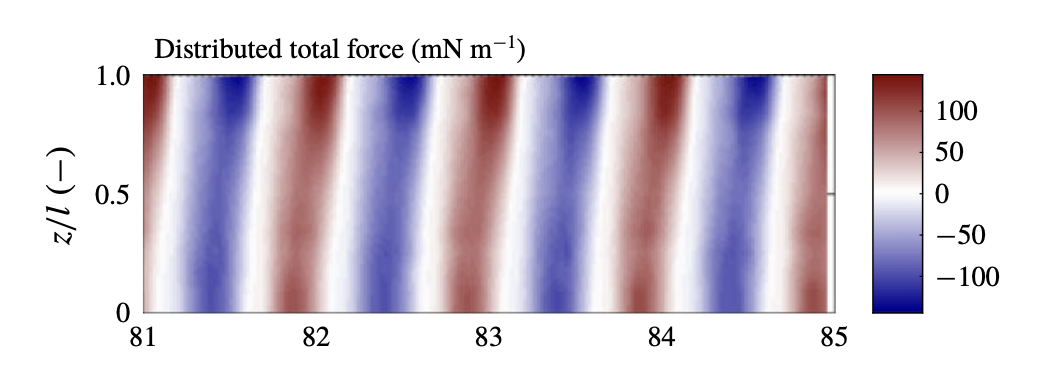}
  \caption{}
\end{subfigure}\\[0.4em]
\begin{subfigure}[b]{0.49\textwidth}
  \centering
  \includegraphics[width=\linewidth]{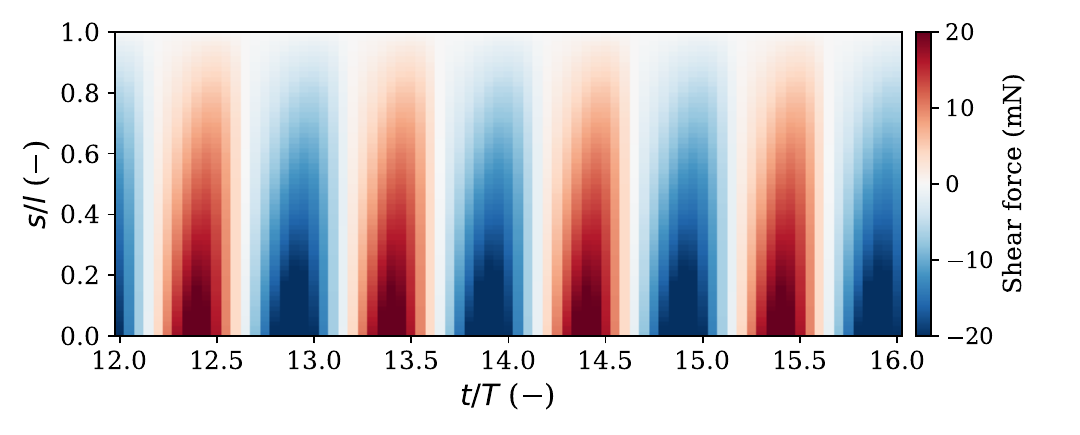}
  \caption{}
\end{subfigure}\hfill
\begin{subfigure}[b]{0.49\textwidth}
  \centering
  \includegraphics[width=\linewidth]{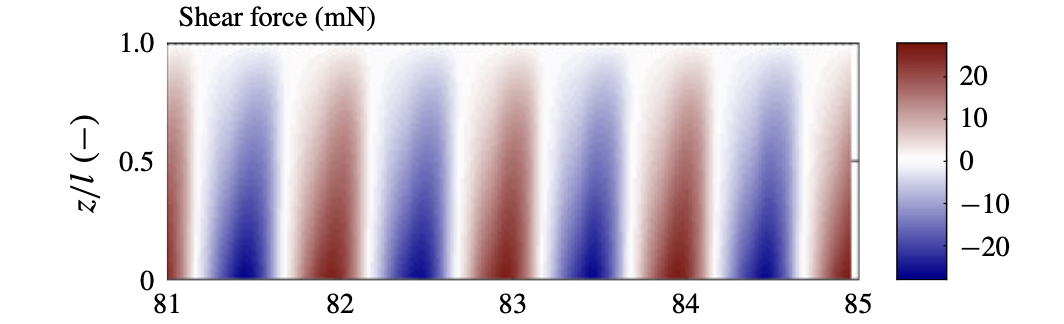}
  \caption{}
\end{subfigure}
\caption{Space--time structure of the hydrodynamic loading over four developed wave periods. (a) Actuator-line force per unit length on the stem in the $C_{dn} = 4.0$ simulation. (b) The corresponding experimental reconstruction, reproduced from Figure~14(a) of \citet{jacobsen_waveinduced_2019}. (c) Simulated internal shear-force estimate $V_x(s,t) = \int_s^l f_x\,\mathrm{d}s$. (d) The measured shear, reproduced from their Figure~14(c). Colour scales are common within each pair, at $\pm100~\mathrm{mN\,m^{-1}}$ for the force and $\pm20~\mathrm{mN}$ for the shear. The experimental clock starts at $t/T = 81$ and the simulation at $t/T = 12$.}
\label{fig:flexveg_heatmaps}
\end{figure}

The experiment makes the force distribution accessible in two forms, namely a distributed total force along the stem reconstructed from the tracked motion and the calibrated structural model, and the internal shear derived from it, whose value at the root is anchored by the transducer measurement. Figure~\ref{fig:flexveg_heatmaps} shows the corresponding quantities of the $C_{dn} = 4.0$ simulation as space--time maps over four wave periods, drawn with the colour scales of the reference figures, which are reproduced alongside for direct comparison.

\noindent
The simulated distributed force reproduces the structure of the experimental reconstruction. The loading is in phase along the entire stem, seen as vertical bands characteristic of the quasi-static regime, and it is largest near the free tip and near the root with the minimum at mid-length, which reflects the competition between the orbital-velocity profile, largest near the tip, and the relative-velocity reduction where the stem itself moves fastest. The simulated extrema of $\pm100~\mathrm{mN\,m^{-1}}$ coincide with the limits of the experimental colour scale, and the shear map inherits the same phase structure with its magnitude growing monotonically from tip to root.

\begin{figure}[!htb]
\centering
\includegraphics[width=0.66\textwidth]{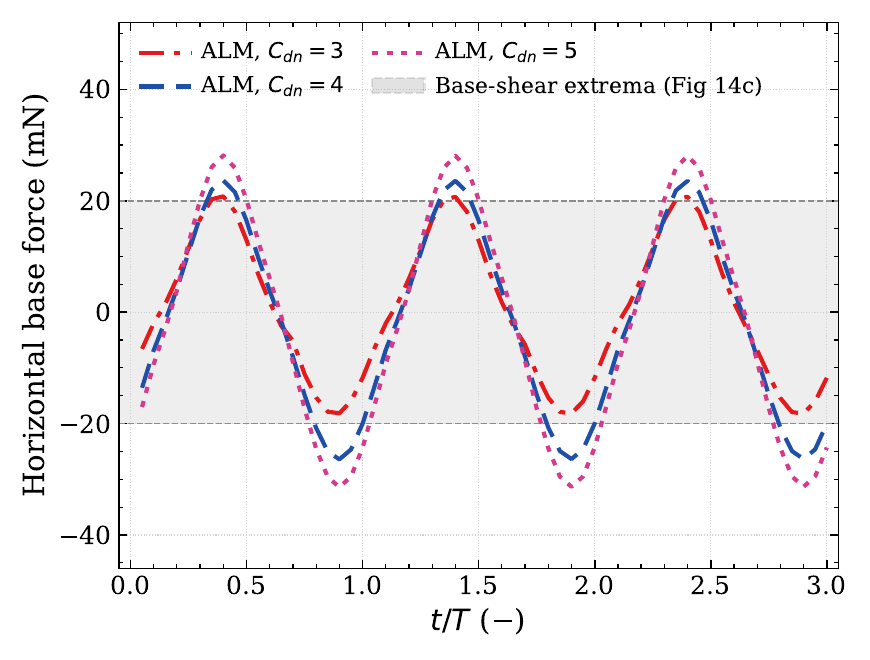}
\caption{Horizontal base force over the final three wave periods for the three coefficients, computed as the integral of the distributed actuator-line force along the stem. The shaded band marks $\pm20~\mathrm{mN}$, the extrema of the base-shear reconstruction of \citet[Figure~14c]{jacobsen_waveinduced_2019}, at which their colour scale saturates, so the band is a lower bound on the measured extremes.}
\label{fig:flexveg_baseforce}
\end{figure}

\noindent
The base force, which is the row $s/l = 0$ of the shear map and physically the quantity measured by the transducer, is shown in Figure~\ref{fig:flexveg_baseforce} for all three coefficients over the final three periods. The experimental reference is the $\pm20~\mathrm{mN}$ extremum of the reconstructed base shear, at which the colour scale of the reference figure saturates, so the measured extremes are bounded below rather than known exactly. The simulated base force oscillates between $+21$ and $-18~\mathrm{mN}$ at $C_{dn} = 3.0$ and between $+24$ and $-26~\mathrm{mN}$ at $4.0$, reaching and modestly exceeding the saturated scale in the same way the reference's own Morison fit does. This force-level agreement contrasts with the displacement comparison, where the range is $15$ to $20\%$ short at the coefficient the experiment best supports, the two differing because force and displacement are related through the full nonlinear balance rather than proportionally.

\subsection{Response across the experimental wave conditions}
\label{sec:flexveg_conditions}
\begin{figure}[!htb]
\centering
\includegraphics[width=0.68\textwidth]{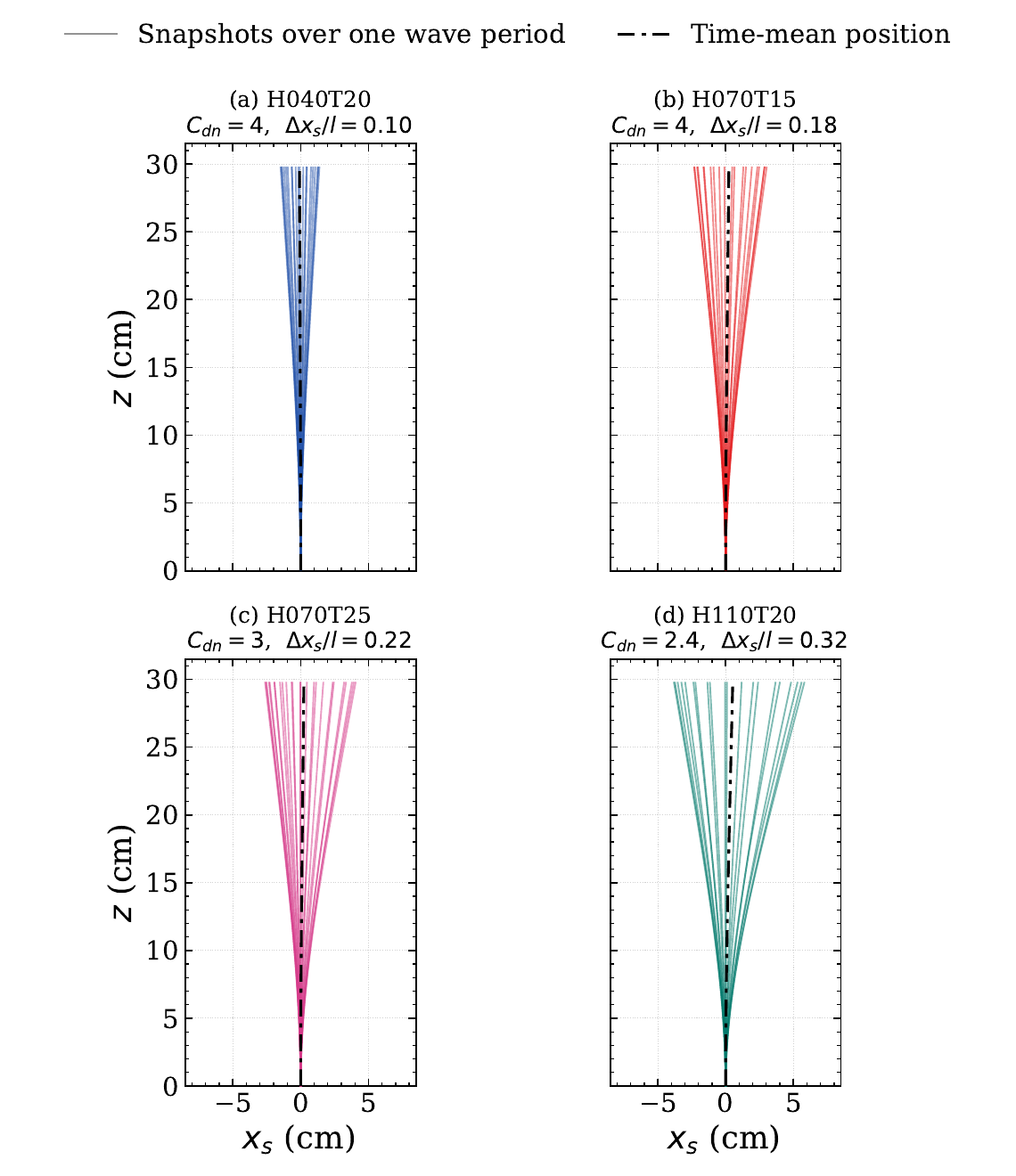}
\caption{Simulated swaying envelopes at the four wave conditions, at the production coefficient of Table~\ref{tab:flexveg_conditions} and with axes true to scale, showing stem centrelines at twenty equispaced instants of one wave period and the time-mean position. The tip oscillation range grows from $0.096\,l$ to $0.32\,l$. No experimental envelope data exist for the three additional conditions.}
\label{fig:flexveg_cond_envelopes}
\end{figure}

\begin{figure}[!htb]
\centering
\includegraphics[width=0.50\textwidth]{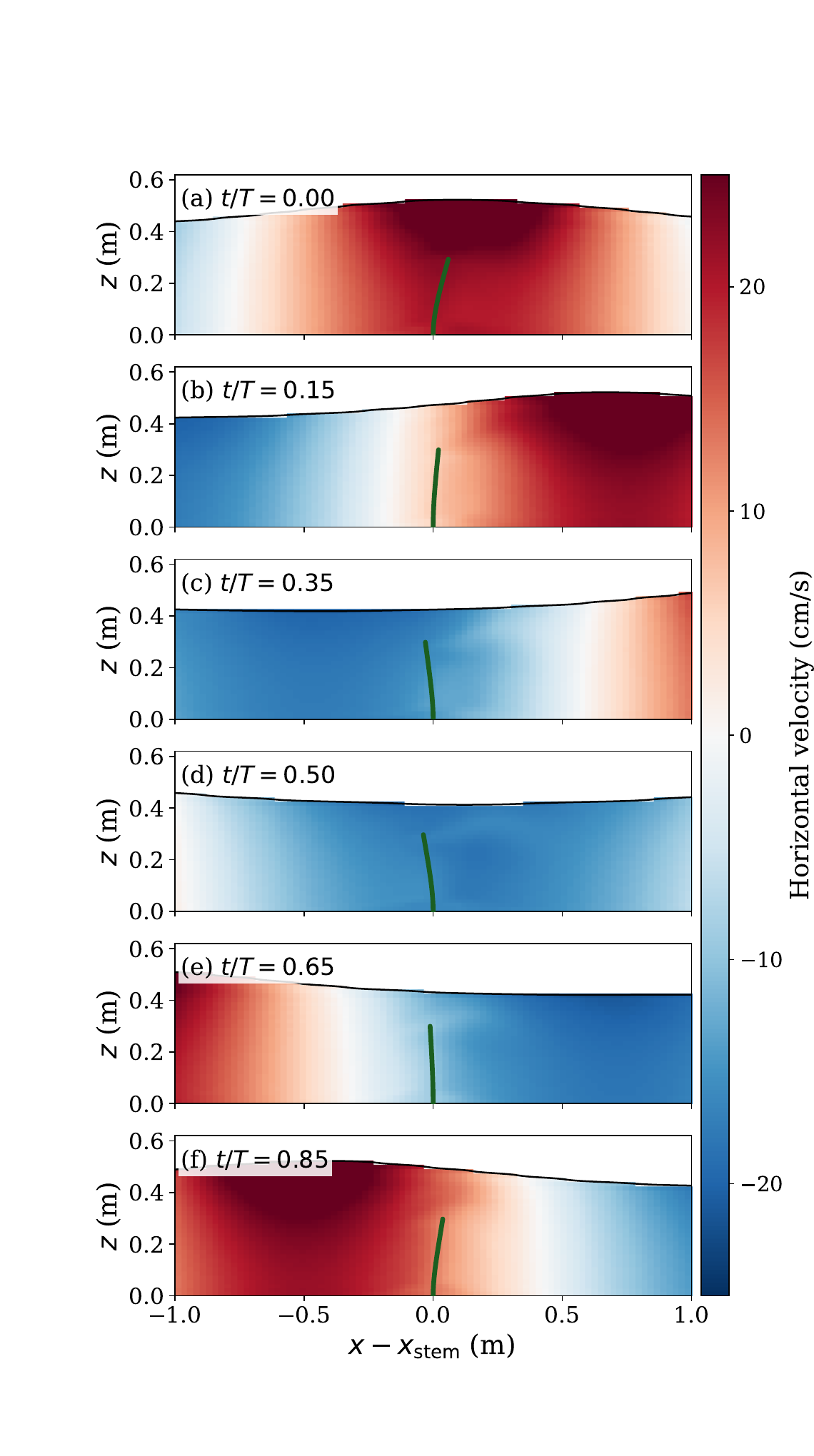}
\caption{Coupled solution at six phases of one developed wave period for the largest-deflection condition H110T20, on the vertical centre plane, showing the horizontal orbital velocity in the water (colour), the free surface (black), and the stem centreline taken from the solver beam-point positions (green). The sequence is anchored at the maximum forward tip deflection and the plotted window covers $\pm1.0~\mathrm{m}$ of the domain around the stem. Every field shown is a solver prediction.}
\label{fig:flexveg_snapshots}
\end{figure}

The three additional conditions extend the comparison from one point of the experimental parameter space to four, spanning a factor of three in Keulegan--Carpenter number and nearly an order of magnitude in Cauchy number, with the coefficient fixed a priori at each condition. Figure~\ref{fig:flexveg_cond_envelopes} shows the simulated envelopes, and Figure~\ref{fig:flexveg_snapshots} the coupled solution at six phases of a developed wave period for the largest-deflection condition. The tip range grows from $0.096\,l$ for the primary case through $0.18\,l$ and $0.22\,l$ to $0.32\,l$ for the largest wave, at which the stem sweeps through a visibly curved, large-deflection cycle. The geometrically exact formulation and the partitioned coupling handle this progression without adjustment, converging with the same robustness at $0.32\,l$ tip excursion as at $0.096\,l$. In that phase sequence the stem is strongly bent under and just after the crest and relaxes back through the trough phases toward the backward extreme, and the pale streak around and downstream of the stem is the local velocity deficit created by the momentum sink, the footprint of the two-way coupling on the flow. That deficit stays confined to the immediate neighbourhood of the stem and does not disturb the incident wave, consistent with the wave-field verification above.

\noindent
Table~\ref{tab:flexveg_cond_metrics} summarises the scalar outcome against the tip oscillation range normalised by the wave excursion, read for each condition from the tip-motion compilation of the reference. Two captured fractions are quoted. The \emph{raw} fraction compares the simulation directly with the measurement and therefore mixes model error with the input mismatch documented above. The \emph{forcing-corrected} fraction removes that mismatch by normalising the simulated range with the excursion of the wave the simulation actually delivered at the stem, $a_{w}^{\mathrm{sim}} = u_w^{\mathrm{sim}}T/2\pi$, and is the fairer measure of the closure itself. On that measure the three conditions simulated at a coefficient matching its fitted cluster capture $71$ to $83\%$ of the measured response, a consistent level across the full $KC$ range, with no additional degradation appearing at large deflection.

\begin{table}[!ht]
\centering
\small
\caption{Tip response across the four wave conditions. The simulated peak-to-trough tip range, normalised by the wave excursion, against the measured value read from Figure~5 of \citet{jacobsen_waveinduced_2019}. The raw captured fraction uses the experimental $a_w$, and the forcing-corrected fraction normalises the simulation by its own achieved wave excursion. The last column is the amplitude ratio of the quasi-static reconstruction test.}
\label{tab:flexveg_cond_metrics}
\setlength{\tabcolsep}{5pt}
\renewcommand{\arraystretch}{1.15}
\begin{tabular}{l c c c c c c c}
\toprule
Condition & $C_{dn}$ & Tip range$/l$ & $\Delta x_s/a_w$ sim & $\Delta x_s/a_w$ meas & raw & corrected & quasi-static \\
\midrule
H040T20 & $4.0$ & $0.096$ & $1.15$ & $1.41$ & $81\%$ & $83\%$ & $0.98$ \\
H070T15 & $4.0$ & $0.18$ & $1.98$ & $2.41$ & $82\%$ & $71\%$ & $0.97$ \\
H070T25 & $3.0$ & $0.22$ & $1.34$ & $1.48$ & $90\%$ & $81\%$ & $0.98$ \\
H110T20 & $2.4$ & $0.32$ & $1.38$ & $2.21$ & $62\%$ & $71\%$ & $0.97$ \\
\bottomrule
\end{tabular}
\end{table}

\noindent
The boundary probe H070T15 stands outside that quantitative set on two independent grounds, its coefficient having been deliberately chosen above its own cluster and its local wave carrying the reflection contamination quantified above, so its fraction is reported for completeness but not counted toward the closure-accuracy claim. At $CaL/KC = 0.241$, just beyond the quasi-static limit, the experiment shows the largest normalised response of the four conditions, which is the onset of the dynamic amplification that the added-mass inertia provides and that the drag-only closure cannot reproduce. The simulation captures $71\%$ of it on the corrected measure, and even that required a coefficient at the upper edge of the fitted cluster. The shortfall is the expected signature of the closure limitation, in the same direction as the leaflet cases at $\mathrm{St} = 0.5$, but a sharper claim is not possible because the numerical flume is least clean here and the measured deficit mixes the closure limitation with a wave-quality artefact.

\begin{figure}[!htb]
\centering
\includegraphics[width=0.92\textwidth]{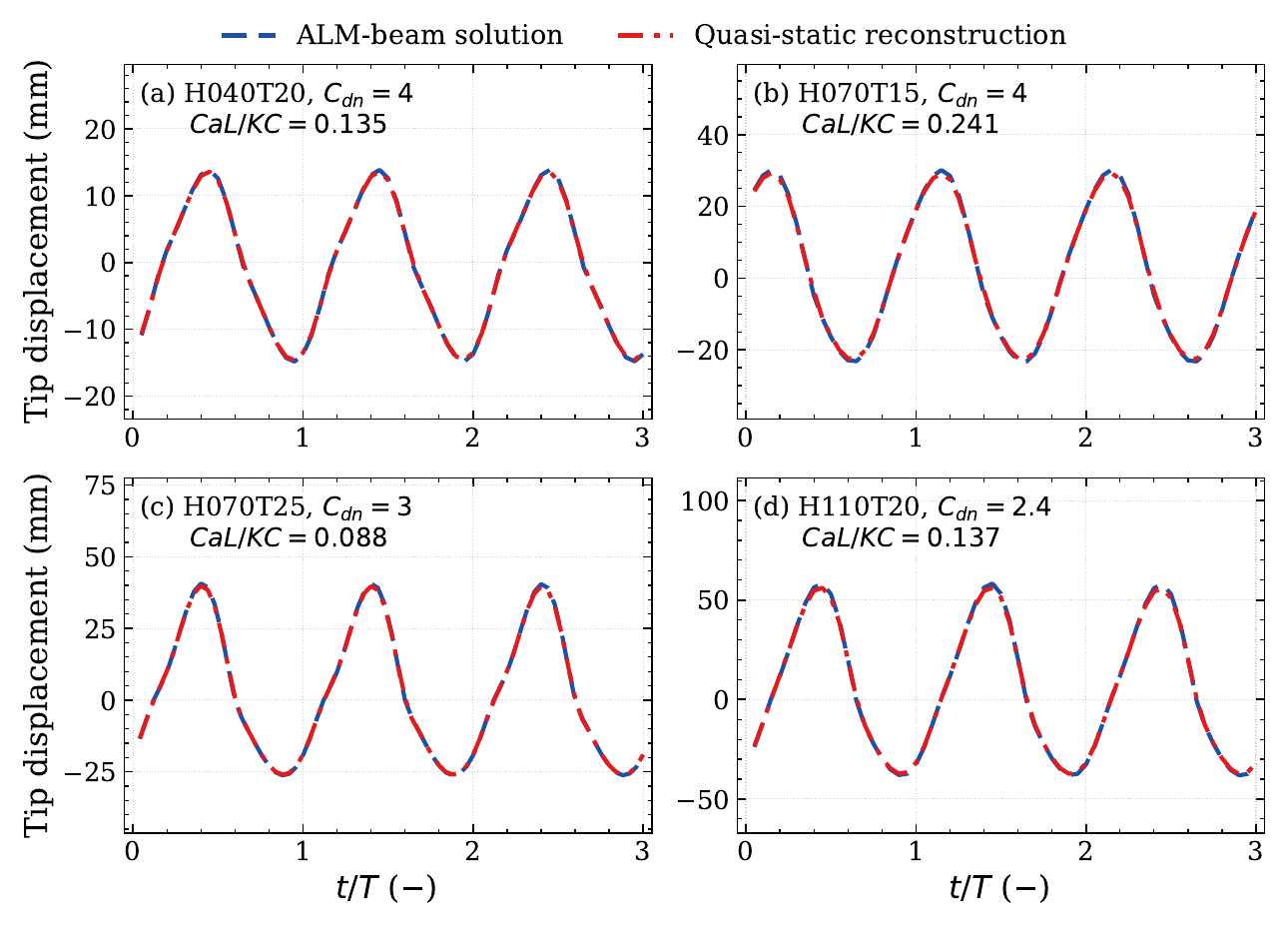}
\caption{Quasi-static consistency test at the four wave conditions. Tip displacement of the coupled solution against the response of an analytic linear cantilever loaded statically with the instantaneous actuator-line force distribution, with no time shift applied. The reconstruction is in phase and captures $97$ to $98\%$ of the amplitude in every case.}
\label{fig:flexveg_quasistatic}
\end{figure}

\noindent
By itself, an amplitude deficit does not indicate where the model needs improving, since the shortfall could originate anywhere along the chain that produces the response, from the wave generation, through the drag closure and its operators, to the beam solver and its discretisation. The final test isolates the structural end of that chain. For each condition, the tip displacement of the coupled solution is compared in Figure~\ref{fig:flexveg_quasistatic} with a quasi-static reconstruction, in which the instantaneous actuator-line load is applied segment by segment as static point loads on an analytic linear Euler--Bernoulli cantilever of the same $EI$, with no dynamics at all and no time shift. The reconstruction takes the simulated \emph{load} as given and asks whether the simulated \emph{response} is simply the static deflection under that load, so if the coupled solution carried structural dynamics, resonant amplification, a phase lag behind the load, or artefacts of the beam discretisation and the partitioned coupling, the two curves would separate in amplitude or in phase.

\noindent
In all four panels the reconstruction lands on the coupled solution in phase and captures $97$ to $98\%$ of its amplitude, at the largest wave and at the boundary-probe condition alike. The response is therefore quasi-static at all four conditions, exactly as the a priori criterion of Section~\ref{sec:flexveg_case_selection} predicts, so that classification is confirmed numerically rather than only assumed. The $60$-segment beam, the geometrically exact formulation, and the load transmission of the partitioned coupling add no dynamics and no lag of their own and reproduce the amplitude to within $2$ to $3\%$ even at large deflection. The residual deficits therefore sit on the load side of the chain, which covers the quasi-steady closure, the operators that feed and apply it, and the accuracy with which the local wave is reproduced. The test takes the simulated load as given, so it cannot certify that this load is correct, but the wave verification of Section~\ref{sec:flexveg_results_wave} bounds the wave contribution separately, which leaves the closure as the leading candidate for the remainder.

\subsection{Discussion}
\label{sec:flexveg_discussion}
The framework reproduces the shape and the timing of the motion accurately at all four wave conditions. The simulated motion consists of a single spatial mode, in agreement with the decomposition of the measured motion, it is in phase along the whole length of the blade, it shows the correct asymmetry within each wave period, and its mode shape matches the measurement at every drag coefficient tested. The time-mean position is almost exactly vertical, the expected response of a slender blade to symmetric oscillatory forcing when its motion is quasi-static. The distributed force agrees with the experimental reconstruction along the stem and over the wave period and in magnitude, the total force at the base matches the value inferred from the transducer, and at the upper end of the measured coefficient range the positive tip excursion of the primary case is reproduced to within $1\%$.

\noindent
The single feature that the framework does not reproduce accurately is the amplitude of the oscillation, recovering $65$ to $89\%$ of the measured amplitude at the primary condition across the two coefficients, four stations, and two experimental datasets, and $71$ to $83\%$ of the measured tip response across the three quasi-static conditions once the target-versus-achieved wave difference is removed. Increasing the drag coefficient does not remove the shortfall. Raising $C_{dn}$ from $3.0$ to $4.0$ moves across the whole range the experiment fitted to its own measured force, and even that raises the reproduced amplitude only from $65\%$ to $81\%$, while recovering the full measured amplitude would require $5.0$, about $25\%$ larger than the highest value the experiment reports. The response grows more slowly than the coefficient because two effects oppose it, the stem following the surrounding water more closely as the drag becomes stronger, which reduces the relative velocity that generates the drag, and the force applied back to the fluid removing momentum from the local flow, the wave-crest velocity at the stem being lower at $C_{dn} = 4.0$ than at $3.0$. The missing $15$ to $20\%$ is therefore due to physics that the closure does not represent rather than to an incorrect coefficient, and the most likely source is the fluid-inertia force, which the reference's own force decomposition places below $10\%$ of the maximum force for this stem. The same limitation was identified in the leaflet benchmark at a Strouhal number of $0.5$, and it is now confirmed against a physical experiment rather than a reference simulation. The quasi-static reconstruction locates it, since the structural side reproduces the coupled response to within $2$ to $3\%$ in amplitude and exactly in phase at all four conditions.

\noindent
Two features of the experiment limit the precision of the comparison across conditions. The published wave kinematics describe the waves the flume actually generated, which exceed their nominal targets by as much as $17\%$, whereas the simulations are driven by the target waves, and the forcing-corrected column accounts for this only to first order because the correction is based on the wave kinematics alone. At H070T15 the outlet absorption also operates close to the limit of its shallow-water validity and admits a weak reflected wave, so the comparison there is qualitative, the shortfall having the expected sign and magnitude but not separating cleanly into the part caused by the closure and the part caused by the imperfect wave.

\noindent
It is finally useful to place the result beside the most detailed simulation of the same experiment available. \citet{prueter_vegetation_2025} coupled a geometrically exact rod to a fully resolved two-phase flow and simulated the same campaign at the primary wave condition, using the thinner Mimic~2 blade whose response lies in the dynamic regime deliberately excluded here. Their model resolves the added-mass force that the present closure omits and reproduces the tip position of that blade with an average error of about $6$ to $7\%$ of the stem length, but it does not recover the measured mean forward drift or the second mode of the dynamic response. Because the two studies use different blades the comparison is indicative rather than exact, but for the quasi-static conditions considered here a drag-only actuator-line model reaches an accuracy of the same order as a fully resolved method achieves on the more demanding dynamic blade, with an error budget that separates clearly into the coefficient range, the wave quality, and the missing closure physics. A direct comparison of computational cost would require both studies to be run with the same hardware, tolerances, durations, and accuracy targets, none of which the published work reports. What can be reported is the cost of the present simulations, approximately six days of wall-clock time on $256$ cores for each wave condition, using about $1.35\times10^{7}$ fluid cells and $6.4\times10^{5}$ time steps, with the small step set by the beam solver rather than by the fluid. The actuator-line representation requires neither a body-conforming mesh around the millimetre-scale cross-section nor any mesh motion, so its cost is expected to grow far more slowly than that of a fully resolved method as the number of stems increases, which is the setting the framework is intended for, accepting a known amplitude error of $15$ to $20\%$ per stem in exchange.


\section{Conclusions}
\label{sec:conclusion}

\noindent
A two-way actuator-line coupling was developed between a geometrically exact Simo--Reissner beam, discretised by a cell-centred finite volume method, and an incompressible finite volume flow solver. The fluid velocity is sampled at a prescribed offset upstream of each beam control volume, a quasi-steady Morison-type drag closure evaluates the hydrodynamic line force, and the equal and opposite reaction is projected back into the fluid momentum equation through a normalised Gaussian kernel. Because the two subsystems share one discretisation, one mesh and field infrastructure, one time loop, and one domain decomposition, the coupling reduces to two transfer operators between non-matching meshes rather than an interface between two independently developed codes. The framework was implemented in OpenFOAM. The beam solver was verified in isolation on a large-deflection cantilever, where it reproduced the measured tip deflections to within $1.74\%$ and matched the predictions of an independent structural model at every load level, and the coupling was then verified against two channel-flow benchmarks and validated against a wave-flume experiment on a flexible vegetation stem. The main conclusions are as follows.

\begin{enumerate}[leftmargin=*]

\item The projection operator conserves the transferred resultant force by construction, because normalising the accumulated kernel weight per actuator node makes the discrete volume integral of the applied source equal the total actuator-line load, whatever the mesh, the kernel width, or the kernel truncation. The same guarantee does not extend to angular momentum, and the associated moment discrepancy scales with the kernel width. Both operators are parallel consistent on a domain-decomposed fluid mesh, with a bounding-box rejection test, a seeded cell-walking search, and an octree fallback limiting the per-step search cost to $\mathcal{O}(N_{b}\log N_{c})$, and two reduction-broadcast pairs as the only collective communication required.

\item In the steady confined channel benchmark, with the normal drag coefficient fixed at $C_{dn} = 26.73$ by a separate steady CFD calculation on the equivalent rigid plate, the predicted steady tip deflection was $5.73~\mathrm{mm}$, within the range of the five published solutions that resolve the same cantilever as a two-dimensional continuum, and inside the $4.7$ to $6.0~\mathrm{mm}$ band formed by three of them. Six sensitivity studies established that the result is converged in the discretisation sense. The beam mesh and the time step change it by less than $0.5\%$ across their whole ranges, the fluid mesh by the same margin once the two under-resolved coarsest meshes are excluded, and tightening the structural and pressure--velocity tolerances brings it to within $0.1\%$ of its asymptotic value.

\item The kernel width and the upstream sampling distance were identified as the two operator settings with a physical rather than a purely numerical influence on the response, changing the steady tip deflection by $20\%$ and $31\%$ respectively across their admissible ranges, and in a confined channel the two act together as an effective calibration of the quasi-steady closure to the local blockage. The pair selected in the channel benchmark was transferred unchanged to the leaflet benchmark, and the same relative kernel-sizing rule was applied in the wave flume, so what the later cases test is the transferred settings rather than the settings themselves.

\item For the leaflet in a sinusoidal channel flow, the drag coefficient was fixed by two steady CFD calibrations, one in free stream and one in the channel, and the conditions under which the in-channel coefficient can be identified directly with the closure input were derived analytically and confirmed to within $0.6\%$ by a probe-line measurement. Across the four Reynolds and Strouhal combinations of the reference study, the forcing period was reproduced exactly, the peak tip excursion was over-predicted at a Strouhal number of $1$ at both Reynolds numbers, and it was under-predicted by $30\%$ at a Strouhal number of $0.5$. That under-prediction was traced to the absence of the added-mass and history terms in the closure rather than to the calibration or the time-step resolution, since an unsteady recalculation of the rigid-plate drag at the two Keulegan--Carpenter numbers returned $C_{d} \approx 107$ in both.

\item In the wave-flume validation, with the drag coefficient taken from the fits published alongside the experiment and the coupling-operator settings inherited from the channel benchmarks, the simulated stem motion reproduced the measured mode shape, the single-mode character of the response, the intra-wave asymmetry, the distributed force along the stem, and the base force. At the upper published coefficient the positive tip excursion matched the measurement to within $1\%$. The oscillation range was recovered to $65$ to $89\%$ at the primary condition across the propagated coefficient range, and to $71$ to $83\%$ across three wave conditions once the difference between the target wave and the wave the flume achieved was removed, spanning a threefold change in Keulegan--Carpenter number and tip excursions up to a third of the stem length, with no additional degradation at large deflection.

\item A quasi-static reconstruction test, in which the instantaneous actuator-line load was applied statically to an analytic cantilever of the same bending stiffness, reproduced the coupled response to $97$ to $98\%$ in amplitude and exactly in phase at every wave condition. The structural model, its discretisation, and the transmission of the load through the partitioned coupling therefore add no dynamics and no lag of their own, and the residual amplitude deficit sits on the load side of the chain. That $15$ to $20\%$ deficit is the quantified cost of the omitted fluid inertia rather than an error in the drag coefficient, since a deliberately out-of-band run showed that closing the gap requires a coefficient about a quarter above the largest value the experiment supports.

\end{enumerate}

\noindent
The quasi-steady drag-only closure is the leading limitation of the framework as it stands, quantified twice, first against a reference numerical solution at a Strouhal number of $0.5$ and then against a physical experiment beyond the quasi-static regime limit. It also requires a coefficient that must be re-established whenever the Reynolds number or the blockage changes, and no such preparatory calculation is possible in a flow that never reaches a steady state. The kernel width and the sampling distance were selected in the channel benchmark with the reference result in view, so the agreement reported there is an assessment against other numerical solutions with two calibrated operator settings rather than a fully blind prediction, and the wave-flume case reused that kernel sizing rather than repeating the study. Three properties of the reference data limit the precision of the validation, namely that the published wave kinematics describe waves exceeding their nominal targets by as much as $17\%$, that at one condition the outlet absorption operates close to the limit of its shallow-water validity and admits a weak reflected wave, and that the blade carries a documented manufacturing pre-tension biasing its measured mean position. The validation also covers a single isolated stem, and canopy configurations, in which the wake of one member loads the next, were not tested. A drag-only regularised line force cannot generate vortex-induced lift either, so vortex-induced vibration of cables and risers lies outside the reach of this closure regardless of the coupling scheme.

\noindent
The single change that would remove the dominant error is the addition of the Morison inertia term of Eq.~\eqref{eq:fext_addedmass} to the line force. The fluid acceleration would be sampled with the operator that already supplies the velocity, and the beam-acceleration part of the added-mass force would be treated on the diagonal of the structural system rather than explicitly, which is also what stabilises partitioned coupling at density ratios near unity \citep{causin_addedmass_2005,forster_artificial_2007}. The $30\%$ deficit at a Strouhal number of $0.5$ and the $15$ to $20\%$ amplitude deficit in the wave flume are the two targets that extension would have to meet. The sampling operator is the next candidate, because sampling a single fluid cell at a prescribed upstream offset is discontinuous in the beam position and is what makes the frozen-force treatment of Section~\ref{sec:almParallelSearch} necessary. A kernel-weighted volume average using the same Gaussian that projects the force \citep{churchfield_advanced_2017,merabet_parametric_2019,muscari_effective_2024}, or a Lagrangian-averaged sample \citep{xie_actuatorline_2021}, would make the sampled velocity smooth in the beam position, remove the Newton-loop oscillation at its root instead of suppressing it, make resampling within the outer corrector loop safe, and remove the ambiguity between the kernel width and the sampling distance identified in Section~\ref{sec:beamtunnel_sensitivity}. Its counterpart is that sampling at the line reads the self-induced velocity deficit of the smeared force, for which lift lines use filtered lifting-line theory \citep{martinez_tossas_filtered_2019} and vortex-based smearing corrections \citep{meyerforsting_vortexbased_2019,kleine_noniterative_2023}, so a drag line would need either an upstream shift of the kernel centroid or an explicitly calibrated deficit correction. With smooth sampling in place, the beam solve could be re-fired at every outer corrector under Aitken dynamic relaxation on the interface force \citep{kuttler_fixedpoint_2008}, or with quasi-Newton acceleration \citep{degroote_performance_2010}, which would replace the present staggered scheme by a strongly coupled iteration, while the rod formulations of \citet{tschisgale_immersed_2020,tschisgale_large_2021} reach added-mass-stable coupling without any global iteration.

\noindent
Three smaller changes to the projection also follow from the results reported here. Summing a segment-integrated three-dimensional Gaussian over all segments within the cutoff, which has a closed form in the tangential coordinate, would remove the seams the present nearest-segment assignment creates on strongly reconfigured stems. Sizing the kernel locally as $\varepsilon(\mathbf{x}) = \max(2\Delta x_{\mathrm{local}}, \varepsilon_{\mathrm{phys}})$ would keep the projection resolved on graded meshes such as the wave-flume mesh. A one-sided or reflected kernel near a boundary would remove the truncation artefact identified in the kernel-width study, where the kernel mass falling outside the domain is currently redistributed along the whole node support. The projection limits its distance evaluations to the cells inside the inflated bounding box of the beam, but inside that box every cell is still tested against every actuator segment, and the box of a long or strongly curved structure can enclose a large part of the domain. Restricting the sweep to a band of cells within $3\varepsilon$ of the beam polyline, found through a spatial index over the segments, is therefore a prerequisite for the canopy and net configurations the method is intended for. Validating those configurations against canopy-scale measurements is the natural next application, and the per-stem accuracy quantified here is what a canopy result would inherit.


\begin{appendices}
\section{Mesh sensitivity of the wave-flume case}
\label{app:flexveg_mesh}

\noindent
The wave-flume case carries two resolution requirements at once, because the mesh has to propagate the incident wave without distorting it and it also has to resolve the Gaussian kernel through which the stem acts on the fluid. This appendix examines both by repeating the largest-deflection condition H110T20, at its production coefficient $C_{dn} = 2.4$, on three successively refined meshes. Everything outside the mesh is held fixed, namely the kernel width at $\varepsilon = 0.015~\mathrm{m}$, the drag coefficients, the sampling offset, the force relaxation, the wave theory and its ramp, the discretisation schemes, the solver tolerances, and the Courant limit. Table~\ref{tab:flexveg_mesh} lists the three meshes. The middle one is the production mesh of Section~\ref{sec:flexveg_setup}, on which every result reported in that section is computed, and the coarse and fine meshes are obtained from it by dividing and multiplying the cell count in each direction by $\sqrt{2}$. Each level therefore carries about $2.8$ times the cells of the one below, the three spanning a factor of eight overall, and the vertical resolution of the largest wave rises from $19$ to $38$ cells per wave height. Because $\varepsilon$ is a physical width and is held fixed, the ratio $\varepsilon/\Delta x$ rises with refinement from $1.77$ to $3.54$, so the study varies the kernel-to-cell ratio at the same time as the mesh.

\noindent
Each case contributes its last three complete wave periods, at $26$ to $32~\mathrm{s}$ on the coarse mesh, $24$ to $30~\mathrm{s}$ on the baseline, and $18$ to $24~\mathrm{s}$ on the fine mesh, the last because the fine run reached $t = 25.6~\mathrm{s}$ within its wall-clock allocation. Every window starts at an integer multiple of the wave period and the paddle forcing is exactly periodic, so the three windows are phase-locked and overlay directly. Peaks and ranges are read after band-limited interpolation of each whole-period window, because the output interval of $0.1~\mathrm{s}$ gives only twenty samples per period and places up to $1.2\%$ of sampling bias on a raw extreme, which is the same order as the mesh differences being measured.

\begin{table}[!ht]
\centering
\small
\caption{The three meshes of the wave-flume sensitivity study, run at condition H110T20 with $C_{dn} = 2.4$ and $\varepsilon = 0.015~\mathrm{m}$ held fixed. The baseline is the production mesh of Section~\ref{sec:flexveg_setup} and the other two are that mesh scaled by $\sqrt{2}$ in each direction. Cell counts are for the whole flume and the spacings are those over the plateau. $H/\Delta z$ is the number of cells per wave height and $\varepsilon/\Delta x$ the kernel width in streamwise cells. The last four columns are the base-force range $\Delta F$, the tip oscillation range $\Delta x_s$, the crest elevation $\eta_c$ at the stem, and the mean water level $\bar{\eta}$ at the stem, each taken over the last three complete wave periods. Both elevations are referenced to the still water level, whereas Section~\ref{sec:flexveg_results_wave} reports elevations about the local mean, so the two differ by $\bar{\eta}$.}
\label{tab:flexveg_mesh}
\footnotesize
\setlength{\tabcolsep}{2.5pt}
\renewcommand{\arraystretch}{1.15}
\begin{tabular}{l c c c c c c c c}
\toprule
Mesh & Cells & $\Delta x \times \Delta y \times \Delta z$ (mm) & $H/\Delta z$ & $\varepsilon/\Delta x$ & $\Delta F$ (mN) & $\Delta x_s$ (mm) & $\eta_c$ (mm) & $\bar{\eta}$ (mm) \\
\midrule
Coarse   & $4.8\times10^{6}$  & $8.5 \times 56.6 \times 5.8$ & $19$ & $1.77$ & $173.2$ & $99.6$  & $65.9$ & $+1.0$ \\
Baseline (prod.) & $1.35\times10^{7}$ & $6.0 \times 40.0 \times 4.1$ & $27$ & $2.50$ & $170.3$ & $97.7$  & $74.4$ & $+9.1$ \\
Fine     & $3.8\times10^{7}$  & $4.2 \times 28.3 \times 2.9$ & $38$ & $3.54$ & $173.6$ & $100.4$ & $73.3$ & $+6.8$ \\
\bottomrule
\end{tabular}
\end{table}

\noindent
The load and the response are converged from the baseline mesh upward. The base-force range is $173.2$, $170.3$ and $173.6~\mathrm{mN}$ on the three meshes, so the baseline sits $1.9\%$ below the fine mesh, and the tip oscillation range is $99.6$, $97.7$ and $100.4~\mathrm{mm}$, a baseline deviation of $2.7\%$. The range of the upstream-sampled velocity that drives the drag agrees to within $2\%$ on all three meshes. One qualification belongs with those numbers. The per-period tip range is constant to $\pm0.1\%$ on the coarse and the baseline meshes, but the fine run is still drifting downward by about $0.4\%$ per period, so its own range is converged only to roughly $\pm1\%$, which is the same order as the deviations quoted. That drift runs toward the baseline value, so a fully settled fine run would improve the agreement rather than worsen it.

\begin{figure}[!ht]
\centering
\includegraphics[width=0.70\textwidth]{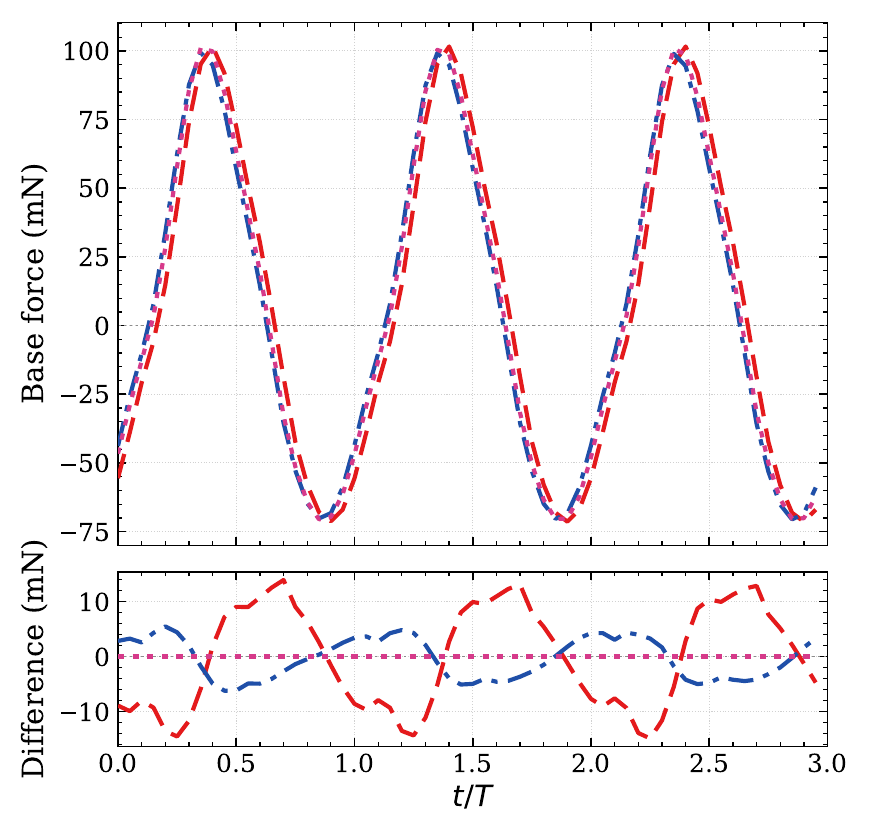}\\[0.5em]
\includegraphics[width=0.95\textwidth]{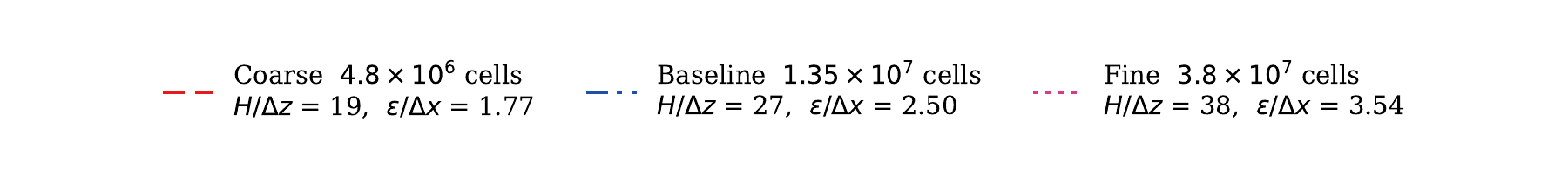}
\caption{Horizontal base force on the stem over three developed wave periods at condition H110T20, on the three meshes, with the difference from the fine mesh below. The legend strip applies to this figure and to Figures~\ref{fig:flexveg_mesh_eta} and~\ref{fig:flexveg_mesh_env}. The coarse curve differs from the fine one by up to $13~\mathrm{mN}$, which is the signature of its $8.85^{\circ}$ phase lag rather than of a load-magnitude error.}
\label{fig:flexveg_mesh_force}
\end{figure}

\noindent
Figure~\ref{fig:flexveg_mesh_force} shows that the visible difference on the coarse mesh is phase rather than magnitude. In the difference panel the coarse curve departs from the fine one by up to $13~\mathrm{mN}$ on a range of $173~\mathrm{mN}$, which reads as a substantial load error, but its range is within $0.24\%$ of the fine mesh and it is the baseline that lies furthest from the fine mesh in magnitude. The whole of the visible difference is a first-harmonic phase lag of $8.85^{\circ}$, which is $2.5\%$ of a period and accounts for $2 \times 81.9~\mathrm{mN} \times \sin(8.85^{\circ}/2) = 12.6~\mathrm{mN}$. The same lag of $8.7$ to $9.6^{\circ}$ appears in the surface elevation, in the sampled velocity, in the base force, and in the tip displacement, and the lags between those four quantities are mesh-independent to under $1^{\circ}$. The coarse mesh therefore propagates the wave slightly too slowly, and the coupling introduces no mesh-dependent lag of its own.

\noindent
The free surface is where the three meshes differ most, and the difference sits almost entirely in the mean level rather than in the wave itself. The mean water level at the stem is $+1.0$, $+9.1$ and $+6.8~\mathrm{mm}$ on the three meshes, so the coarse mesh develops essentially no wave setup, and this is the least converged quantity of the study. Referenced to the still water level the coarse mesh then appears to lose $10.1\%$ of its crest, at $65.9$ against $73.3~\mathrm{mm}$, but most of that gap is the missing setup. Referenced to the local mean, which is how a wave gauge reports an elevation and how Section~\ref{sec:flexveg_results_wave} reports it, the crest runs $64.9$, $65.4$ and $66.5~\mathrm{mm}$ and the trough $-42.7$, $-44.2$ and $-44.1~\mathrm{mm}$, so the wave at the stem is reproduced to within $2.4\%$ on every mesh. The crest fraction on that datum is $0.627$, $0.624$ and $0.627$, so the crest-trough asymmetry of the Stokes wave is captured identically at all three resolutions. Figure~\ref{fig:flexveg_mesh_env} shows the same picture along the flume, the coarse envelope lying below the other two at the crest and at the trough alike, which is a shift of the mean rather than a distortion of the wave.

\begin{figure}[!ht]
\centering
\includegraphics[width=0.67\textwidth]{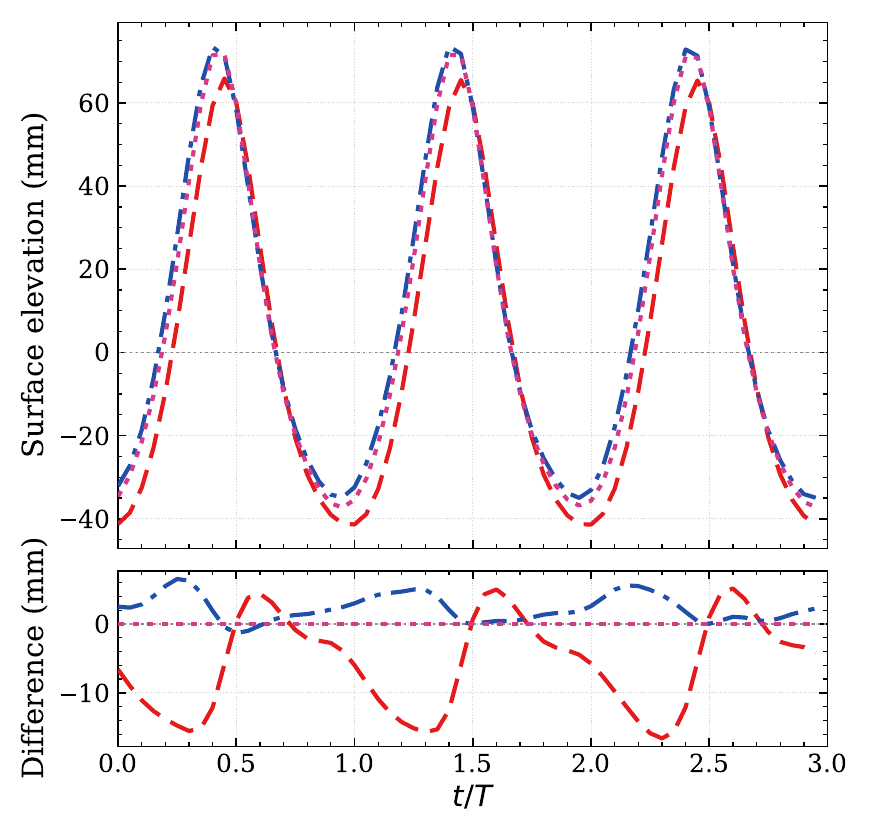}
\caption{Free-surface elevation at the stem over three developed wave periods on the three meshes, referenced to the still water level, with the difference from the fine mesh below. The coarse curve sits below the other two through the whole cycle, which is its missing wave setup; about the local mean the three crests agree to within $2.4\%$. Line styles follow the legend of Figure~\ref{fig:flexveg_mesh_force}.}
\label{fig:flexveg_mesh_eta}
\end{figure}

\begin{figure}[!ht]
\centering
\includegraphics[width=0.67\textwidth]{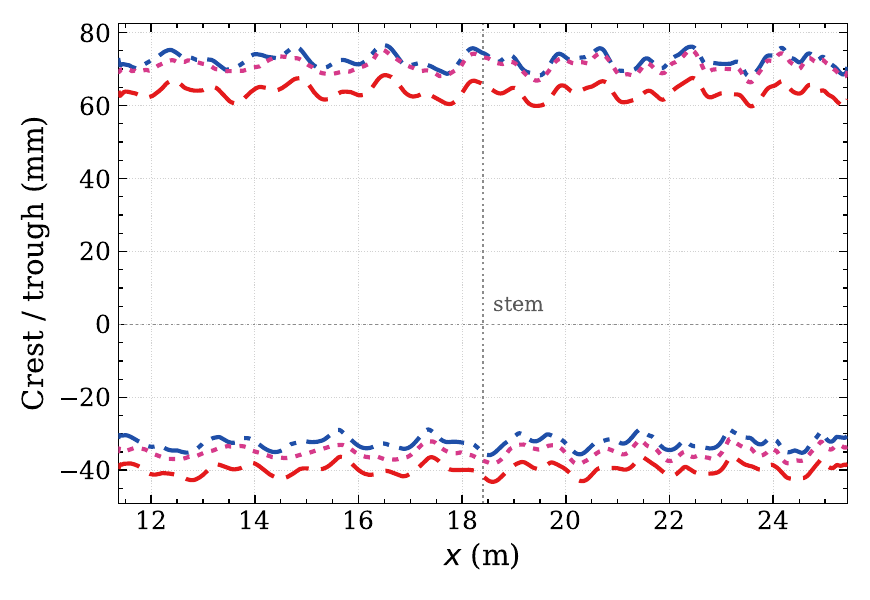}
\caption{Crest and trough envelope along the plateau on the three meshes, referenced to the still water level, with the stem location marked. The coarse envelope lies below the other two at the crest and at the trough alike, at every station rather than only at the stem, which is the signature of its missing wave setup. The $\pm5\%$ modulation along the flume is the standing-wave pattern of a weak outlet reflection. Line styles follow the legend of Figure~\ref{fig:flexveg_mesh_force}.}
\label{fig:flexveg_mesh_env}
\end{figure}

\noindent
The flume-averaged wave height converges monotonically, at $103.5$, $104.7$ and $106.0~\mathrm{mm}$ against the $110~\mathrm{mm}$ target, so all three meshes fall short of the target by $3.6$ to $5.9\%$. That deficit is not numerical damping. The local wave height is flat along the flume within a $\pm5\%$ modulation whose measured spatial period is $1.92~\mathrm{m}$ against the $1.942~\mathrm{m}$ of $\lambda/2$ from linear dispersion, which is the signature of a weak reflection off the outlet with an amplitude coefficient of about $0.05$, the same on all three meshes.

\noindent
Two limits of the study should be stated. Only three levels are available and the fine run is not fully settled, so no Richardson extrapolation was attempted. The refinement is also simultaneous in the three directions with $\varepsilon$ held fixed, so the vertical resolution of the free surface is not separated from the streamwise celerity error or from the change in $\varepsilon/\Delta x$, and an isolated vertical refinement is the run that would separate them. What the study does establish is that the load and the response, which are the quantities compared against the experiment in Section~\ref{sec:flexveg}, are insensitive to the mesh at the $2\%$ level over an eightfold change in cell count, and that the wave at the stem is reproduced to within $2.4\%$ on all three meshes once the mean level is removed. The quantity that does need the resolution is the wave setup, which the coarsest mesh fails to develop, and which enters the response only through the $1\%$ change it makes to the orbital velocity.

\end{appendices}


\backmatter

\bmhead{Data availability}
\noindent
The coupling library presented here is publicly available at
\url{https://github.com/solids4foam/moorFV}, and the finite volume beam solver it builds on
is available at \url{https://github.com/solids4foam/beamFoam}.

\bmhead{Declarations}
\noindent
The authors declare no competing interests.

\bibliography{bibliography}

\end{document}